\documentstyle[12pt,epsfig]{article}

\newcommand{\be}{\begin{equation}}
\newcommand{\ee}{\end{equation}}
\newcommand{\bea}{\begin{eqnarray}}
\newcommand{\eea}{\end{eqnarray}}
\def\nn{\nonumber}

\def\evsq{{${\rm eV^2}$ }}
\def\L{{\rm L }}

\def\pmutau{{${\rm P_{\mu \tau}}$ }}
\def\pmue{{${\rm P_{\mu e}}$ }}
\def\pmumu{{${\rm P_{\mu \mu}}$ }}
\def\pmutaum{{${\rm P_{\mu \tau}^{m}}$ }}
\def\pmuem{{${\rm P_{\mu e}^{m}}$ }}
\def\pmumum{{${\rm P_{\mu \mu}^{m}}$ }}
\def\numutonutau{{${\rm {\nu_\mu \to \nu_\tau}}$ }}
\def\numutonue{{${\rm {\nu_\mu \to \nu_e}}$ }}
\def\numutonumu{{${\rm {\nu_\mu \to \nu_\mu}}$ }}
\def\nuetonutau{{${\rm {\nu_e \to \nu_\tau}}$ }}

\newcommand{\st}{\mbox{$\sin^{2}2\theta$}}

\begin{document}

\begin{flushright}
HRI-P-04-10-002 \\
hep-ph/0411252
\end{flushright}


\begin{center}

{\Large \bf Earth Matter Effects at Very Long Baselines and the
Neutrino Mass Hierarchy}
\\[20mm]

\vspace*{1.cm}

{\begin{center} {{\sf
                Raj Gandhi $^{a, \,\!\!\!}$
\footnote[1]{\makebox[1.cm]{Email:}
                \sf raj@mri.ernet.in},
                Pomita Ghoshal $^{a, \,\!\!\!}$
\footnote[2]{\makebox[1.cm]{Email:}
                \sf pomita@mri.ernet.in},
                Srubabati Goswami $^{a,b, \,\!\!\!}$
\footnote[3]{\makebox[1.cm]{Email:}
                \sf sruba@mri.ernet.in,sruba@ph.tum.de},
           Poonam Mehta $^{a,c, \,\!\!\!}$
\footnote[4]{\makebox[1.cm]{Email:}
                \sf mpoonam@mri.ernet.in,
                    poonam.mehta@weizmann.ac.il}
 \&  S Uma Sankar $^{d, \,\!\!\!}$
\footnote[5]{\makebox[1.cm]{Email:}
                \sf uma@phy.iitb.ac.in}
                }}
\end{center}}

\vskip 0.5cm

{\it
\begin{center}
       $^{a}$Harish-Chandra Research Institute,
 Chhatnag Road, Jhunsi,\\
Allahabad 211 019, India\\[4mm]
       $^{b}$Technische Universit\"at M\"unchen,
       James--Franck--Strasse,\\
       D--85748 Garching, Germany\\[4mm]
       $^{c}$Department of Particle Physics,
Weizmann Institute of Science,\\
Rehovot 76 100, Israel
\\[4mm]
       $^d$Department of Physics, Indian Institute of Technology,
Powai,\\
Mumbai 400 076, India.
\end{center}}
\end{center}

\vskip 2cm

\begin{abstract}
We study matter effects which arise in the muon neutrino oscillation
and survival probabilities relevant to  atmospheric neutrino and
very long baseline ($> 4000$ Km) beam experiments. The
inter-relations between the three probabilities ${\mathrm {P_{\mu
e}}}$, \pmutau and ${\mathrm {P_{\mu \mu}}}$ are examined. It is
shown that large and observable sensitivity to the neutrino mass
hierarchy  can be present in \pmumu and ${\mathrm {P_{\mu \tau}}}$.
We emphasize that at baselines $>$ 7000 Km, matter effects in
\pmutau are important under certain conditions and can be large. The
muon survival rates in experiments with very long baselines thus
depend on matter effects in both \pmutau and ${\mathrm {P_{\mu
e}}}$. We also indicate where these effects provide sensitivity to
$\theta_{13}$ and identify ranges of energies and baselines where
this sensitivity is maximum.
The effect of parameter degeneracies in the three probabilities at these baselines
and energies is studied in detail and large parts
of the parameter space are identified which are free from these degeneracies.
In the second part of the paper, we focus on using the matter effects
studied in the first part as a means of determining the mass hierarchy
via atmospheric neutrinos. Realistic event rate calculations are performed for a
charge discriminating 100 kT iron calorimeter which demonstrate the
possibility of realising this very important goal in neutrino physics.
It is shown that for atmospheric neutrinos, a careful selection of energy and baseline
ranges is necessary in order to obtain a statistically significant signal,
and that the effects are largest in bins where matter effects
in both \pmue and \pmutau combine constructively.
Under these conditions, up to a $4\sigma$ signal for matter effects
is possible (for $\Delta_{31}>0$) within a timescale appreciably shorter
than the one anticipated for neutrino factories.

\vskip 1cm


\end{abstract}

\vskip 1cm \setcounter{footnote}{0}

\section{Introduction}

It is fair to say that over the last few years there
has been a qualitative shift in the  nature of the  goals to be pursued
in neutrino physics. This has been the result of steadily accumulating
evidence in favour of non-zero neutrino mass and flavour oscillations.
Results from atmospheric neutrino experiments
\cite{k,sk1,sk2,nu04a,mac,soudan,imb}, corroborated by the
accelerator beam based {\bf KEK to Kamioka (K2K)} experiment
\cite{k2k} have provided firm evidence for
${\nu_\mu\rightarrow\nu_\tau}$ oscillations with maximal (or almost
maximal) mixing. The solar neutrino results
\cite{nu04b,btc,jna,wha,mal,sfu,qra,sna}, when combined with the
results of the reactor based {\bf KamLAND} experiment \cite{keg},
have established the LMA-MSW solution \cite{msw} as the most
favoured explanation for the solar neutrino deficit. For recent
global analyses of solar, reactor, accelerator and atmospheric,
data, see \cite{sruba,bahcall,maltoni,concha}. In Table
\ref{tablelimits}, we summarize the best-fit values and $3\sigma$
intervals of allowed values of important oscillation parameters
gleaned from experiments so far obtained from global three flavour
neutrino analysis \cite{sruba,maltoni}.

%
%
\begin{table}[htb]
\begin{center}
\begin{tabular}{|| c || c || c ||}
\hline \hline
&&\\
\hspace{0.5cm}{\sf {Parameter}}\hspace{0.5cm} &
\hspace{0.5cm}{\sf {Best-fit value}}\hspace{0.5cm} &
\hspace{0.5cm}{\sf {$3\sigma$ allowed range}}\hspace{0.5cm}
\\
&& \\
        \hline\hline
        $\Delta_{21} [10^{-5} {\mathrm{eV}}^2]$ & 8.3 & 7.2 -- 9.1   \\
        $\Delta_{31} [10^{-3} {\mathrm{eV}}^2]$ & 2.2 &  1.4 -- 3.3 \\
        $\sin^2 \theta_{12} $ & 0.30 &  0.23 -- 0.38 \\
        $\sin^2 \theta_{23} $ & 0.50 &  0.34 -- 0.68 \\
        $\sin^2 \theta_{13} $ & 0.00 &  $\le$ 0.047
          \\
          \hline
\hline
\end{tabular}
\caption[]{\footnotesize{Best-fit values and $3\sigma$ intervals for
three flavour neutrino oscillation parameters from global data
including solar, atmospheric, reactor (KamLAND and CHOOZ) and
accelerator (K2K) experiments \cite{sruba,maltoni}. Here
$\Delta_{ij} \equiv m^2_i - m^2_j$. }} \label{tablelimits}
\end{center}
\end{table}

The shift has been from a search for understanding the particle
physics and/or the astrophysics driving the solar and atmospheric neutrino
deficits to one where we seek to make increasingly  precise
measurements of neutrino mass and mixing matrix parameters.
Experiments planned to yield results over the next ten to fifteen years thus
reflect this change of emphasis. A significant number of the planned
projects are long baseline\footnote{By ``long baseline'' we actually mean the
L/E range of about  50-500 Km/GeV.
For accelerator experiments, this translates to baselines
conventionally termed ``long'', but for the lower reactor neutrino energies, the
 baselines are actually 1-2 Km.}
endeavours using either
(a) {\sl a conventional proton beam} colliding with a target
to produce pions which then decay to give
muon neutrinos, or (b) {\sl superbeams}, which are essentially
technologically upgraded versions of present conventional
beams, or, finally, (c)
{\sl  reactor sources} with both near and far detectors for reduced
 systematic errors.

To begin with, we enumerate and briefly describe the
planned projects in the three categories above which are expected to
give results over the next decade or decade and a half. For a recent detailed
study of their capabilities  we refer the reader to \cite{lin} and
references therein.

\begin{itemize}

\item
{\bf Conventional} beam experiments will have as their primary
goal the improvement
in the precision of the atmospheric oscillation parameters,
$\Delta_{31}$ and $\st_{23}$.
The {\bf Main Injector Neutrino Oscillation Search (MINOS)} project \cite{mil},
 located in the US,  with a baseline from
Fermilab to
 Soudan mine in the US state of Minnesota (735 Km),  will
utilize a 5.4 kT magnetized iron calorimeter and initially an
$<{\mathrm{E_{\nu}}}>\simeq$ 3 GeV  to primarily measure muon
 survival events. By measuring the absolute event rate
and the energy
 distribution for muons produced via charged current (CC)
scattering, $\Delta_{31}$
may be determined to within about 10-20$\%$
of the current best
 fit value. In addition, by measurements in the electron
appearance channel,
MINOS may provide a possibly improved upper bound on
$\st_{13}$, by a
 factor of 1.5 to 2.
 In Europe, the two experiments planned in this category
are the {\bf Imaging Cosmic And Rare Underground Signals (ICARUS)}
 project \cite{pap}, a 2.35 kT liquid Argon detector, and
the {\bf Oscillation Project with Emulsion-tRacking Apparatus (OPERA)}
\cite{ddu}, a 1.65 kT emulsion cloud chamber which will
ride the 732 Km CERN to Gran Sasso
baseline and will be powered by the CNGS beam.
The beam energy is higher,
resulting in neutrinos which are kinematically
capable of producing $\tau$
leptons ($<{\mathrm{E_{\nu}}}>\simeq$ 17 GeV). In addition,
these detectors can identify muons and electrons.
It is thus anticipated that
in addition to an improved precision in  $\Delta_{31}$, an
improvement in the current bound on $\st_{13}$
may be possible at the same levels as anticipated
with MINOS.

\item
{\bf Superbeam} experiments utilise the same basic principle
used in conventional
 beam experiments, but incorporate substantial
technological
 improvements.
This includes higher power beams and the idea of an ``off-axis''
location for the detector. One of the planned projects is the 295 Km
{\bf Tokai to Kamioka (T2K)} project \cite{yit}, with {\bf
Super-Kamiokande (SK)} as the far detector of total mass 50 kT and
$<{\mathrm{E_{\nu}}}>\simeq$ 0.76 GeV. Similarly, the {\bf NuMI
Off-Axis $\nu_{e}$ Appearance experiment (NO$\nu$A)} \cite{day} is
planned for location in the US, with a probable 810 Km
 baseline terminating in a 30 kT
calorimeter and $<{\mathrm{E_{\nu}}}>\simeq$ 2.2 GeV.
 Both experiments primarily aim at
heightened sensitivity to $\st_{13}$ via the electron
 appearance channel.
It is anticipated that the upper bound on this parameter
 will be improved by a factor of four over a five  year
running period.

As discussed in \cite{lin}, the combination of conventional
and superbeam
 experiments over the next 10-12 years will improve the
precision
 on $\Delta_{31}$ by an order of magnitude, while the improvement
 in $\st_{23}$ will be much more modest, i.e by a factor
of about two.

\item
Planned experiments in the {\bf reactor} category \cite{reactors}
include {\bf KASKA} in Japan \cite{kaska},
one in Diablo Canyon, USA \cite{mhs} and
another in Daya Bay, China \cite{mhs} and an upgraded version of
{\bf CHOOZ} \cite{map}, called {\bf Double-CHOOZ}
\cite{dcho} in France.
Reactor experiments detect the electron anti-neutrino flux
by the inverse beta-decay process, and focus on improved
measurements of $\st_{13}$, using a near detector to lower
systematics. A factor of six improvement in the present upper
bound on this parameter is expected.

\end{itemize}

To summarize, the above experiments will, over the next 10-12 years,
greatly improve the precision on  $\Delta_{31}$,
effect a very modest improvement in the existing measurements for
$\st_{23}$, and improve the upper bound on $\st_{13}$
by a factor of two to six, depending on the experiment. We
note that given their insensitivity to matter
effects\footnote{All baselines for the listed experiments are
around or below 800 Km.},
they will not be able to conclusively determine the sign of
$\Delta_{31}$ and thus will not establish whether neutrino masses
follow a normal hierarchy or an inverted one{\footnote{Combined results
from T2K \cite{yit} and the NuMI Off-Axis \cite{day} experiment may be able
to infer the neutrino mass hierarchy. However, first results
on this would be available about 6 years after the NO$\nu$A far detector
is completed, i.e., around 2017.
In addition, inferring the hierarchy may be complicated by ambiguities
resulting from uncertainties in $\st_{13}$ and $\mathrm{\delta_{CP}}$.}}.
This may thus leave one of the major questions of neutrino physics
unresolved over the timescale considered here (10-12 years).
Besides requiring a baseline long enough to allow matter
  oscillations to develop, the resolution of this issue,
in general, requires a detector which can distinguish the sign of the
lepton produced in a CC interaction. Over a longer term,  progress in this
  direction may be possible via the proposed mega-ton water Cerenkov detectors \cite{uno}.
Even though they do not have the capability to distinguish the charge
  on an event by event basis, once enough statistics are collected
  ($\approx 2$ mega-ton year exposures), it may be possible to use the
differences in total and differential CC cross-sections between
neutrinos and anti-neutrinos to obtain a statistical
determination of the sign of  $\Delta_{31}$ \cite{bern,kaj,smi}.
Additionally, if neutrino factories are built, they will be able to
resolve this question in a definitive fashion \cite{nufac}.
This paper examines  the possibility of resolving
this issue using atmospheric muon neutrinos over the
same time-frame as the long baseline program described above
(shorter compared to the time scale of neutrino factories).

In the first part of this paper, we study  the physics related to
the sensitivity of the muon neutrino survival probability to matter.
In general, \pmumu = 1 - \pmue - ${\mathrm {P_{\mu \tau}}}$, and one
normally assumes that the dependence of \pmumu on matter effects
arises primarily from its \pmue component while matter effects in
\pmutau remain small. This is because $\nu_{\mu}$ and $\nu_{\tau}$
have the same coherent interactions with the Earth's matter. In
\cite{us1} recently, it was shown that contrary to expectations, the
\numutonutau oscillation probability {\sl{\pmutau can also undergo
significant change}} (for instance, a reduction as high as $\sim$
70\% or an increase of $\sim$ 15\%) {\sl{at very long baselines}}
($> 6000$ Km) over a broad band of atmospheric (GeV) neutrino
energies {\sl{due to matter effects}}. Given the fact that the
atmospheric neutrino flux is a sharply falling function of energy in
the few GeV range (${\mathrm{\frac{dN_\nu}{dE}}} \propto
{\mathrm{E}}^{-3.6}$) and that the production of $\tau$ leptons via
CC is kinematically suppressed here, a direct observation of this
effect via appearance in an atmospheric neutrino experiment may be
difficult. However, large matter effects in \pmutau can cause
correspondingly large changes in ${\mathrm {P_{\mu \mu}}}$, and we
explore the consequences of this in our paper.
 In particular, in the next section, we systematically  discuss the
inter-relation between the three probabilities
${\mathrm {P_{\mu e}}}$, \pmumu and \pmutau and
study the ranges of energies and pathlengths where they
combine in a synergistic manner to give large effects.
We also examine the problem of
parameter degeneracies in these
oscillation probabilities with reference to very
long baselines.

The remaining part of the
paper studies observational consequences of
these effects for
muon survival rates with particular emphasis on
resolving the hierarchy issue in an atmospheric neutrino
setting, using a detector capable of lepton
charge discrimination.
Earlier studies of this were undertaken in
\cite{ban,berna,tab} and,
more recently, in \cite{pet,indu,sawg}.
We also note that the muon survival rate is a
major constituent of the signal in the existing SK
\cite{sk1,sk2} detector, the planned mega-ton Water
Cerenkov detectors like {\bf{Underground Nucleon
decay and Neutrino Observatory (UNO)}} or
{\bf{Hyper-Kamiokande}} \cite{yit,uno,bnl-hs}, and
several detectors considered for future long baseline
facilities \cite{rubbia}, for which the discussion below may be of relevance.

\section{Discussion of the Matter Probabilities :
${\mathrm {P^{m}_{\mu e}}}$, ${\mathrm {P^{m}_{\mu \tau}}}$ and
\pmumum} \label{sec:prob}

Analytical expressions for oscillation probabilities for neutrino
propagation in vacuum and Earth's matter have been extensively
studied in the literature\cite{bar,ki2,bil,zag,pant,fogli,oh2,xin,
fre1,oh1,otasato,shrock,barger,kim,bcr,har,akh}.
The neutrino flavour states are linear superpositions of the mass
eigenstates with well-defined masses :
\begin{equation}
\vert \nu_\alpha \rangle \;= \sum_i U_{\alpha i}\;\vert \nu_i \rangle \ ,
\end{equation}
where $U$ is a $3 \times 3$ unitary matrix known as the
Pontecorvo-Maki-Nakagawa-Sakata mixing matrix. We use the standard
parametrization of $U$ in terms of three mixing angles and a
Dirac-type phase (ignoring Majorana phases), {\em viz.,}
\be U =
\pmatrix{
          c_{12}c_{13} & s_{12}c_{13} & s_{13}e^{-i\delta_{\rm CP}}  \cr
 -c_{23}s_{12} - s_{23}s_{13}c_{12}e^{i\delta_{\rm CP}} & c_{23}c_{12} -
s_{23}s_{13}s_{12}e^{i\delta_{\rm CP}}&  s_{23}c_{13}\cr
  s_{23}s_{12} - c_{23}s_{13}c_{12}e^{i\delta_{\rm CP}}& -s_{23}c_{12} -
c_{23}s_{13}s_{12}e^{i\delta_{\rm CP}} & c_{23}c_{13} \cr} \ ,
\label{mns}
\ee
where ${\mathrm{c}}_{ij} \equiv \cos \theta_{ij}$,
${\mathrm{s}}_{ij} \equiv \sin \theta_{ij}$ and $\delta_{\rm CP}$ is the CP-violating
phase.

For the purpose of our discussion in the first three subsections
here, besides the approximation of constant density, we set
$\Delta_{21}\equiv\Delta_{\mathrm {sol}}=0$. Consequently the mixing
angle $\theta_{12}$ and the CP phase $\delta_{\rm CP}$ drop out of
the oscillation probabilities. This approximation simplifies the
analytical expressions and facilitates the qualitative discussion of
matter effects. We have checked that this  works well (upto a few
percent) at the energies and length scales relevant here. However,
all the plots we give in this paper are obtained by numerically
solving the full three flavour neutrino propagation equation
assuming the Preliminary Reference Earth Model (PREM) \cite{prem}
density profile for the earth. We use $\Delta_{31}=0.002$ eV$^2$ and
$\sin^2 2\theta_{23}=1$ unless otherwise mentioned. Further, the
numerical calculations assume $\Delta_{21} = 8.3 \times 10^{-5}$
eV$^2$, $\sin^2\theta_{12} = 0.27$ \cite{keg} and $\delta_{\rm
CP}=0$.
 The effect of varying the CP phase over the entire range (0 to 2$\pi$)
will be taken up in Subsection \ref{subsec:degeneracy},
where we include analytic expressions which take into account
small sub-leading effects and discuss the associated parameter
degeneracies.

\subsection{\bf{Review of $\mathrm{P}_{\mu {\mathrm e}}$ in matter}}
\label{subsec:pmue}

We first review $\nu_\mu \to \nu_e$ oscillations in matter. In
vacuum, the $\nu_\mu \to \nu_e$ oscillation probability is
\be
{\mathrm {P^{v}_{\mu e}}} =
\sin^2 \theta_{23} \sin^2 2 \theta_{13}
\sin^2 \left(1.27 \Delta_{31} {\mathrm L}/{\mathrm E} \right),
\label{eq:pmuevac}
\ee
where ${\mathrm{\Delta_{31} \equiv m_3^2-m_1^2}}$ is expressed in
${\mathrm eV}^2$, L in Km and E in GeV. In the constant density
approximation, matter effects can be taken into account by replacing
$\Delta_{31}$ and $\theta_{13}$ in Eq.~\ref{eq:pmuevac} by their
matter dependent values, {\it i.e.},
\be
{\mathrm{ P^{m}_{\mu e} }} =
{\mathrm{ \sin^2 \theta_{23} \sin^2 2 \theta^m_{13}
\sin^2 \left(1.27 \Delta^m_{31} L/E \right)}}.
\label{eq:pmuemat}
\ee
Here ${\mathrm{ \Delta^m_{31} }}$ and
${\mathrm {\sin 2 \theta^m_{13} }}$ are given
by
\bea
{\mathrm{ \Delta^m_{31} }} &=&
{\mathrm{
\sqrt{(\Delta_{31} \cos 2 \theta_{13} - A)^2 +
(\Delta_{31} \sin 2 \theta_{13})^2} }}
\nn \\
\nn \\
{\mathrm {\sin 2 \theta^m_{13} }}
&=& \sin 2 \theta_{13}
{\mathrm {\frac{\Delta_{31}}
{\Delta^m_{31}} }}
\label{eq:dm31}
\eea
where,
$${\mathrm { A = 2\sqrt{2}~G_F~n_e~E = 0.76 \times 10^{-4}
~\rho(gm/cc) ~E (GeV) }}$$

The resonance condition{\footnote{Note that this condition is
sensitive to the sign of $\Delta_{31}$. $\Delta_{31} > 0$ gives rise
to matter enhancement in case of neutrinos, while for anti-neutrinos
(since ${\mathrm {A \to -A}}$) one gets a suppression due to matter
effects. The situation is reversed for $\Delta_{31} < 0$. }} is
${\mathrm {A = \Delta_{31} \cos 2 \theta_{13}}}$, which gives

\be {\mathrm {E_{res}}} = \frac{\mathrm{\Delta_{31} \cos 2
\theta_{13}}} {\mathrm{2 \sqrt{2} G_F n_e}} \label{eq:eres}
\ee

Naively, one would expect ${\mathrm {P^{m}_{\mu e}}}$ to be maximum
at ${\mathrm{E = E_{res}}}$ since ${\mathrm{\sin 2 \theta^m_{13} =
1}}$. But this is not true in general because at this energy
${\mathrm{\Delta^m_{31}}}$ takes its minimum value of $\Delta_{31}
\sin 2 \theta_{13}$ and ${\mathrm {P^{m}_{\mu e}}}$ remains small
for pathlengths of ${\mathrm{L \leq 1000}}$ Km. If \L is chosen
suitably large so as to satisfy ${\mathrm {(1.27 \Delta_{31} \sin 2
\theta_{13} L/E) \geq \pi/4}}$, then ${\mathrm {P^{m}_{\mu e}}}$ can
attain values $\geq 0.25$ for $\sin^2 2\theta_{23} = 1$. For
$\Delta_{31} = 0.002$ \evsq and $\sin^2 2 \theta_{13} = 0.1$, one
needs $\L \geq 6000$ Km to satisfy the above condition.

\begin{figure}[ht]
{\centerline{ \hspace*{2em} \epsfxsize=12cm\epsfysize=8.0cm
                     \epsfbox{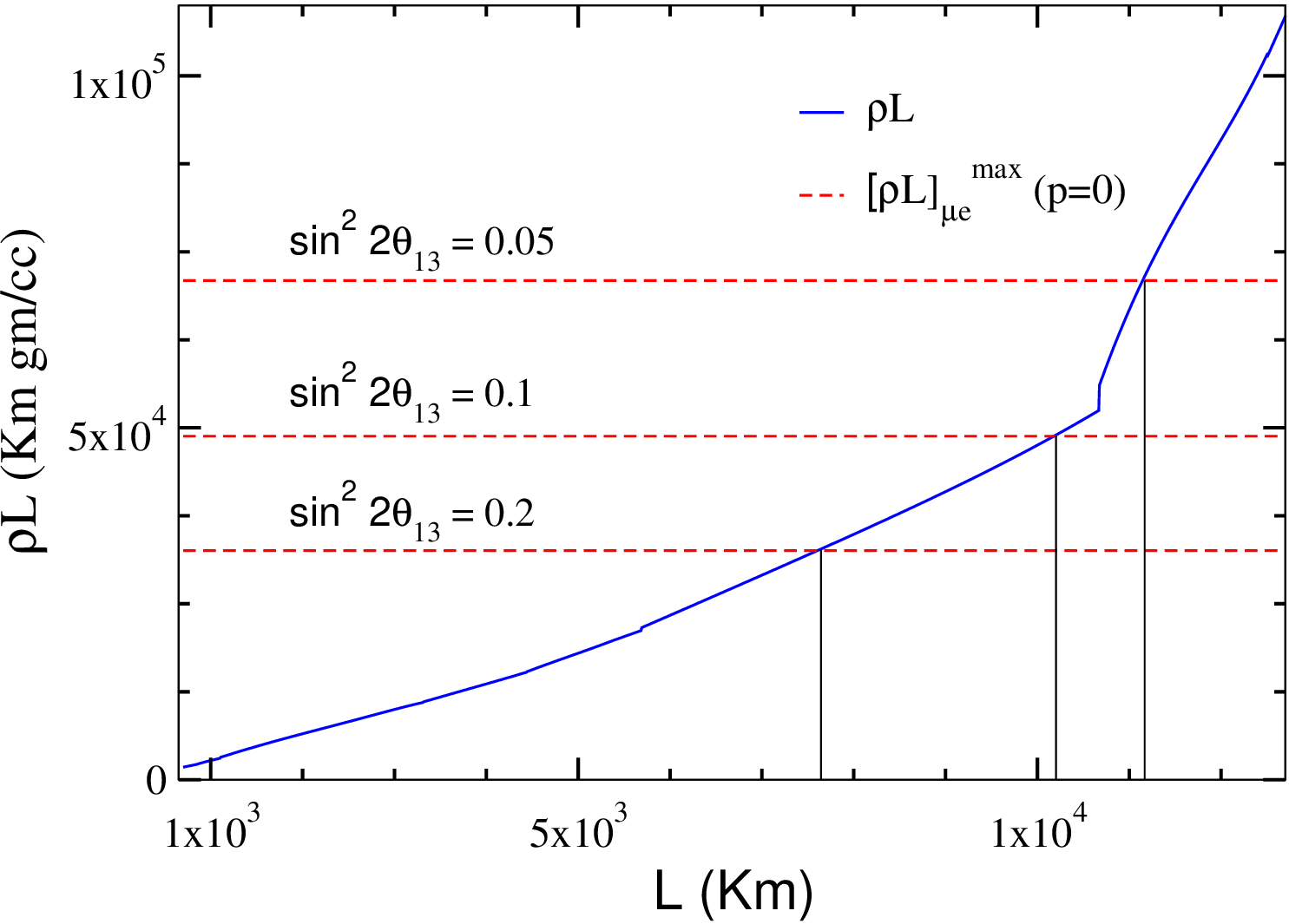}
} \caption[]{\footnotesize {${\mathrm{\rho L}}$ plotted vs L.
Horizontal lines correspond to ${\mathrm{[\rho L]^{max}_{\mu e}}}$
for different values of $\theta_{13}$ calculated using
Eq.~\ref{eq:muecondtn} for p=0.} } \label{fig1} }
\end{figure}

\noindent In particular, ${\mathrm {P^{m}_{\mu e}}}$ is maximum when
both
$${\mathrm {\sin 2 \theta^m_{13} = 1}}$$ and
$${\mathrm {\sin^2 \left(1.27 \Delta^m_{31} L/E \right)=1
}} ~~{\rm or,}~~
{\mathrm {\left(1.27 \Delta^m_{31} L/E \right)=[(2p+1)
\pi/2]}}$$
are satisfied.
This occurs when ${\mathrm {E_{res} = E^{{m}}_{{peak}}}}$.
This gives the condition
\cite{us1,ban,fre3}:
\be
{\mathrm {[\rho L]_{\mu e}^{max} }} \simeq
{\mathrm {
\frac{(2p+1) \pi 5.18 \times 10^3} { \tan 2\theta_{13}} ~{Km ~gm/cc} }}.
\label{eq:muecondtn}
\ee

\noindent Here, ${\mathrm p}$ takes integer values. This condition
is independent of $\Delta _{31}$ but depends sensitively on
$\theta_{13}$.
$\rho$ in Eq.~7 is the average density of matter along the
path of travel.
For trajectories passing through earth's
core, $\rho$ strongly depends on the pathlength. In Figure
\ref{fig1}, we plotted $\rho$L vs L for pathlengths
varying between 700 Km to 12740 Km based on PREM profile
of earth density distribution.
{\it In this paper, the symbol $\rho$ refers to the value of
earth matter density obtained from Figure 1, for any
value of L}.
We identify the particular values of ${\mathrm{\rho L}}$ which
satisfy Eq.~\ref{eq:muecondtn} with ${\mathrm {p}} = 0$ for three
different values of $\sin^2 2 \theta_{13}$. {\sl These occur at \L
$\simeq$ 10200 Km, 7600 Km and 11200 Km for $\sin^2 2 \theta_{13} =
0.1, 0.2$ and $0.05$ respectively}. This identifies  the baselines
at which ${\mathrm{P^{m}_{\mu e}}}$ is maximized. Additionally, the
relatively wide spacing between them demonstrates the  sensitivity
to $\theta_{13}$.
At and around the resonant energies and these baselines (depending
on the value of $\theta_{13}$) ${\mathrm {P^{m}_{\mu e}}}$
significantly impacts not only ${\mathrm {P_{\mu \mu}}}$, but also
${\mathrm {P_{\mu \tau}}}$, as we discuss below. Note that for
higher values of p, the baselines for maximum matter effect are
greater than the Earth's diameter. Hence ${\mathrm {p}} = 0$ is the
only relevant value of p in this case.

\subsection{\bf{Matter effects in \pmutau}}

In vacuum we have
\bea
{\mathrm {P^{v}_{\mu \tau}}} &=&
{\mathrm {\cos^4 \theta_{13} \sin^2 2 \theta_{23} \sin^2
\left(1.27 \Delta_{31} L/E \right)}},
\nonumber \\
& = &
{\mathrm {\cos^2 \theta_{13} \sin^2 2 \theta_{23}
\sin^2 \left(1.27 \Delta_{31} L/E \right)}}
\nonumber \\
&&
-~
{\mathrm{\cos^2 \theta_{23} P^{v}_{\mu e}}}
\label{eq:pmutauvac}
\eea
Including the matter effects changes this to
\bea
{\mathrm{P^{m}_{\mu \tau} }}& = &
{\mathrm{
\cos^2 \theta^m_{13} {\mathrm{\sin^2 2 \theta_{23}}}
\sin^2\left[1.27 (\Delta_{31} + A + \Delta^m_{31}) L/2E \right]}}
\nonumber \\
&+&
{\mathrm{
\sin^2 \theta^m_{13} {\mathrm{\sin^2 2 \theta_{23}}}
\sin^2\left[1.27 (\Delta_{31} + A - \Delta^m_{31}) L/2E \right]}}
\nonumber \\
&-&
{\mathrm { \cos^2 \theta_{23} P^{m}_{\mu e} }}
\label{eq:pmutaumat1}
\eea

\begin{figure}[!hb]
\vskip 1cm {\centerline{\hspace*{2em}
\epsfxsize=10cm\epsfysize=6.0cm
                     \epsfbox{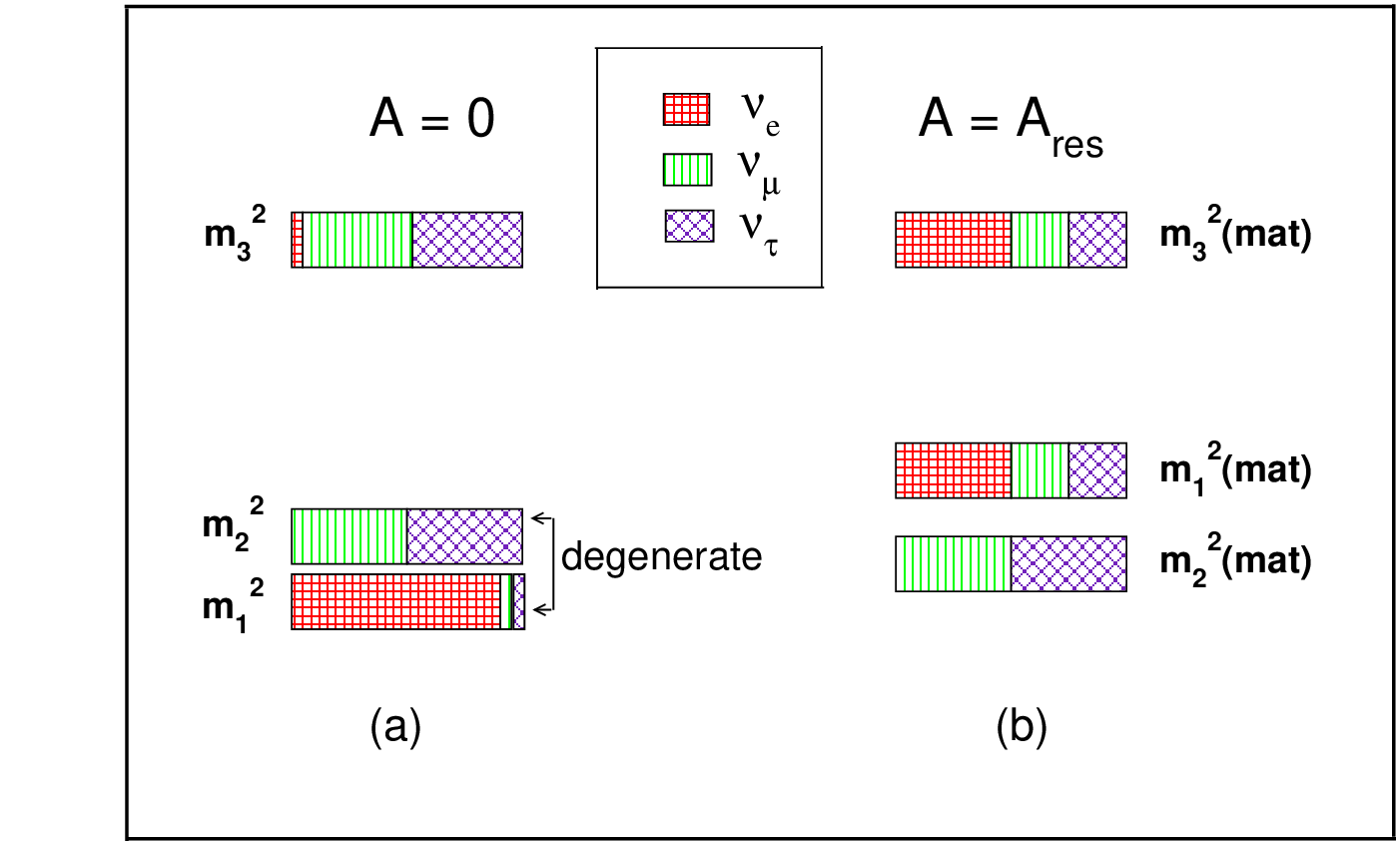}
} \caption[]{\footnotesize The flavor composition of mass
eigenstates in vacuum (A $=$ 0) and in matter at resonance (${\rm
A=A_{res}}$). The vacuum parameters used are $\sin^2
2\theta_{23}=1.0$ and $\sin^2 2\theta_{13}=0.1$. } \label{fig2} }
\end{figure}

%
Compared to ${\mathrm {P^{m}_{\mu e}}}$, these expressions
have a more complex matter dependence. This occurs due to the more
complicated change of $\nu_\mu$ and
$\nu_\tau$ in the flavour content of the matter dependent mass
eigenstates.
Labeling the vacuum mass eigenstates as  $\nu_1$, $\nu_2$ and
$\nu_3$, in the approximation where $\Delta_{21} = 0$,
$\nu_1$ can be chosen to be almost entirely $\nu_e$ and $\nu_2$
to have {\bf no} $\nu_e$ component.
Inclusion of the matter term ${\mathrm {A}}$
leaves $\nu_2$ untouched but gives a non-zero
matter dependent mass to $\nu_1$, thereby breaking the
degeneracy of the two mass states.
As the energy increases, the $\nu_e$ component of
${\mathrm {\nu_1^m}}$ decreases and the  $\nu_\mu,\nu_\tau$ components
increase such that at resonance energy they are $50 \%$.
Similarly, increasing energy increases the ${\mathrm{\nu_e}}$
component of ${\mathrm{\nu_3^m}}$ (and reduces the $\nu_\mu, \nu_\tau$
components) so that at resonance it becomes $50 \%$.
Thus in the resonance region, all three matter dependent
mass eigenstates ${\mathrm{\nu_1^m, \nu_2^m}}$ and
${\mathrm{\nu_3^m}}$ contain significant
$\nu_\mu$ and $\nu_\tau$ components. Hence, the matter
dependent \numutonutau oscillation probability will depend
on all three matter modified mass-squared differences.
In Figure \ref{fig2} we show the flavour
composition of the three mass states
in vacuum and in matter at resonance.

\begin{figure}[!h]
{\centerline{\hspace*{2em} \epsfxsize=12cm\epsfysize=16.0cm
                     \epsfbox{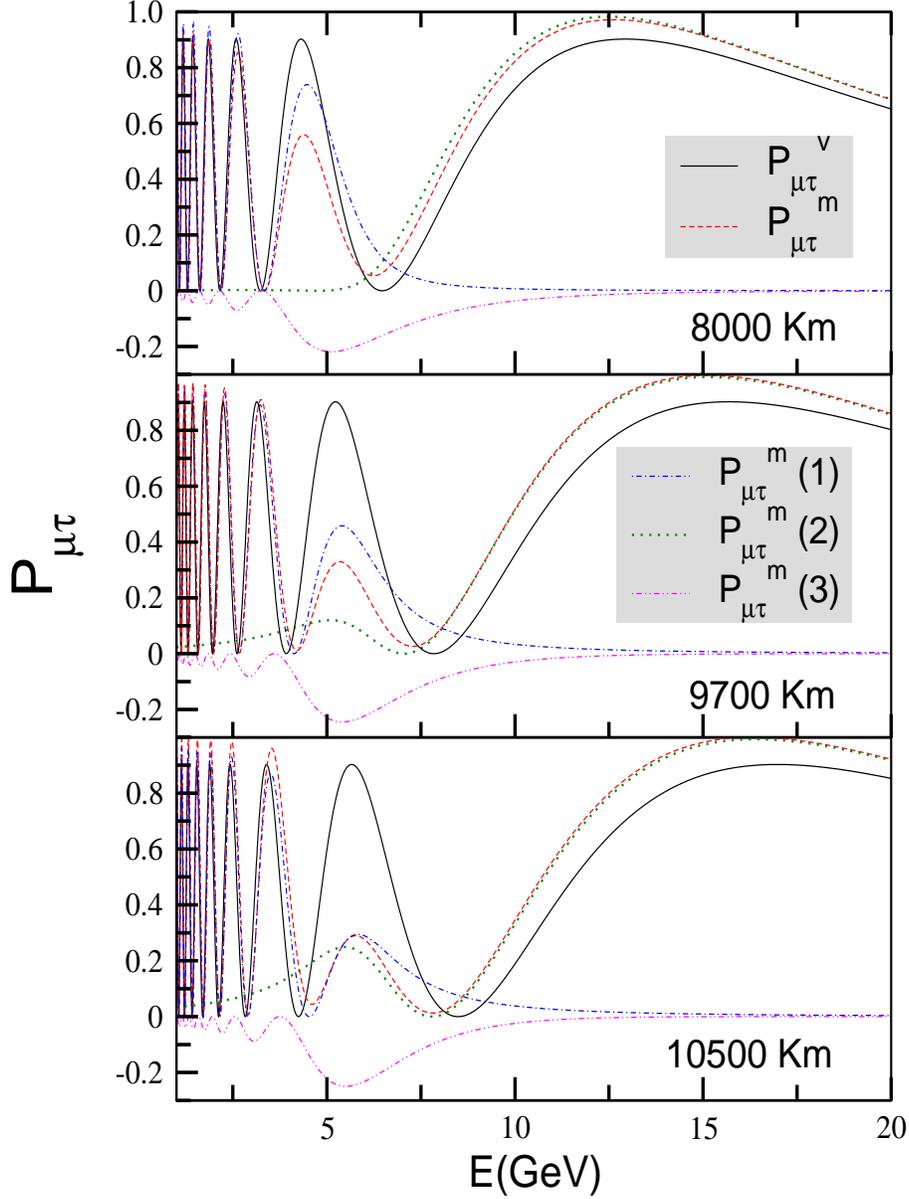}
} \caption[]{\footnotesize The variation of ${\mathrm{P_{\mu
\tau}}}$ in matter (${\mathrm{P^m_{\mu \tau}}}$) and in vacuum
(${\mathrm{P^v_{\mu \tau}}}$) with neutrino energy ${\mathrm{E}}$
for three different baselines of $8000$ Km, $9700$ Km and $10500$ Km
respectively. Here we have used an approximate average density
$\simeq 4.5$ gm/cc. Each of the terms in Eq.~{\ref{eq:pmutaumat1}}
are also shown in the three plots: ${\mathrm{P^m_{\mu \tau}(1)}}$ is
term 1, ${\mathrm{P^m_{\mu \tau}(2)}}$ is term 2 and
${\mathrm{P^m_{\mu \tau}(3)}}$ is term 3 respectively. }
\label{fig3} }
\end{figure}

We are interested in finding ranges of energy and pathlengths
for which there are large matter effects in
${\mathrm {P_{\mu \tau}}}$, {\it i.e.,}
for which
${\mathrm{\Delta P_{\mu \tau} = P^{m}_{\mu \tau} - P^{v}_{\mu \tau}}}$
is large.
To this end, it is useful to examine the set of probability
plots given in Figure \ref{fig3}.
The plots display the variation of
${\mathrm{P^{m}_{\mu \tau}}}$ with
neutrino energy E for baselines of $8000$ Km,
$9700$ Km and $10500$ Km respectively, with each of the terms in Eq.~{\ref{eq:pmutaumat1}}
plotted separately.
${\mathrm {P^m_{\mu \tau}}}(1)$,
${\mathrm {P^m_{\mu \tau}}}(2)$
and ${\mathrm {P^m_{\mu \tau}}}(3)$
in the figure refer to the first, second and third term in
Eq.~{\ref{eq:pmutaumat1}} respectively,
with the third term plotted incorporating its negative sign.

Additionally,  the plots display the full vacuum probability,
${\mathrm {P^{v}_{\mu \tau}}}$. The approximate average density at
these baselines is ${\mathrm{\rho}} \simeq 4.5$ gm/cc, hence one
gets ${\mathrm{E_{res} \simeq 5}}$ GeV from Eq.~{\ref{eq:eres}}.
The vacuum peaks, of course, shift with baseline and we see that a
significantly broad peak gradually positions itself above the
resonant energy at the baselines shown in the figure. We use this
fact in (i) below.

We show that appreciable changes in ${\mathrm{P^{m}_{\mu \tau}}}$  occur
for two different sets of conditions, leading in one case to
a sharp decrease from a  vacuum maximum and in another
to a smaller but extended  increase over a broad range of energies.
Both are discussed below in turn.

\begin{enumerate}
\item[]
{\bf (i) Large decrease in ${\mathrm{P^{m}_{\mu \tau}}}$
in the resonance region :}\\
As is evident in Figure {\ref{fig3}},
at energies appreciably below resonance, term 1
 in  Eq.~\ref{eq:pmutaumat1}
(i.e. the one with ${\mathrm{\cos^2 \theta^m_{13}}}$) is
very nearly equal to ${\mathrm{P^{v}_{\mu \tau}}}$. This is because
${\mathrm{\theta_{13}^m=\theta_{13}}}$,
${\mathrm{A<< \Delta_{31}, \Delta^m_{31} \simeq \Delta_{31}}}$. Term $2$,
 the ${\mathrm{\sin^2 \theta^m_{13}}}$ term, is nearly
 zero. Similarly, term $3$ is also very small.
As we increase the energy and approach resonance,
${\mathrm{\cos^2 \theta^m_{13}}}$ begins to decrease sharply,
deviating from its vacuum values,
while ${\mathrm{\sin^2 \theta^m_{13}}}$ increases rapidly.
However, {\it if resonance is in the vicinity of  a vacuum peak},
which is the case at and around the baselines shown in
Figure {\ref{fig3}},
then the decrease in the
${\mathrm{\cos^2 \theta^m_{13}}}$ term has a much stronger impact on
${\mathrm{P^{m}_{\mu\tau}}}$ than the increase in the
${\mathrm{\sin^2 \theta^m_{13}}}$ term,
since the latter starts out at zero while the former is initially
close to its peak value ($\sim 1$).
As a result, ${\mathrm{P^{m}_{\mu\tau}}}$ falls sharply.
This fall is enhanced by the third term in Eq.~\ref{eq:pmutaumat1},
which is essentially
${\mathrm{0.5 \times P^{m}_{\mu e}}}$
(which is large due to resonance),
leading to a large overall drop in ${\mathrm{P^{m}_{\mu \tau}}}$
from its vacuum value.
Note that the requirement that we be at a vacuum peak
to begin with forces
${\mathrm{\Delta P_{\mu \tau}}}$
to be large and negative
, with the contributions from the first and the third term
reinforcing each other.

The criterion for a maximal  matter effect,
${\mathrm{E_{res} \simeq E^{{v}}_{{peak}} }}$,
leads to the following condition:
\be
{\mathrm{[\rho L]_{\mu \tau}^{max}}} \simeq
{\mathrm{(2p+1) \,\pi\, 5.18 \times 10^{3}
\,(\cos2\theta_{13}) ~{ Km~gm/cc}}}.
\label{eq:mutaucondtn}
\ee
Unlike Eq.~\ref{eq:muecondtn},
which has a $\tan 2 \theta_{13}$ in its denominator,
Eq.~\ref{eq:mutaucondtn} has
a much
weaker dependence on $\theta_{13}$.

\noindent
From Eq.~\ref{eq:pmutauvac} and Eq.~\ref{eq:pmutaumat1}, we write
\bea
{\mathrm{{\Delta P}_{\mu \tau}}}
&=&
{\mathrm {P^{m}_{\mu \tau}}} -
{\mathrm {P^{v}_{\mu \tau}}}
\nonumber \\
&=&
{\mathrm \sin^2 2\theta_{23}
\lbrack~
- \cos^4 \theta_{13}
\sin^2 (1.27 \Delta_{31} {\mathrm L}/{\mathrm E}) }
\nonumber\\
&+&
{\mathrm{
\cos^2 \theta^{m}_{13}
\sin^2 [1.27 (\Delta_{31}+{\mathrm A}+\Delta^{m}_{31}){\mathrm L}/2{\mathrm E}}]
}
\nonumber \\
&+&
{\mathrm{\sin^2 \theta^{m}_{13} \sin^2
[1.27 (\Delta_{31}+{\mathrm A}-\Delta^{m}_{31}){\mathrm L}/2{\mathrm E}}] }
\nonumber\\
&-&
{\mathrm{\sin^2 \theta^{m}_{13} \cos^2 \theta^{m}_{13}
\sin^2 (1.27 \Delta^{m}_{31} {\mathrm L}/{\mathrm E})}}
~\rbrack
\label{eq:delpmutau1}
\eea
Incorporating the ${\mathrm{E_{res} \simeq E^{{v}}_{{peak}} }}$
condition (Eq.~\ref{eq:mutaucondtn})
leads to
\bea
{\mathrm{\Delta P_{\mu \tau}}}  & \simeq &
{\mathrm {\cos^4\left[\sin2\theta_{13}(2p+1)\frac{\pi}{4}\right] - 1}}
\label{eq:delpmutau}
\eea
%
\begin{figure}[h]
{\centerline{\hspace*{2em} \epsfxsize=14cm\epsfysize=12.0cm
                     \epsfbox{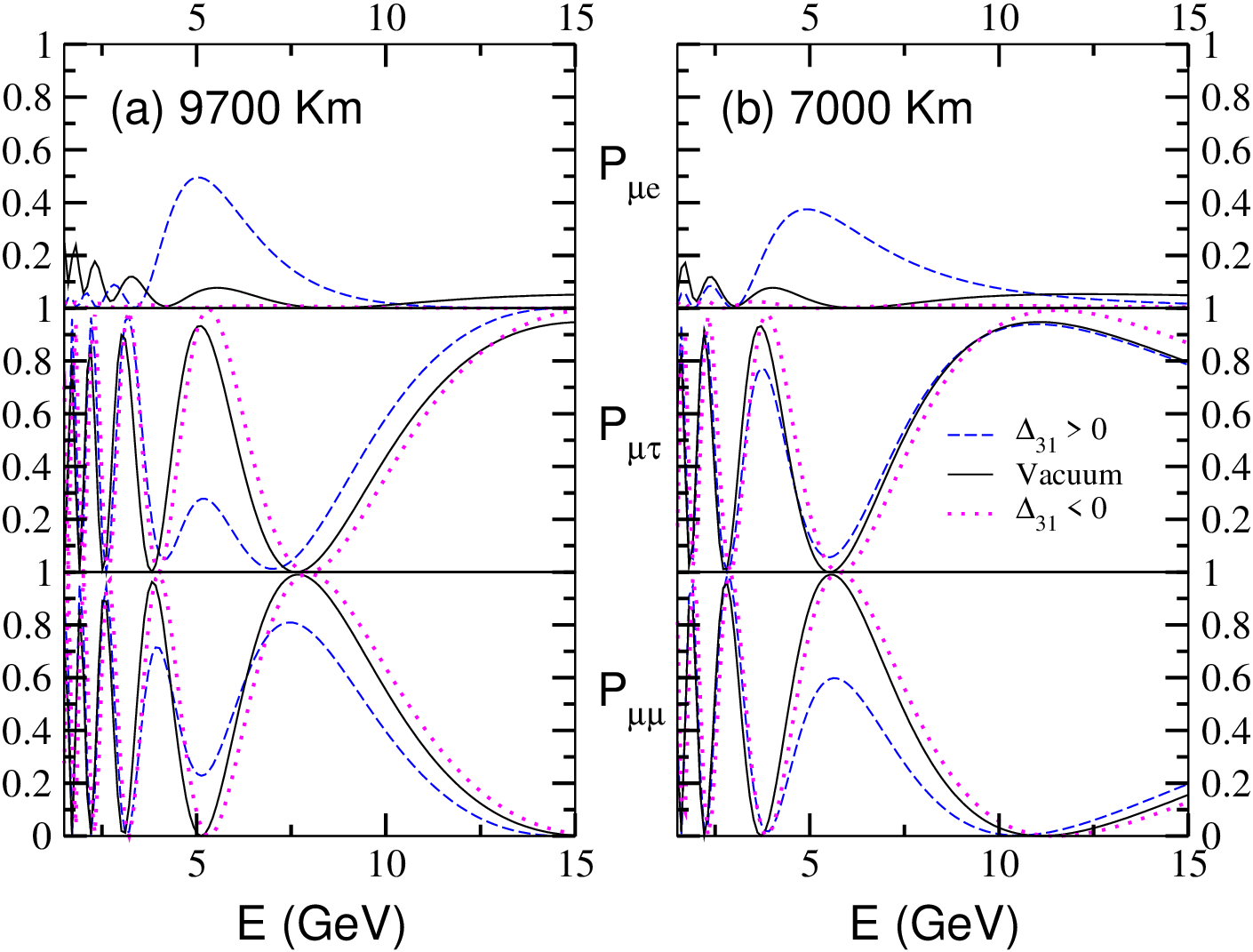}
} \caption[]{\footnotesize {\pmutau and \pmumu plotted as a function
of neutrino energy, E (in GeV) in presence of matter and in vacuum
for both signs of $\Delta_{31}$ for two different baseline lengths,
{\bf{(a)}} for \L= 9700 Km and {\bf{(b)}} for \L= 7000 Km. These
plots use $\Delta_{31} = 0.002$ eV$^2$ and $\sin^2 2\theta_{13}
=0.1$.
 }}
\label{fig4} }
\end{figure}

We note that, in general, ${\mathrm{\Delta P_{\mu \tau}}}$
 (at ${\mathrm{E_{res} \simeq E^{{v}}_{{peak}} }}$ )
will be larger for higher values of both ${\mathrm {p}}$ and
$\theta_{13}$. For ${\mathrm {p}} = 1$ and $\sin^2 2\theta_{13} =
0.1$, ${\mathrm {E_{res} \simeq E^{v}_{peak}}}$ occurs at $\sim$
9700 Km (from Eq.~\ref{eq:mutaucondtn}) and ${\mathrm{\Delta P_{\mu
\tau}}} \sim -0.7$ (from Eq.~\ref{eq:delpmutau}). For ${\mathrm p} =
0$, Eq.~\ref{eq:mutaucondtn} gives $\mathrm{ L_{\mu \tau}^{max}}$
$\sim$ 4400 Km for $\sin^2 2\theta_{13} = 0.1$. However,
${\mathrm{\Delta P_{\mu \tau}}}$ is roughly one-tenth of the
${\mathrm {p}} = 1$ case. In general, for a given baseline, the
choice of an optimal p is also dictated by the constraint that the
vacuum peak near resonance have a breadth which makes the effect
observationally viable. Note that for ${\mathrm{\sin^2
2\theta_{13}}}$ = 0.05 and 0.2, Eq.~\ref{eq:mutaucondtn} gives the
distances of maximum matter effect as $\sim$ 9900 and 9300 Km for
${\mathrm {p}} = 1$. Because of the weaker dependence on
$\theta_{13}$ here compared to ${\mathrm{P^{m}_{\mu e}}}$
(Eq.~\ref{eq:muecondtn}), the distances for various values of
$\theta_{13}$ are bunched together in the vicinity of 9500 Km.

\par  In Figure \ref{fig4}(a), we show all three matter and vacuum
probabilities for 9700 Km.
In these plots $\Delta_{31}$
is taken as 0.002 eV$^2$ which gives
${\mathrm {E_{res} \simeq E^{v}_{peak}}}$ at 5 GeV.
The middle panel of
Figure \ref{fig4}(a) shows that near this energy
${\mathrm{P^{m}_{\mu \tau}}}$
($\sim$ 0.33)  is appreciably lower compared to
${\mathrm{P^{v}_{\mu \tau}}}$ ($\sim$ 1).
Thus the drop due to matter effect is
0.67, which agrees well with that
obtained in the paragraph above using the approximate expression
Eq.~\ref{eq:delpmutau}.

\begin{figure}[h]
{\centerline{\hspace*{2em} \epsfxsize=10cm\epsfysize=8.0cm
                     \epsfbox{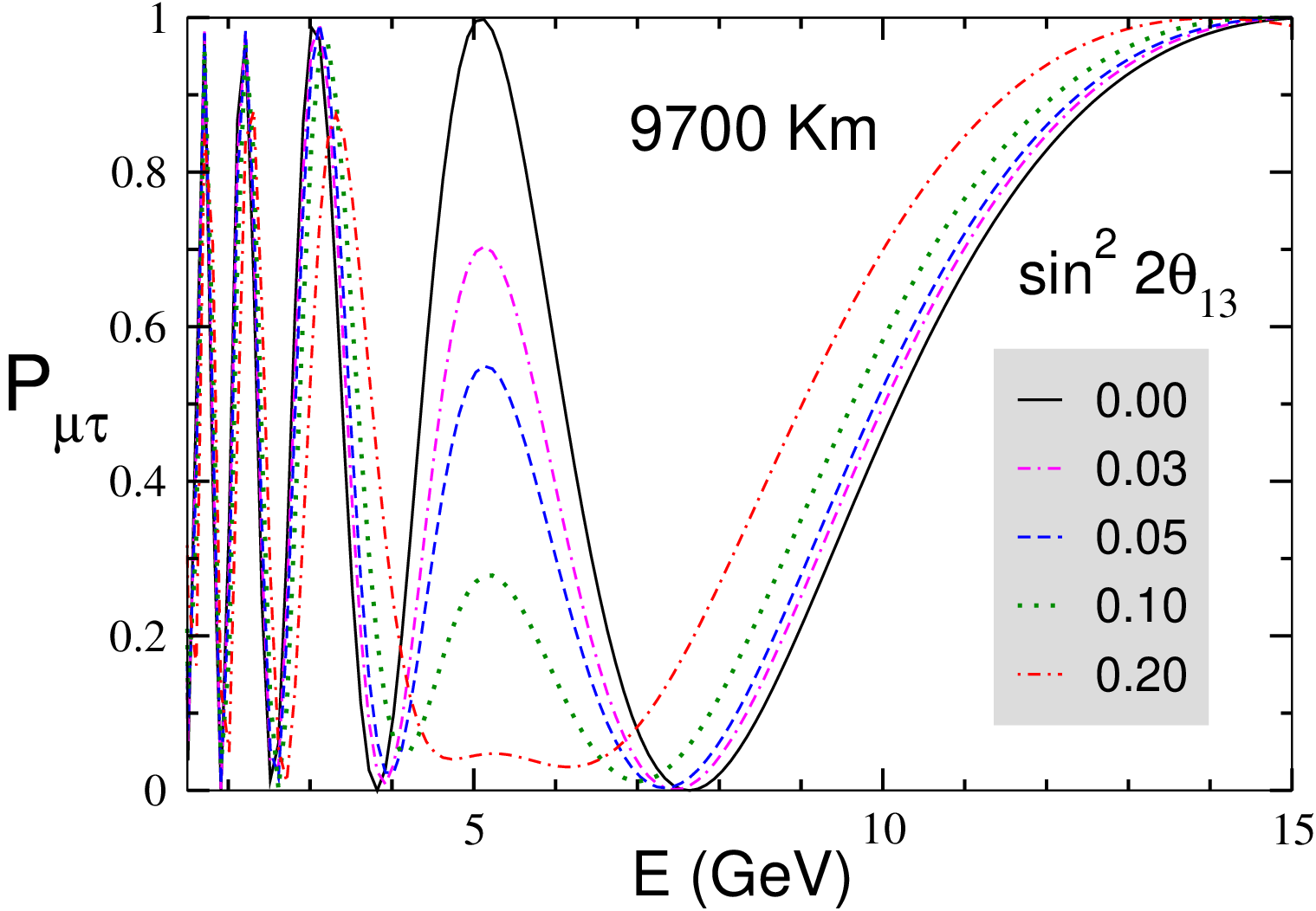}
} \caption[]{\footnotesize {The plot shows sensitivity to
$\theta_{13}$ in case of \pmutau for a baseline length of 9700 Km.
The value of $\Delta_{31}$ is taken to be $0.002$ eV$^2$. } }
\label{fig5} }
\end{figure}

\par In Figure \ref{fig5} we show the $\theta_{13}$
sensitivity of ${\mathrm{P^{m}_{\mu \tau}}}$ at 9700
Km{\footnote{\pmutaum develops good sensitivity to $\theta_{13}$
only at around L $=$ 9700 Km. As shown in the middle panel of
\ref{fig4}(b), even at L$=$ 7000 Km, its sensitivity to matter
effects is relatively poor \cite{shrock}.}}. In particular, at
${\mathrm{E_{res} \simeq E^{{v}}_{{peak}}}}$ the strong dependence
on $\theta_{13}$ is also governed by Eq.~\ref{eq:delpmutau} above.
Thus we have significant change in ${\mathrm{P^{m}_{\mu \tau}}}$ due
to matter effects, even for small values of $\theta_{13}$. This is
illustrated in Figure \ref{fig5}, where we see that, even for as
small a value of $\sin^2 2\theta_{13}=0.03$, matter effects at
maximal resonance
cause a change in ${\mathrm{P^{m}_{\mu \tau}}}$ of about $30 \%$.
This amplification of matter effects at resonance occurs for
similar pathlengths for all allowed values of $\theta_{13}$,
as given in (Eq.~\ref{eq:mutaucondtn}). This is to be contrasted with the case
of $\nu_{\mu} \rightarrow \nu_{e}$ oscillations,
where the observation of resonance amplification
is possible only for $\theta_{13}$ close to the
present upper bound.
Actual observation of these maximal resonance
matter effects in the $\nu_{\mu} \rightarrow \nu_{\tau}$ oscillations
will be very difficult, because of the smallness of cross-sections
and the difficulty in reconstructing the $\tau$. With a very high intensity source,
such as a neutrino factory and a specialized $\tau$ detector,
it may be possible. A study of this will be reported elsewhere
\cite{us2}.


\item[] {\bf (ii) Increase in \pmutaum
away from resonance :}\\
We now discuss the second scenario for which ${\mathrm {P^{m}_{\mu
\tau}}}$ can differ appreciably from ${\mathrm {P^{v}_{\mu \tau}}}$.
This happens away from resonance, and is evident in Figure
\ref{fig4}(a) (central panel) in the energy range $7.5$ GeV - $ 15$
GeV. This effect is an enhancement rather than a drop, {\it i.e.},
${\mathrm {\Delta P_{\mu \tau}}}$ is now positive, unlike the
previous case.

$\Delta$\pmue is small in most
of the latter part of the energy region under
consideration and does not contribute in an
important way overall (Figure \ref{fig4}(a)
(top panel)).
The dominant contribution to this enhancement arises
from the
${\mathrm{\sin^2 \theta_{13}^m}}$ term in
${\mathrm{{P}^{m}_{\mu\tau}}}$
(Eq.~\ref{eq:pmutaumat1})
which is large for ${\mathrm{E}} >> {\mathrm{E_{res}}}$.
Since ${\mathrm{(\Delta_{31} + A - \Delta^m_{31}) \simeq 2\,\Delta_{31}}}$ for these energies,
we obtain a enhancement($\sim 15\%$) which follows the vacuum curve.
The difference between the two curves, vacuum and matter,
largely reflects the difference between the $\cos^4\theta_{13}$
multiplicative term in the vacuum expression
Eq.~\ref{eq:pmutauvac}
and the ${\mathrm{\sin^2 \theta_{13}^m}}$
multiplicative term in
Eq.~\ref{eq:pmutaumat1}. While this effect is
smaller compared to the effect discussed in (i) above, it occurs
over a broad energy band and can manifest itself in energy
integrated event rates.

\end{enumerate}

\subsection{\bf{Matter effects in \pmumu}}
In vacuum \pmumu is given by

\bea
{\mathrm {P^{v}_{\mu \mu}}}
&=&
1 - {\mathrm {P^{v}_{\mu \tau}}} - {\mathrm {
P^{v}_{\mu e}}}
\nonumber \\
&=&
1 - \cos^4 \theta_{13} \sin^2 2 \theta_{23}
\sin^2 \left(1.27 \Delta_{31} {\mathrm L}/{\mathrm E} \right)
\nonumber \\
&-&
\sin^2 \theta_{23} \sin^2 2\theta_{13}
\sin^2 \left(1.27 \Delta_{31} {\mathrm L}/{\mathrm E} \right)
\label{eq:pmumuvac}
\eea
Including the matter effects changes this to
\bea
{\mathrm{P^{m}_{\mu \mu} }}& = &
1 - {\mathrm {
P^{m}_{\mu \tau}}} - {\mathrm {
P^{m}_{\mu e}}}
\nonumber \\
&=&
{\mathrm{1 -
\cos^2 \theta^m_{13} {\mathrm{\sin^2 2 \theta_{23}}}
\sin^2\left[1.27 (\Delta_{31} + A + \Delta^m_{31}) L/2E \right]}}
\nonumber \\
&-&
{\mathrm{
\sin^2 \theta^m_{13} {\mathrm{\sin^2 2 \theta_{23}}}
\sin^2\left[1.27 (\Delta_{31} + A - \Delta^m_{31}) L/2E \right]}}
\nonumber \\
&-&
 {\mathrm{\sin^4 \theta_{23}}}
{\mathrm { \sin^2 2\theta^m_{13}
\sin^2 \left(1.27 \Delta^m_{31} {\mathrm L}/{\mathrm E} \right) }}
\label{eq:pmumumat1}
\eea

The deviation of ${\mathrm{P_{\mu \mu}^{m}}}$
from ${\mathrm{P_{\mu \mu}^{v}}}$ clearly results from the
combined effects in
${\mathrm{P_{\mu \tau}^{m}}}$ and ${\mathrm{P_{\mu e}^{m}}}$. In order to quantify the extent of deviation, we define,
\begin{eqnarray}
{\mathrm {\Delta P_{\mu \mu} }}
&=&
- {\mathrm{\Delta{P_{\mu \tau}} - \Delta{P_{\mu e}}}}
\label{eq:deltapmumueqn}
\end{eqnarray}

Below we illustrate the various conditions which can give rise to
a significant change in ${\mathrm {P_{\mu \mu}}}$
due to matter effects arising in both ${\mathrm {P_{\mu \tau}}}$  and
${\mathrm {P_{\mu e}}}$ :

\begin{enumerate}
\item[]{\bf (a)} {\bf Large, negative ${\mathrm{\Delta{P_{\mu \tau}}}}$ and
positive ${\mathrm{\Delta{P_{\mu e}}}}$}

Large and negative ${\mathrm{\Delta{P_{\mu \tau}}}}$ corresponds
to the case (i) discussed in Subsection 2.2.
In this case, ${\mathrm {\Delta P_{\mu e}}}$ is positive, so
the signs of the two changes are opposite
and  hence the two terms
(Eq.~\ref{eq:deltapmumueqn}) do not contribute in
consonance.
However, the resulting {\it increase} in \pmumu is still
significant ($\sim 20 \%$), given the magnitude of the
change ($\sim 70 \%$) in ${\mathrm {P_{\mu \tau}}}$.
This is visible in the bottom panel of Figure \ref{fig4}(a),
in the energy range 4-6 GeV.
\vskip.7cm
\begin{figure}[ht]
\centerline{
\epsfxsize=8.5cm\epsfysize=6.5cm\epsfbox{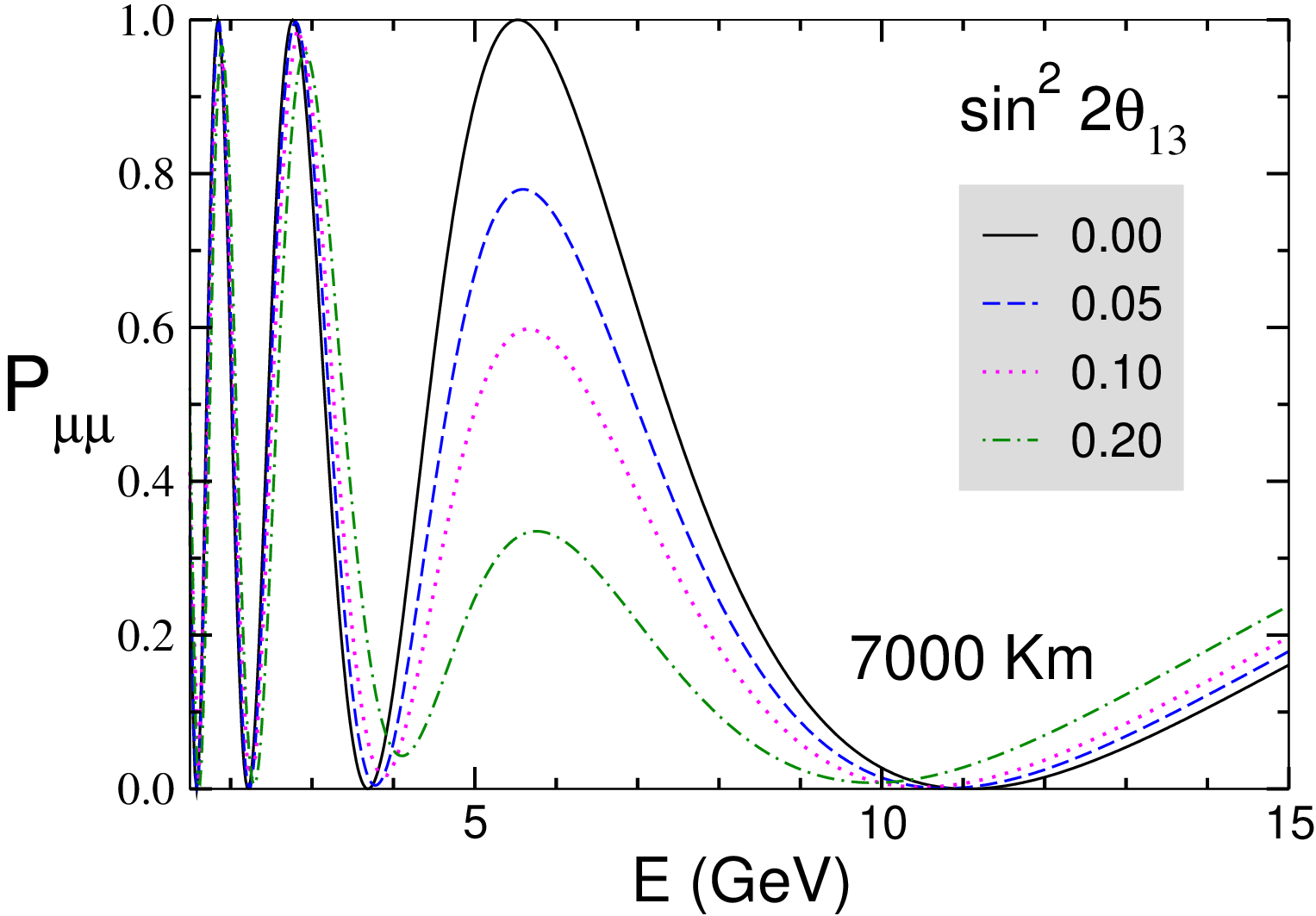}
        \hspace*{1.5ex}
\epsfxsize=8.5cm\epsfysize=6.5cm
                     \epsfbox{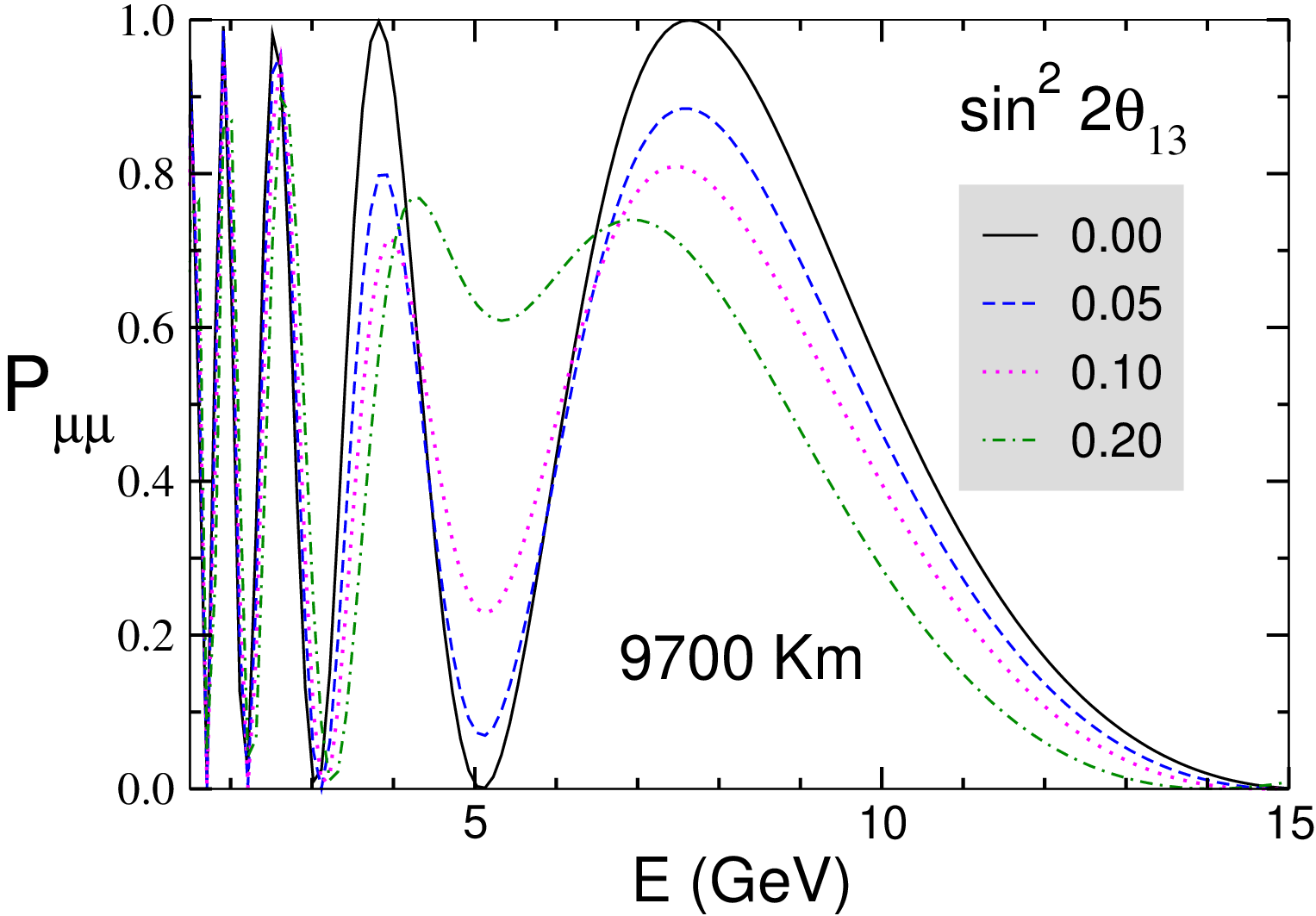}
}
\hskip 5cm
{\bf (a)}
\hskip 7cm
{\bf (b)}
\caption
{\footnotesize
Figure {\bf (a)} and {\bf (b)} depict \pmumu in matter
plotted against the neutrino energy, E(in GeV)
for different values of $\theta_{13}$ and two
baseline lengths, viz, 7000 Km and 9700 Km,
respectively.
}
\label{fig6}
\end{figure}
\begin{figure}[ht]
\centerline{
\epsfxsize=8.5cm\epsfysize=6.5cm\epsfbox{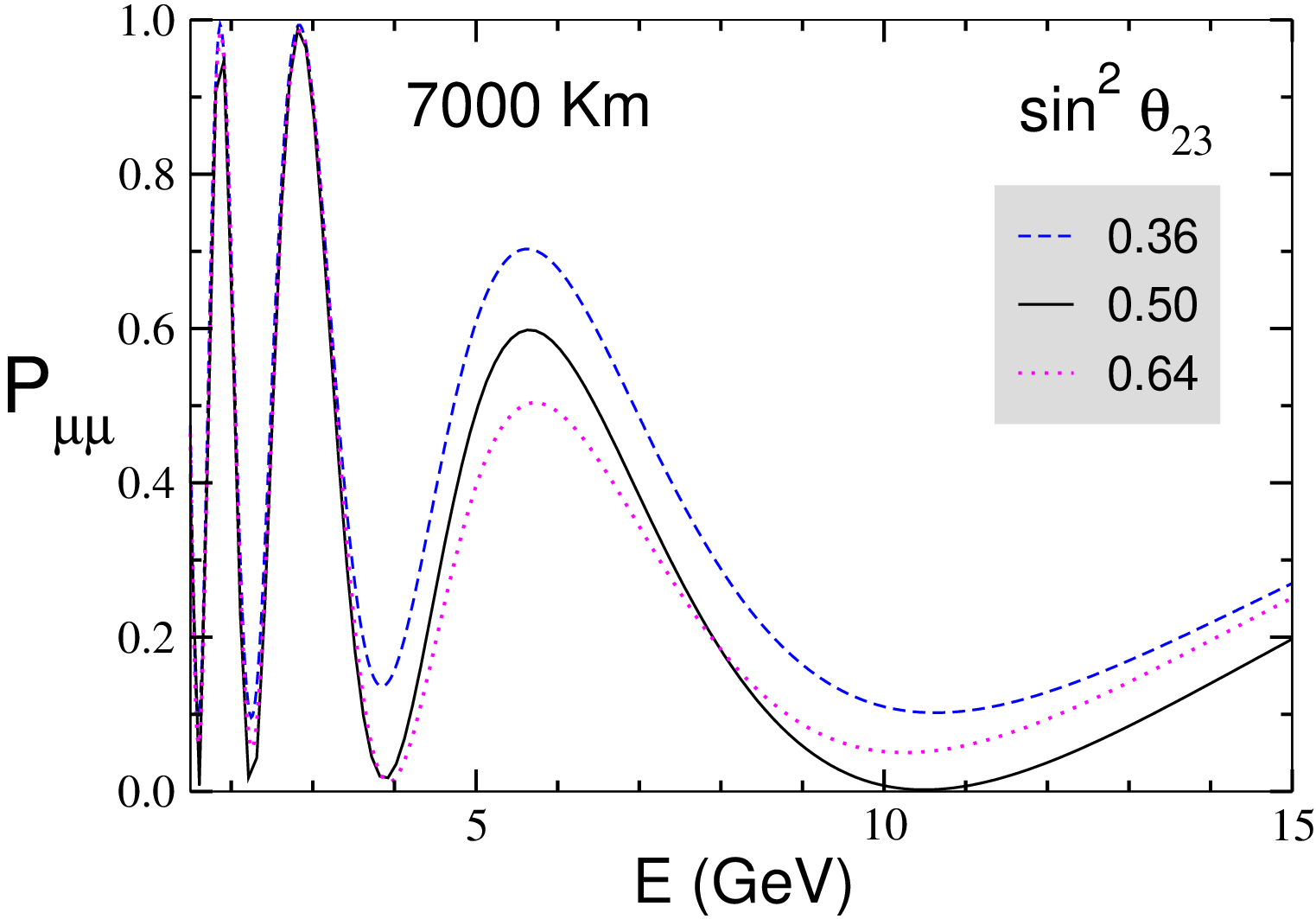}
        \hspace*{1.5ex}
\epsfxsize=8.5cm\epsfysize=6.5cm
                     \epsfbox{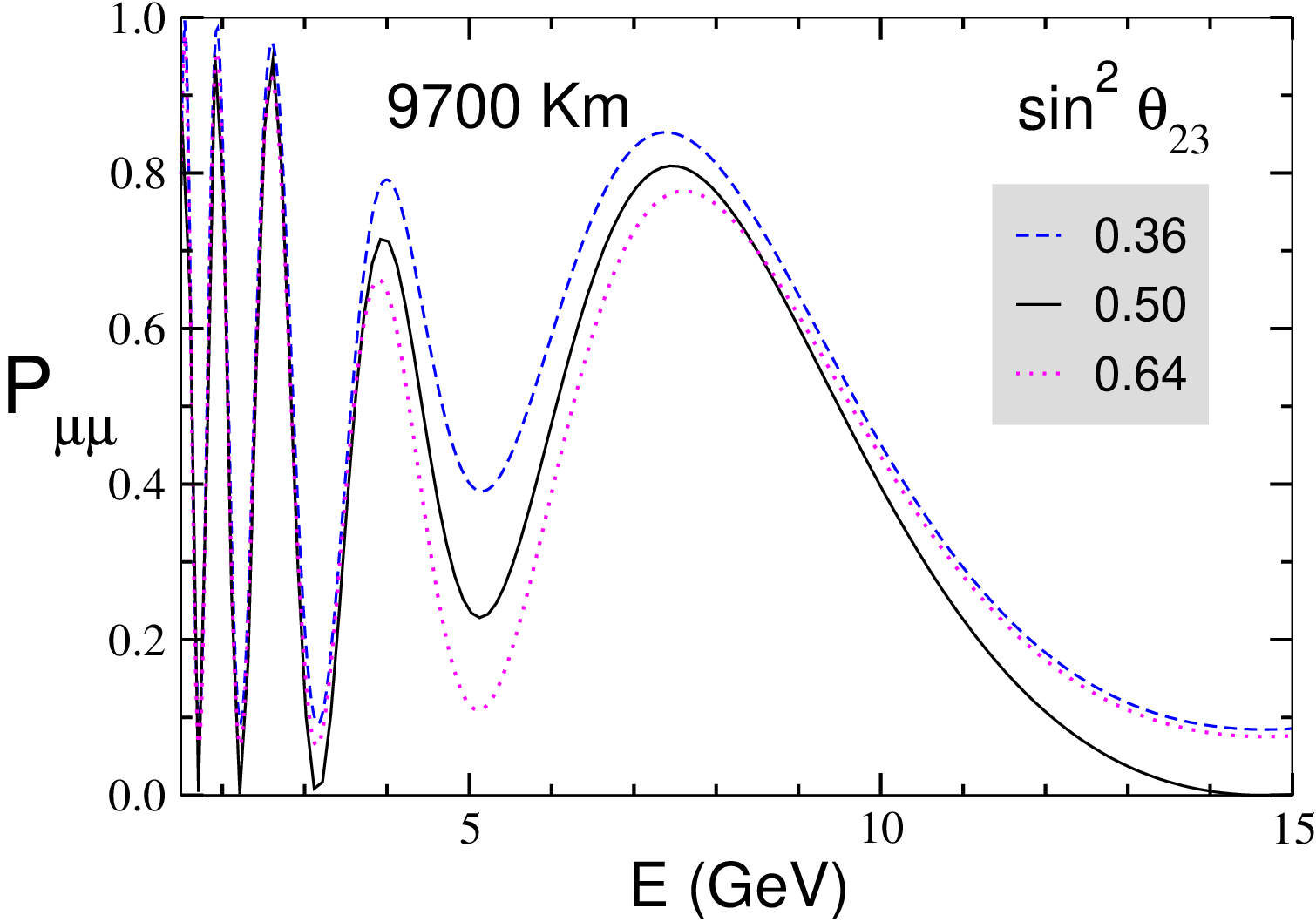}
}
\hskip 5cm
{\bf (a)}
\hskip 7cm
{\bf (b)}
\caption
{\footnotesize
Figure {\bf (a)} and {\bf (b)} depict \pmumu in matter plotted vs E (in GeV)
for different values of $\theta_{23}$ and two baseline lengths, viz,
7000 Km and 9700 Km respectively, for
$\sin^2 2\theta_{13}=0.1$. The value of $\Delta_{31}$ is taken to be
$0.002$ eV$^2$.
}
\label{fig7}
\end{figure}


\item[]{\bf (b)} {\bf Positive
${\mathrm{\Delta{P_{\mu \tau}}}}$ and small ${\mathrm{\Delta{P_{\mu e}}}}$}

This case corresponds to a significant {\it drop} in
${\mathrm{P_{\mu \mu}^{m}}}$, which
is seen in Figure
\ref{fig4}(a) (bottom panel) in the energy range
$\sim 7-15$ GeV. The enhancement in
${\mathrm{P^{m}_{\mu \tau}}}$ (corresponding
to the case (ii) discussed in Subsection 2.2)
is reflected in the decrease in
${\mathrm{P_{\mu \mu}^{m}}}$ compared to its vacuum value.

\item[]{\bf (c)} {\bf Small
${\mathrm{\Delta{P_{\mu \tau}}}}$ and large, positive
${\mathrm{\Delta{P_{\mu e}}}}$
}

This situation occurs when a minimum in the vacuum
value of ${\rm P_{\mu \tau}}$ resides in the proximity of a
resonance. The condition for a minimum in ${\rm P_{\mu \tau}}$ is ${\rm 1.27
\Delta_{31} L/E = p \pi}$. In this region, the rapidly changing
${\rm \sin^2 \theta_{13}^m}$ and ${\rm \cos^2 \theta_{13}^m}$
increase ${\mathrm {P_{\mu \tau}}}$
from its vacuum value of ${\rm 0}$ to about ${\rm 0.1}$.
At the same time, ${\mathrm {P_{\mu e}}}$
undergoes an increase of $0.3$, thus leading to a net change of
${\rm P_{\mu \mu} = 0.4}$
(the three panels of Figure \ref{fig4}(b)).
Note that the above condition corresponds to a vacuum peak of
${\rm P_{\mu \mu}}$ and the $40 \%$ drop of its value makes this
energy range (4-10 GeV) suitable for searching for matter effects.
Substituting ${\mathrm E}$ as ${\mathrm {E_{res}}}$ gives the distance
for maximum matter effect in ${\rm P_{\mu \mu}}$ as

\bea
{\mathrm{[\rho L]_{\mu \mu}^{max} \simeq {p\,\pi\,\times10^{4}
\,(\cos2\theta_{13})}~{Km~gm/cc}}}.
\label{eq:mumucondtn}
\eea
For p=1, this turns out to be $\sim$
7000 Km{\footnote{Note that this length is close to
the "magic baseline" \cite{barg,magic}.}}.
This effect \cite{shrock} is shown in
the bottom panel of Figure \ref{fig4}(b).
The large ($40\%$ at its peak) drop in
\pmumum seen in this figure
derives its strength mainly from the resonant enhancement
in ${\mathrm {P^{m}_{\mu e}}}$.

\noindent
Defining ${\mathrm{{\Delta P}_{\mu \mu}}}={\mathrm{P^{m}_{\mu \mu}}}-{\mathrm{P^{v}_{\mu \mu}}}$
and incorporating the condition that ${\mathrm{E_{res} \simeq E^{{v}}_{{peak}} }}$ of \pmumu
 (Eq.~\ref{eq:mumucondtn})
leads to
\bea
{\mathrm{\Delta P_{\mu \mu}}}  & \simeq &
{\mathrm {-\sin^2\left[\sin2\theta_{13}p\frac{\pi}{2}\right]}}
- {\mathrm {0.25\sin^2\left[\sin2\theta_{13}
p{\pi}\right]}}.
\label{eq:delpmumu}
\eea

\end{enumerate}

The ${\rm \theta_{13}}$ sensitivity of ${\rm{P^m_{\mu \mu}}}$
for L = 7000 Km and for L = 9700 Km is shown in Figure \ref{fig6}.
In Figure \ref{fig6}(a), we see that for L $=$ 7000 Km,
the maximum $\theta_{13}$ sensitivity is in the energy range 4-10 GeV.
Our ability to observe this drop in ${\mathrm {P^{m}_{\mu \mu}}}$
depends on the statistics. The condition for 3$\sigma$ observability may be stated as
\be
{\mathrm{N_\mu (\theta_{13} = 0) - N_\mu (\theta_{13}) \geq
3 \sqrt{N_\mu(\theta_{13})}
}}.
\label{eq:muevent}
\ee

Figure \ref{fig7} depicts the sensitivity of \pmumum to $\theta_{23}$ for the
same distances, showing a similar inverse relation between the survival probability
and the value of $\theta_{23}$. Note that, for these baselines, ${\rm
P_{\mu \mu}}$ changes by as much as $20 \%$, for the
currently allowed range of $\theta_{23}$. A criterion similar
to Eq.~\ref{eq:muevent} can be applied to these sensitivities in situations
where their observability is feasible. This is likely to happen in
long baseline experiments with high luminosity sources.

The width of the effects
discussed above for \pmumu is significant, ranging from
$4-6$ GeV in case (a), $7-15$ GeV in case (b)
(Figure \ref{fig4}(a)) and $4-10$ GeV in case (c)
(Figure \ref{fig4}(b)).
We have checked that they persist over a range of baselines
(6000 - 10500 Km), making them observationally feasible, as is
demonstrated later.
\par
In general the resonance has a width, and this fact affects observability.
Below we give an expression for ${\mathrm{\Delta P_{\mu \mu}}}$
which incorporates the resonance width and thus
quantifies the deviation between ${\mathrm{E_{res}}}$ and ${\mathrm {E^{v}_{peak}}}$
which can exist while still keeping the effect observable.
To include the width of the resonance, we write
${\mathrm {A = \Delta_{31} (\cos 2 \theta_{13} + q \sin 2 \theta_{13})}},
$
where q varies from -1 to 1. With this parametrization,
\bea
{\mathrm {
\Delta P_{\mu\mu}}} &=& {\mathrm {
 - Q_{+} \sin^2 \left[ 1.27 \frac{\Delta_{31}L}{E} \left( 1 -
\sin 2 \theta_{13} \frac{\sqrt{q^2 + 1} - q}{2} \right) \right] }}
\nonumber \\
&-&
{\mathrm {Q_{-}
\sin^2 \left[ 1.27 \frac{\Delta_{31}L}{E} \left( 1 +
\sin 2 \theta_{13} \frac{\sqrt{q^2 + 1} + q}{2} \right) \right]
}}
\nonumber \\
&-&
{\mathrm {1/[4(q^2+1)]
\sin^2 \left[ 1.27 \frac{\Delta_{31}L}{E} \left(
\sin 2 \theta_{13} \sqrt{q^2 + 1}  \right) \right]}}
\nonumber \\
&+& {\mathrm {\sin^2 (1.27 \Delta_{31}L/E)}}
\label{eq:deltapmumuwidth}.
\eea
Here, ${\mathrm {Q_{\pm} = ({\sqrt{q^2 + 1}\pm q})/({2 \sqrt{q^2 +1}})}}$.
In obtaining
Eq.~\ref{eq:deltapmumuwidth}
we have  approximated
${\mathrm {\cos 2 \theta_{13},~ \cos^2 \theta_{13} \simeq 1}}$.

\subsection{\bf{Degeneracies in the determination of oscillation
parameters at very long baselines}}
\label{subsec:degeneracy}

In our analytic discussion of matter effects on oscillation probabilities,
we have used approximate expressions with the solar mass squared difference
$\Delta_{21}$ set to zero. While this is adequate for a broad understanding of matter effects,
it is interesting to look at a more accurate picture with $\Delta_{21} \neq 0$ and analyze
the issue of parameter degeneracies associated with the inclusion of sub-leading effects.
In this situation, the determination of neutrino mass and mixing parameters
is complicated by the presence of degeneracies in the
oscillation probabilities, which are inherent in a three generation analysis
due to the presence of the non-zero CP phase ${\mathrm{\delta_{CP}}}$.
The degeneracies, extensively
studied in the literature for baselines less than 3000 Km,
are {\bf {the $({\mathrm{\delta_{CP}}}, \theta_{13})$
ambiguity, the sign$(\Delta_{31})$ or mass hierarchy
degeneracy and the $(\theta_{23}, \pi/2-\theta_{23})$ or
atmospheric angle degeneracy}}, combining to
give an overall eight-fold degeneracy \cite{lind,barg,mina,yasu}.
It is relevant to discuss briefly the
effect of parameter degeneracies in the context of very long
baselines, as are considered in this paper.
\\\\

\begin{figure}[ht]
\centerline
{
\epsfxsize=8.5cm\epsfysize=6.5cm\epsfbox{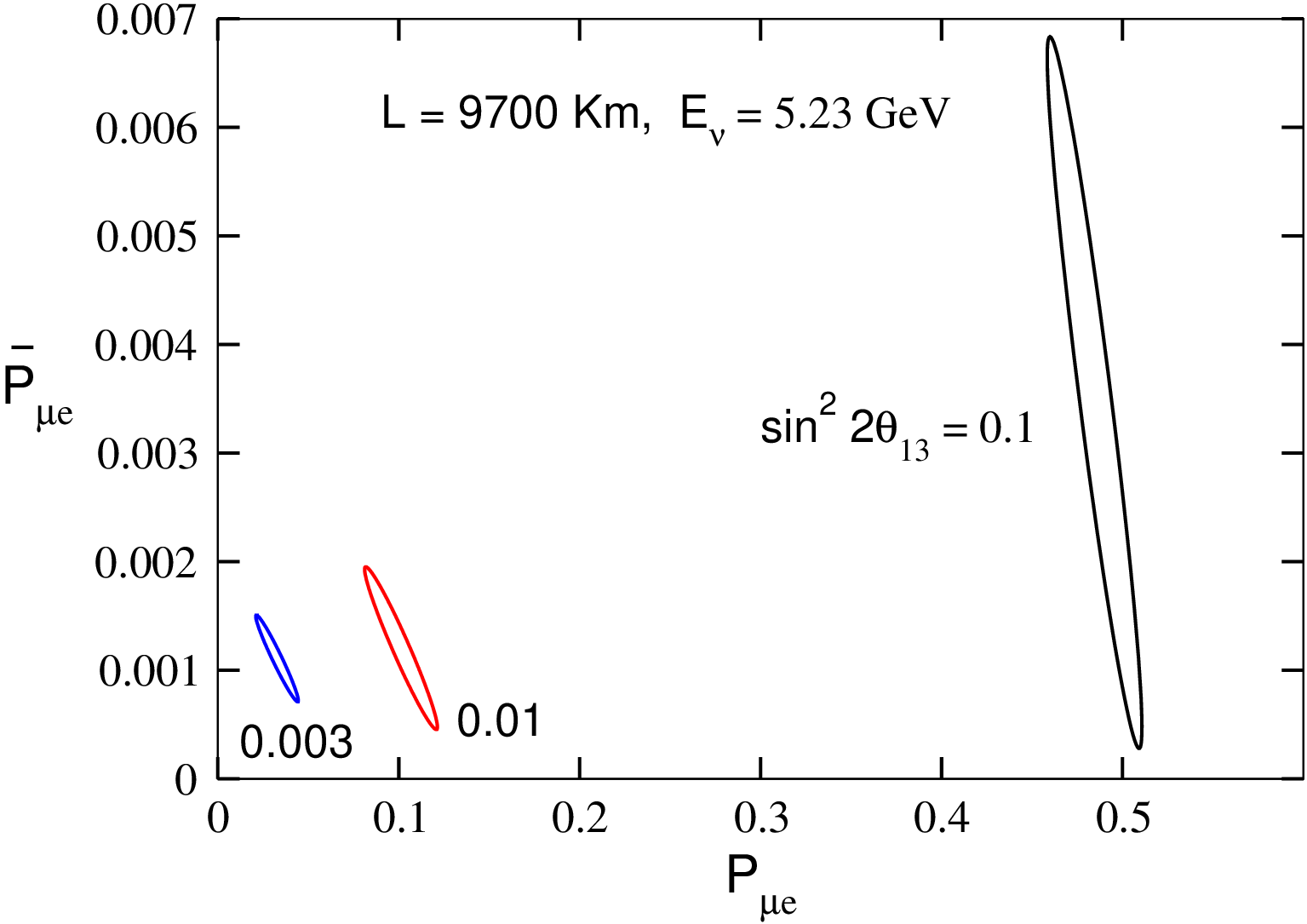}
        \hspace*{0.5ex}
\epsfxsize=8.5cm\epsfysize=6.5cm
                     \epsfbox{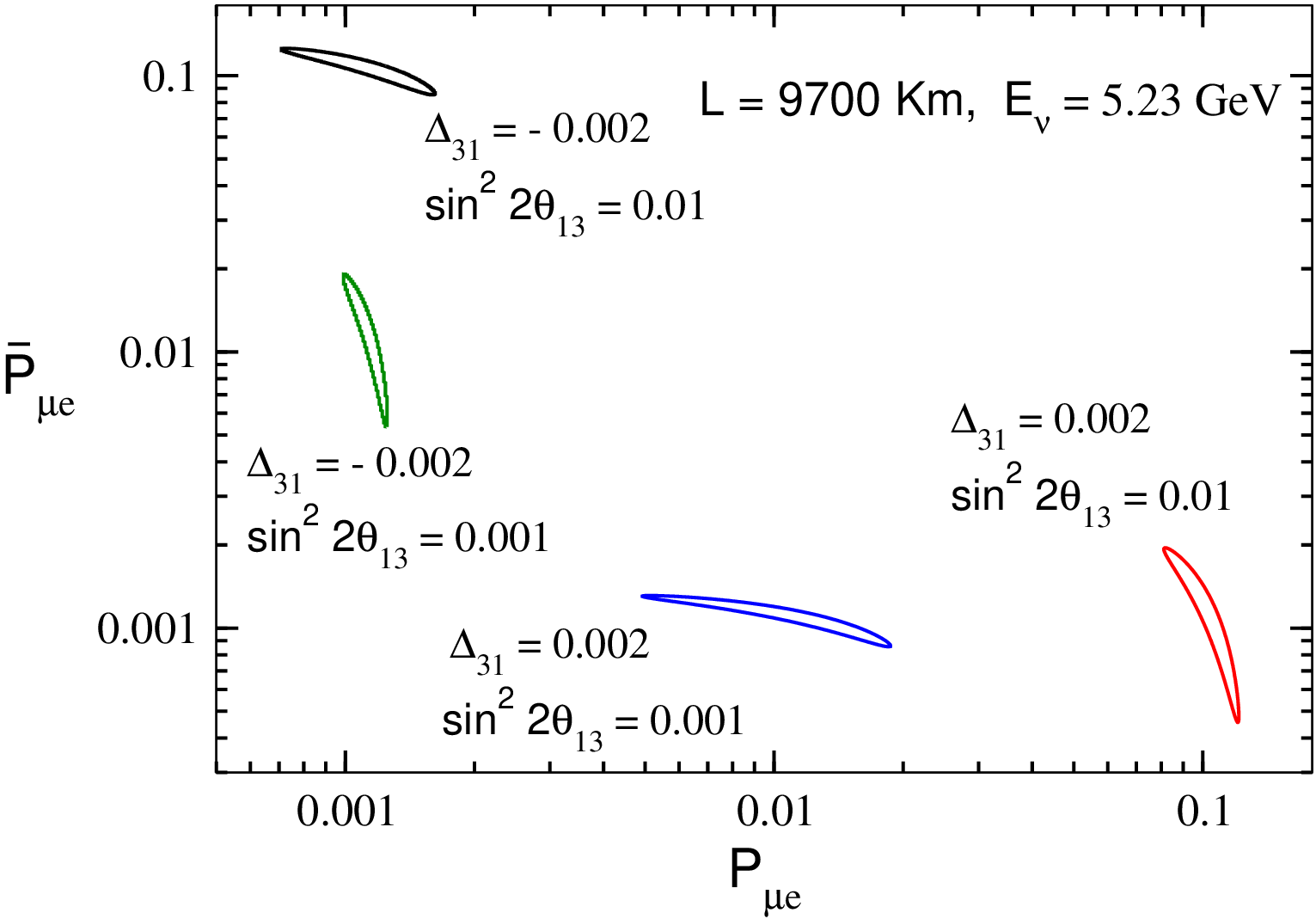}
}
\hskip 5cm
{\bf (a)}
\hskip 7cm
{\bf (b)}
\caption
{\footnotesize {\bf (a)}
CP trajectory orbits in (
${\mathrm P}_{\mu e}$,
$\bar {\mathrm P}_{\mu e}$
)
biprobability space with E = 5.23 GeV, L = 9700 Km
(${\mathrm {1.27 \Delta_{31} L/E}} = 3\pi/2$).
Orbits with different values of $\theta_{13}$ do not
overlap, showing that the $({\mathrm{\delta_{CP}}}, \theta_{13})$
degeneracy is resolved for these values of L and E.
{\bf (b)} CP trajectory plots in
(${\mathrm P}_{\mu e}$, $\bar {\mathrm P}_{\mu e}$)
biprobability space for two different values of
$\theta_{13}$ and both signs of $\Delta_{31}$.
Orbits with positive and negative $\Delta_{31}$ do not coincide,
showing that the sign$(\Delta_{31})$
degeneracy is resolved for this L and E till $\theta_{13}$ as
low as $\sin^2 2\theta_{13}=0.001$.
}
\label{fig8}
\end{figure}

The analytic treatment of degeneracies is based on approximate
expressions for the oscillation probabilities in matter of constant
density. These are computed by series expansions in the small
parameters - the solar mass squared difference $\Delta_{21}$ and the
mixing angle $\theta_{13}$. CP trajectory orbits in biprobability
space are widely used for depicting degeneracies, and are
conventionally plotted using the analytic expressions. For the
purpose of the biprobability calculations here, analytic expressions
become progressively inadequate beyond 4000 Km, or more precisely,
for values of L and E such that ${\mathrm L}/{\mathrm E} \ge 10^4$
Km/GeV. Additionally, the small $\theta_{13}$ expansion also fails
for relatively large values of $\theta_{13}$ (close to the present
CHOOZ bound). In view of this, we have plotted the CP trajectories
for a sample long baseline using the full numerical solution of the
evolution equation with earth matter effects. However, the analytic
expressions remain useful for a qualitative understanding of the
features and interdependence of parameter degeneracies, and we use
them for this purpose whenever necessary.
\\\\

\begin{figure}[ht]
\centerline
{
\epsfxsize=8.5cm\epsfysize=6.5cm \epsfbox{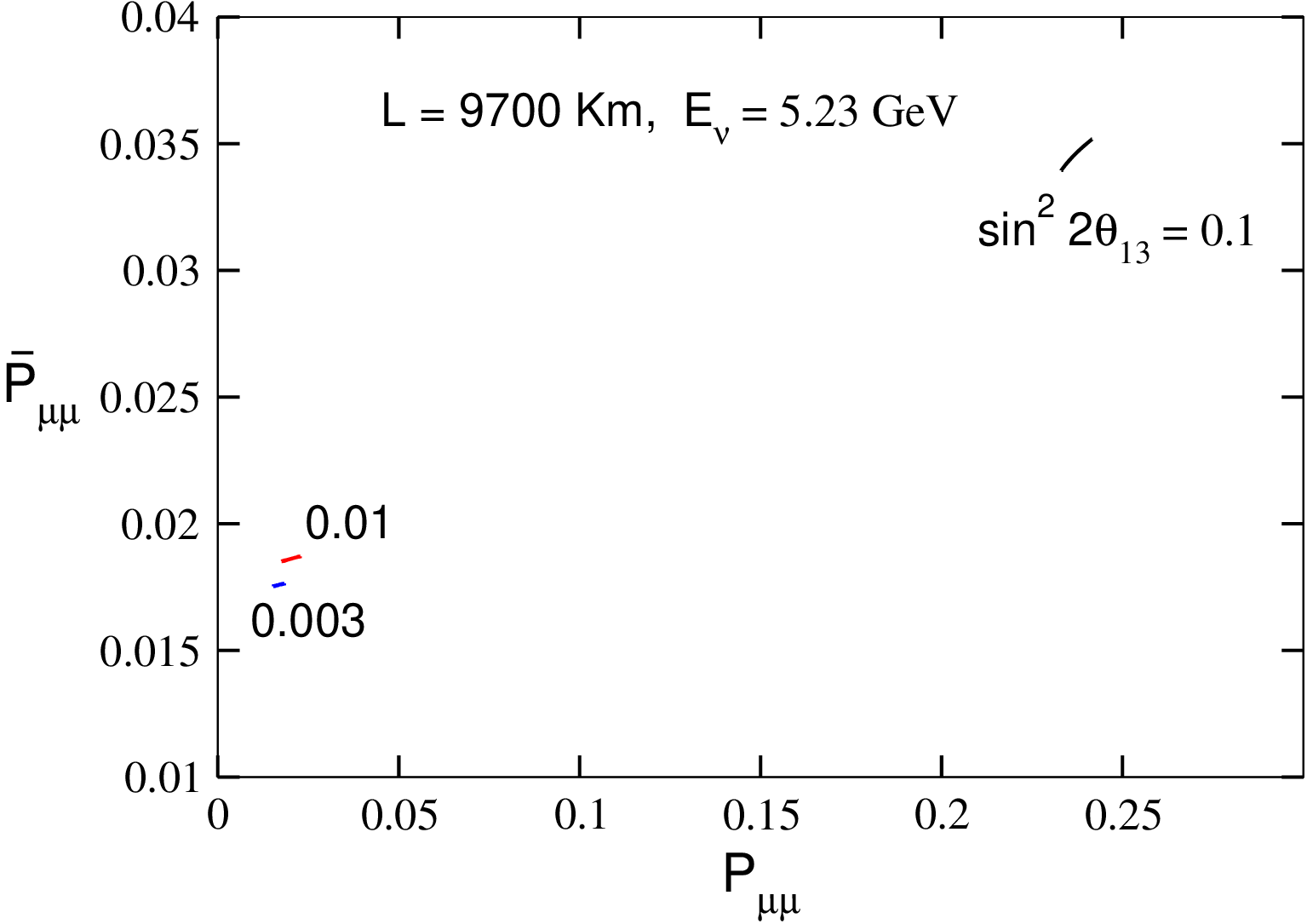}
        \hspace*{0.5ex}
\epsfxsize=8.5cm\epsfysize=6.5cm
             \epsfbox{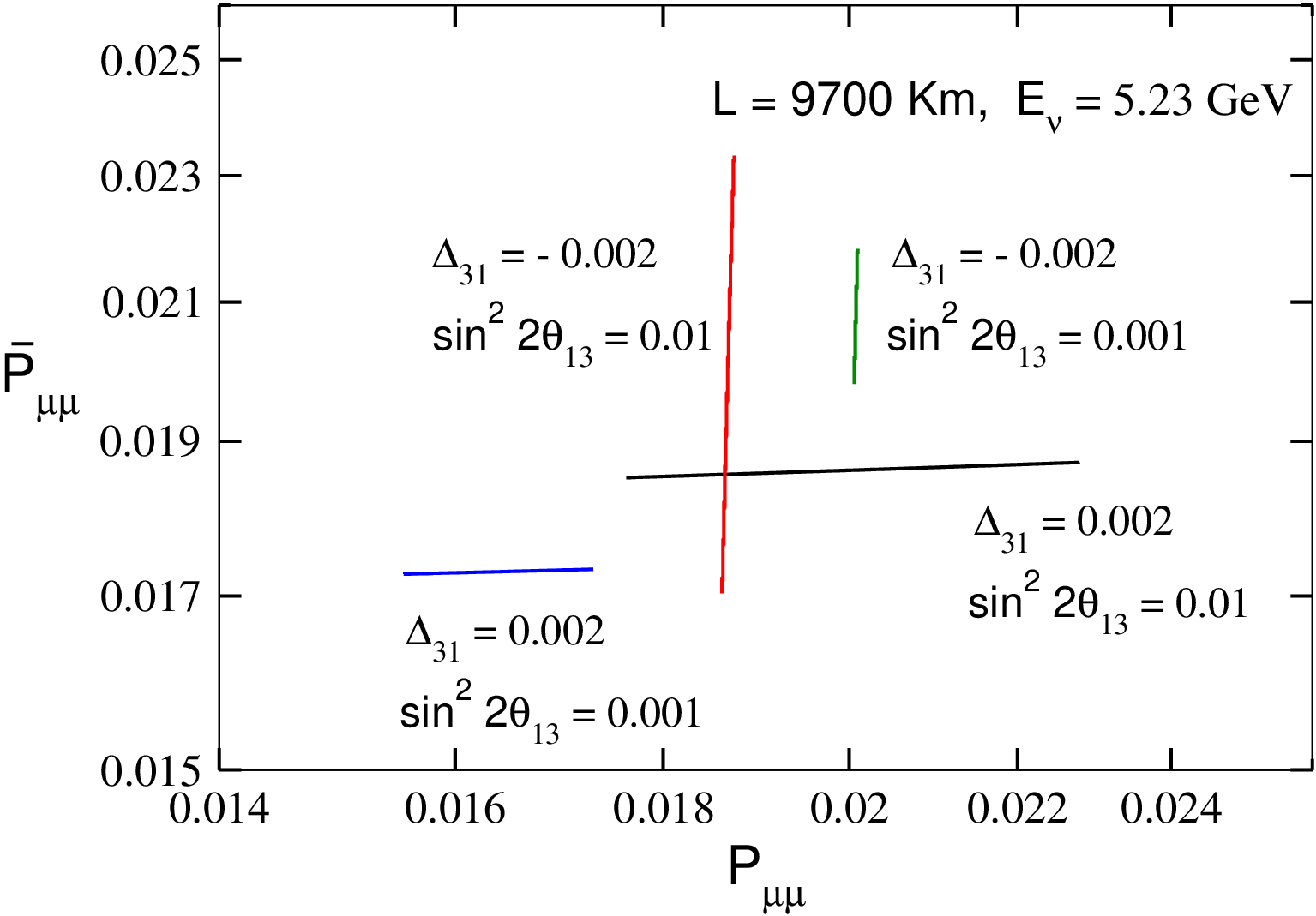}
} \hskip 5cm {\bf (a)} \hskip 7cm {\bf (b)} \caption {\footnotesize
Same as Figure \ref{fig8}(a) and \ref{fig8}(b) but in (${\mathrm
P}_{\mu \mu}$, $\bar {\mathrm P}_{\mu \mu}$) biprobability space. In
Figure {\bf (b)} the sign$(\Delta_{31})$ degeneracy is present for
$\sin^2 2\theta_{13}=0.01$. } \label{fig9}
\end{figure}

The approximate analytic expressions for
oscillation probability between two flavors $\alpha$, $\beta$
are of the form \cite{kim,mina}
\bea
{\mathrm P_{\alpha\beta}} =
{\mathrm X_{\alpha\beta} \cos {\mathrm{\delta_{CP}}} + {\mathrm Y}_{\alpha\beta} \sin {\mathrm{\delta_{CP}}} + {\mathrm Z}_{\alpha\beta}},
\eea
\noindent
where ${\mathrm X_{\alpha\beta}}$, ${\mathrm Y_{\alpha\beta}}$
and ${\mathrm Z_{\alpha\beta}}$ are functions of the neutrino
mass squared differences and mixing parameters but independent
of the CP phase ${\mathrm{\delta_{CP}}}$.
Note that ${\mathrm Z_{\alpha\beta}}$ contains the
dominant contribution to the probabilities
given earlier (Eq.~{\ref{eq:pmuevac}},
Eq.~{\ref{eq:pmuemat}}, Eq.~{\ref{eq:pmutauvac}},
Eq.~{\ref{eq:pmutaumat1}}, Eq.~{\ref{eq:pmumuvac}}, Eq.~{\ref{eq:pmumumat1}}).
It can be shown that the generic form of
a CP trajectory in biprobability space (the orbit traced in
(${\mathrm P_{\alpha\beta}}$,
${\mathrm {\bar P}_{\alpha\beta}}$)
space as ${\mathrm{\delta_{CP}}}$ varies from 0 to 2$\pi$) is an ellipse
\cite{mina}, which collapses to a line under certain conditions.

\noindent Here ${\mathrm {\bar P}_{\alpha\beta}}$ is the CP
conjugated (anti-neutrino) probability. Specific expressions for
${\mathrm {P^{m}_{\mu e}}}$, ${\mathrm {P^{m}_{\mu \tau}}}$ and
${\mathrm {P^{m}_{\mu \mu}}}$ may be found in \cite{akh}. The small
$\Delta_{21}$, small $\theta_{13}$ series expansions (to second
order in these parameters) are compact and hence easier to analyze
for degeneracies. Below we list the expressions for the coefficients
${\mathrm {X_{\alpha\beta}}}$ and ${\mathrm {Y_{\alpha\beta}}}$ in
this approximation for ${\mathrm {P^{m}_{\mu e}}}$ and ${\mathrm
{P^{m}_{\mu \tau}}}$ :
\bea
{\mathrm {X_{\mu e}}} =
{\mathrm {2 \alpha \sin \theta_{13}
\sin 2\theta_{12} \sin 2\theta_{23}
\cos \Delta \frac{\sin {\hat A} \Delta}
{{\hat A}}
\frac{\sin ({\hat A}-1) \Delta}{({\hat A}-1)}}}
~~~~~~~~~~~~~~~~~~ \nonumber \\
{\mathrm {Y_{\mu e}}} = -
{\mathrm {2 \alpha \sin \theta_{13} \sin 2\theta_{12} \sin 2\theta_{23}
\sin \Delta \frac{\sin {\hat A} \Delta}{{\hat A}}
\frac{\sin ({\hat A}-1) \Delta}{({\hat A}-1)} }}
~~~~~~~~~~~~~~~~ \nonumber \\
{\mathrm {X_{\mu \tau}} }
=
{\mathrm{
-
\frac{2}{{\hat A}-1}}}
{\mathrm {
2 \alpha \sin \theta_{13} \sin 2\theta_{12} \sin 2\theta_{23}
\cos 2\theta_{23} \sin \Delta}}
~~~~~~~~~~~~~~~~~~~~~~~ \nonumber \\
~~~~~~~~~~~~~\times
{\mathrm {
[{\hat A}\sin \Delta -
\frac{\sin {\hat A} \Delta}{{\hat A}}
\cos ({\hat A}-1)\Delta]}}
~~~~~~~~~~~~~~\nonumber \\
{\mathrm {
Y_{\mu \tau}}} = {\mathrm{
2 \alpha \sin \theta_{13} \sin 2\theta_{12} \sin 2\theta_{23}
\sin \Delta \frac{\sin {\hat A} \Delta}{{\hat A}}
\frac{\sin ({\hat A}-1) \Delta}{({\hat A}-1)}}}
~~~~~~~~~~~~~~~~~
\label{eq:deltacoeff}
\eea

\noindent
where $\Delta = 1.27 \Delta_{31} {\mathrm L}/{\mathrm E}$,
${\mathrm{{\hat A}}}={\mathrm{A/\vert \Delta_{31} \vert}}$,
$\alpha = \vert \Delta_{21} \vert/\vert \Delta_{31} \vert$.
The above expressions are for the normal mass hierarchy.
For the inverted hierarchy, the transformations
${\mathrm{{\hat A}}} \rightarrow {\mathrm{-{\hat A}}}$, $\alpha \rightarrow -\alpha$ and
$\Delta \rightarrow -\Delta$ are required.
The coefficients for
${\mathrm{P^{m}_{\mu \mu}}}$ are given by
${\mathrm X_{\mu \mu}} = -{\mathrm X_{\mu e}}-{\mathrm X_{\mu \tau}}$,
and similarly for
${\mathrm Y_{\mu \mu}}$.
Note that ${\mathrm Y_{\mu \mu}=0}$ and hence the survival
probability is independent of the CP-odd term.
 Next we discuss in turn each of the possible degeneracies.
\\\\
\begin{figure}[ht]
\centerline
{
\epsfxsize=8.5cm\epsfysize=6.5cm
            \epsfbox{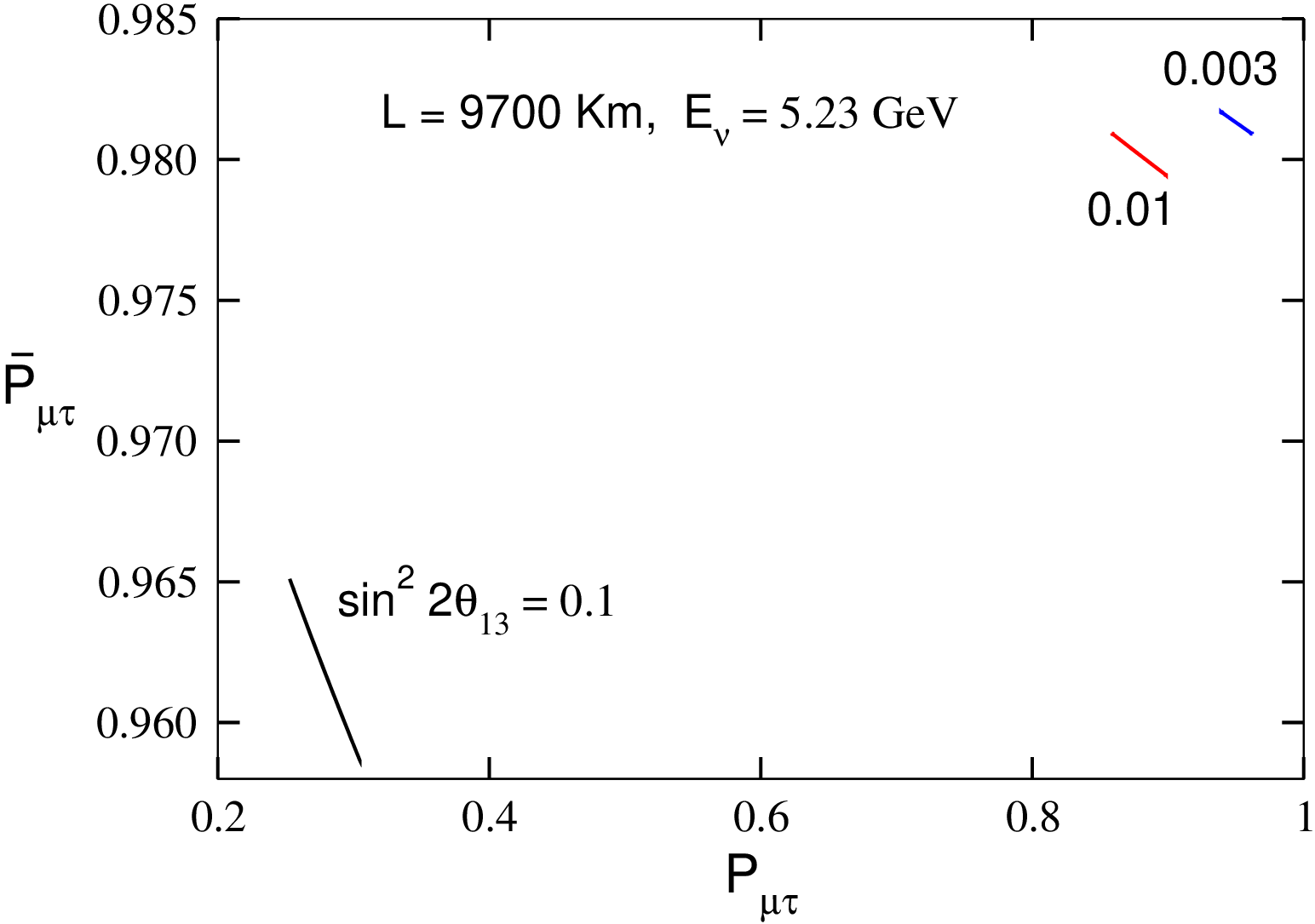}
        \hspace*{0.5ex}
\epsfxsize=8.5cm\epsfysize=6.5cm
            \epsfbox{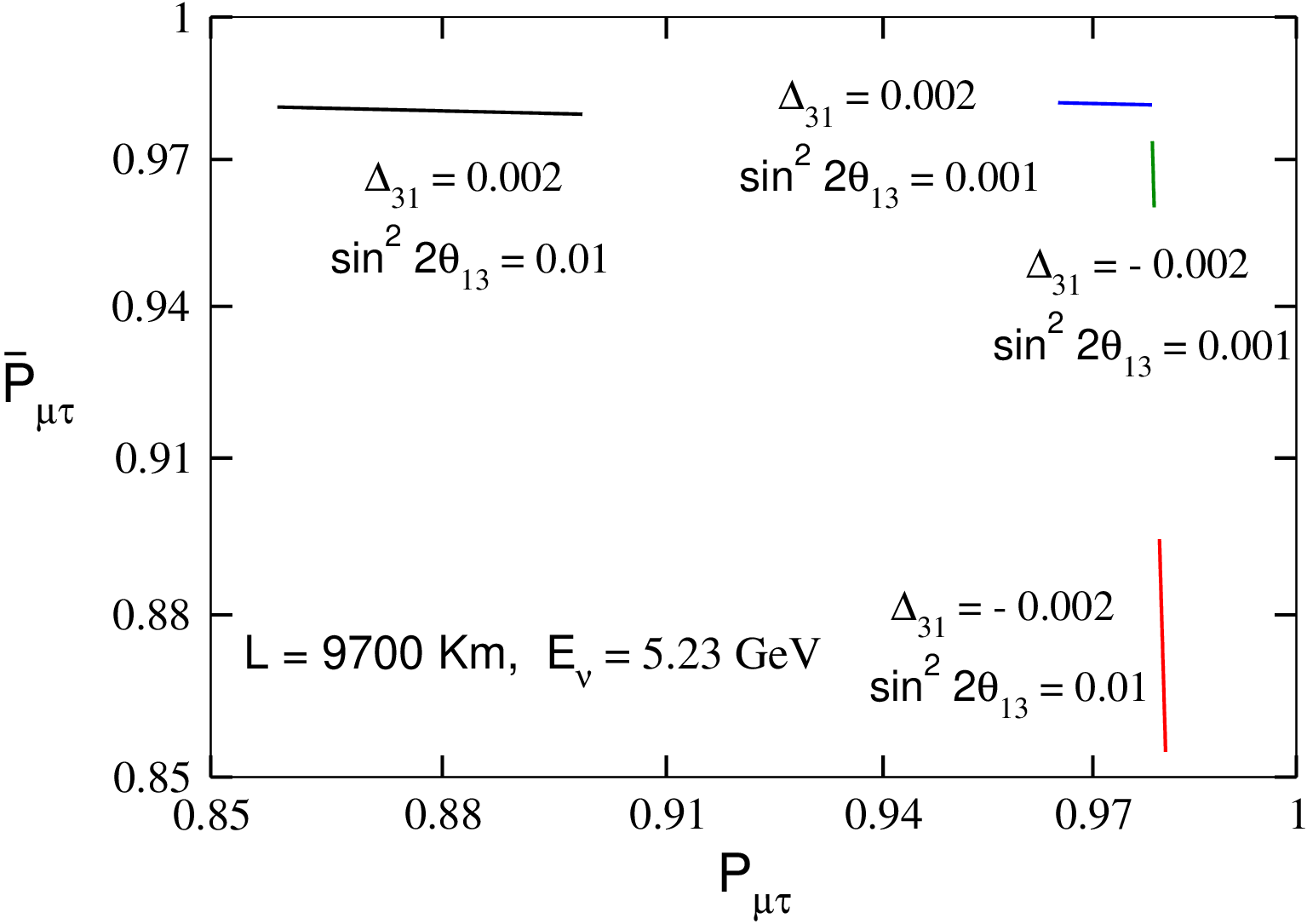}
} \hskip 5cm {\bf (a)} \hskip 7cm {\bf (b)} \caption {\footnotesize
Same as Figure \ref{fig8}(a) and \ref{fig8}(b) in (${\mathrm P}_{\mu
\tau}$, $\bar {\mathrm P}_{\mu \tau}$) biprobability space. }
\label{fig10}
\end{figure}

\begin{itemize}
\item
The {\bf{$({\mathrm{\delta_{CP}}}, \theta_{13})$ degeneracy}} arises
when different pairs of values of the parameters
${\mathrm{\delta_{CP}}}$ and $\theta_{13}$ give the same neutrino
and anti-neutrino oscillation probabilities, assuming other
parameters to be known and fixed.
This may be expressed as \bea
{\mathrm{P_{\alpha\beta}({\mathrm{\delta_{CP}}}, \theta_{13})}} &=&
{\mathrm{P_{\alpha\beta}({\mathrm{\delta_{CP}'}}, \theta_{13}')}} \nonumber \\
{\mathrm{{\bar P}_{\alpha\beta}({\mathrm{\delta_{CP}}},
 \theta_{13})}}
&=&
{\mathrm{{\bar P}_{\alpha\beta}({\mathrm{\delta_{CP}'}}, \theta_{13}')}}
\label{eq:delcpth13}
\eea

This ambiguity manifests itself in the intersection
of CP orbits with different values of $\theta_{13}$. The intersection points
indicate the different values of ${\mathrm{\delta_{CP}}}$ and  $\theta_{13}$
for which the above Eq.~{\ref{eq:delcpth13}} are satisfied.
For values of L and E such that
$\Delta = n\pi/2$, Eq.~\ref{eq:deltacoeff}
predict that either $\sin {\mathrm{\delta_{CP}}}$ or $\cos {\mathrm{\delta_{CP}}}$ drops
out of the expression for ${\mathrm{P^{m}_{\mu e}}}$, reducing the
CP trajectory to a straight line.
Further, it can be shown that the $({\mathrm{\delta_{CP}}}, \theta_{13})$
ambiguity reduces in this case
 to a simple $({\mathrm{\delta_{CP}}}, \pi-{\mathrm{\delta_{CP}}})$ or $({\mathrm{\delta_{CP}}}, 2\pi-{\mathrm{\delta_{CP}}})$
ambiguity which does not mix different values of $\theta_{13}$
\cite{barg}.
This fact holds true at long baselines also, as is depicted in
Figure {\ref{fig8}}(a)
for L = 9700 Km, E = 5.23 GeV (giving $\Delta = 3\pi/2$).
 Here the CP orbits appear as widely separated narrow
ellipses, showing that the $({\mathrm{\delta_{CP}}}, \theta_{13})$
ambiguity is effectively resolved{\footnote{
Here we have given a theoretical discussion of degeneracies in probabilities, and therefore not
taken into account possible error bars
in the experimental results for event rates, which, if large enough, may cause points on
different $\theta_{13}$ orbits to overlap.}}.
CP orbit diagrams for ${\mathrm{P^{m}_{\mu \mu}}}$ and ${\mathrm{P^{m}_{\mu \tau}}}$ are given in
Figure {\ref{fig9}}(a) and {\ref{fig10}}(a),
illustrating a similar lifting of the degeneracy involving
$\theta_{13}$ in both cases.
The approximate expressions
 (Eq.~\ref{eq:deltacoeff}) show that the muon neutrino survival
probability is a function only of the CP-even term $\cos {\mathrm{\delta_{CP}}}$,
while ${\mathrm{P^{m}_{\mu \tau}}}$ only depends on $\sin {\mathrm{\delta_{CP}}}$
when $\theta_{23}$ is maximal, so their CP orbits resemble straight
lines instead of ellipses.
\\\\

\begin{figure}[hb]
\centerline
{
\epsfxsize=8.5cm\epsfysize=6.5cm
        \epsfbox{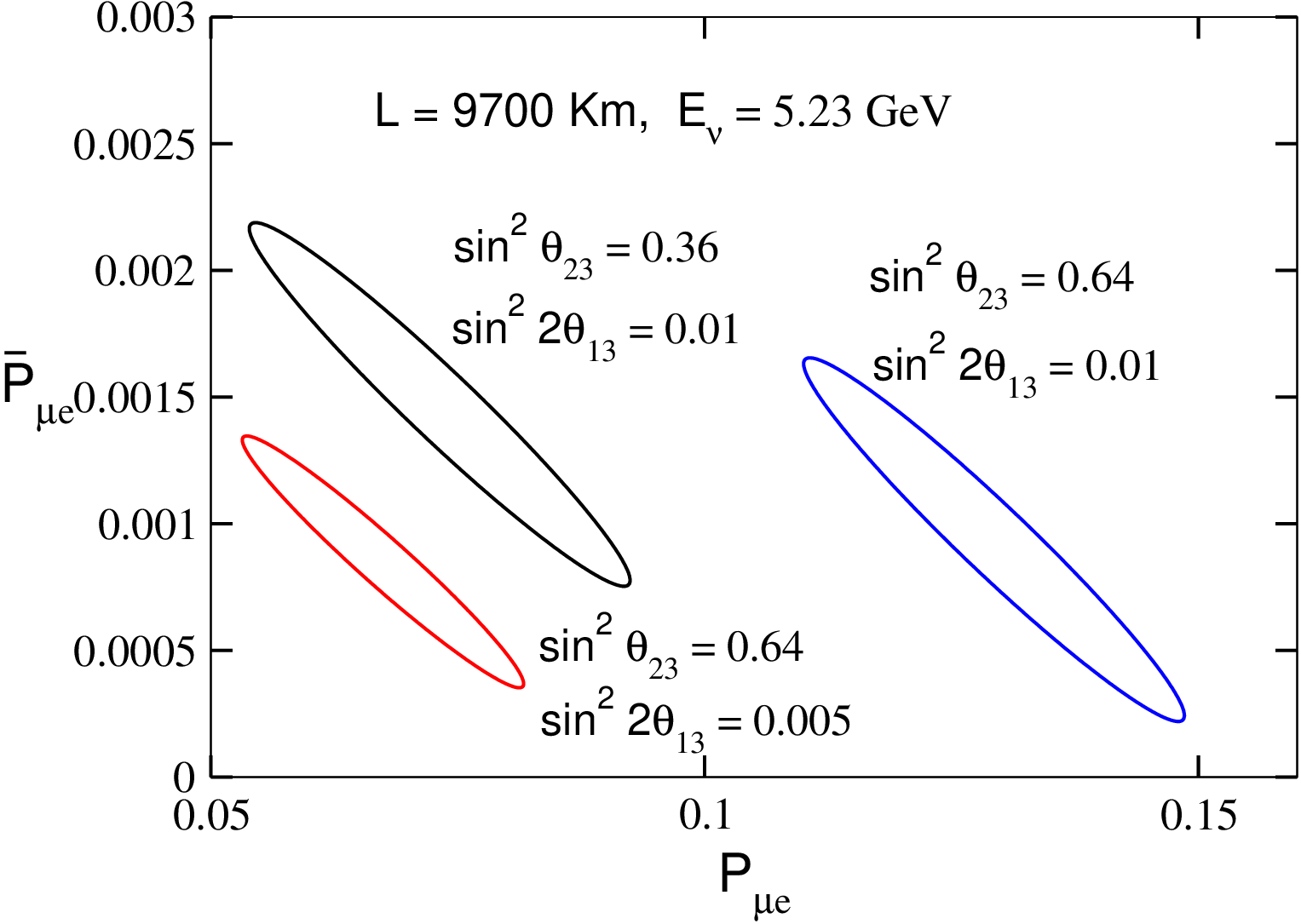}
        \hspace*{0.5ex}
\epsfxsize=8.5cm\epsfysize=6.5cm
             \epsfbox{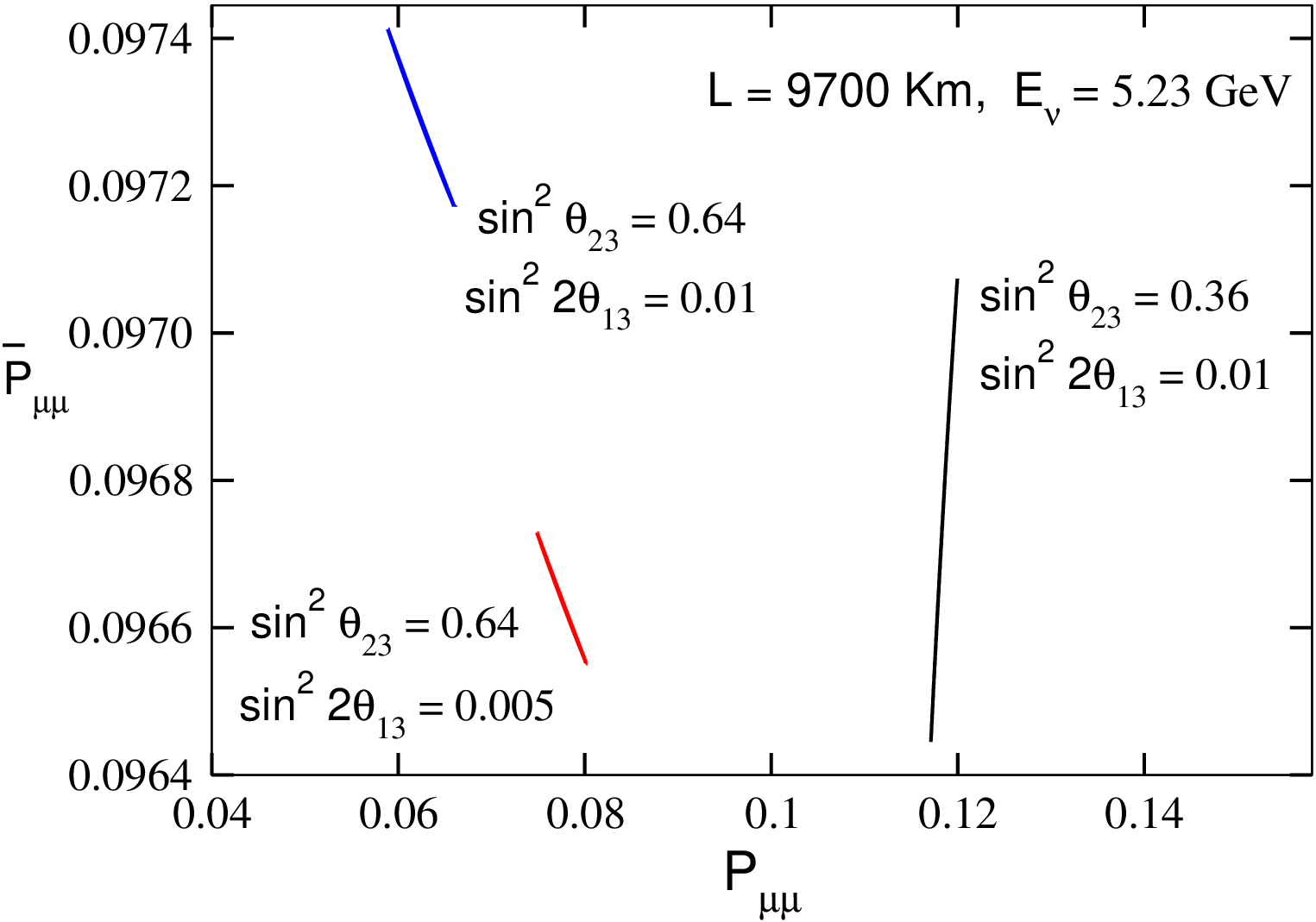}
}
\hskip 5cm
{\bf (a)}
\hskip 7cm
{\bf (b)}
\caption
{\footnotesize
{\bf (a)}
 CP trajectory plots in (${\mathrm P}_{\mu e}$, $\bar {\mathrm P}_{\mu e}$) biprobability space
for different values of $\theta_{13}$ and complementary values of
$\theta_{23}$. {\bf (b)} Same as Figure \ref{fig11}(a) but in
(${\mathrm P}_{\mu \mu}$, $\bar {\mathrm P}_{\mu \mu}$)
biprobability space. } \label{fig11}
\end{figure}

\vskip1.0cm
\begin{figure}[ht]
{\centerline{\hspace*{2em}
\epsfxsize=8cm\epsfysize=6.0cm
                     \epsfbox{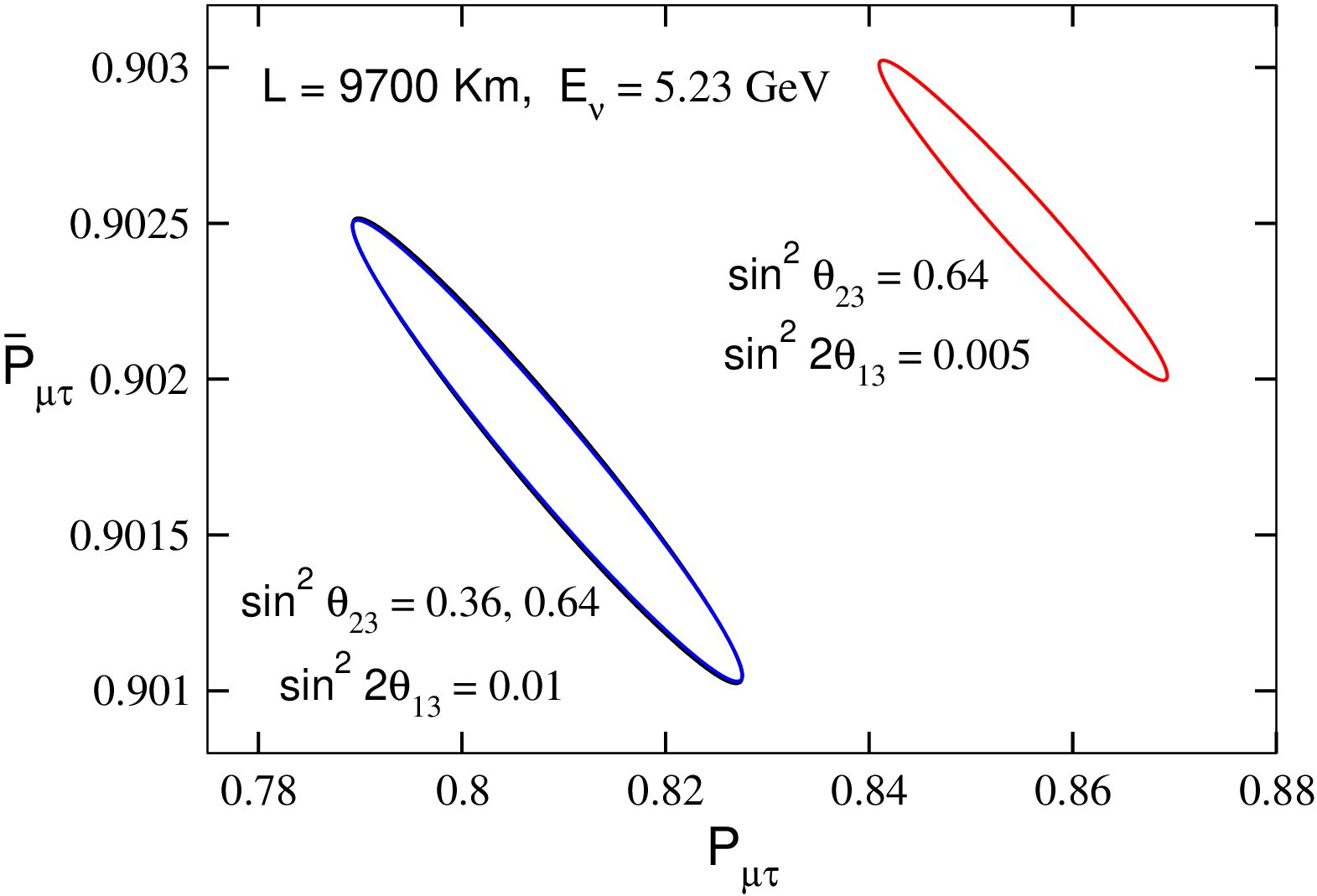}
}
\caption[]{\footnotesize
 Same as Figure \ref{fig9}(a) but in (${\mathrm P}_{\mu \tau}$, $\bar {\mathrm P}_{\mu \tau}$) biprobability space.
}
\label{fig12}
}
 \end{figure}
\item
The {\bf{mass hierarchy degeneracy}} occurs due to identical solutions
for P and ${\mathrm {\bar P}}$ for different pairs of ${\mathrm{\delta_{CP}}}$
and $\theta_{13}$
 with opposite signs of $\Delta_{31}$ (again fixing other parameters):
\bea
{\mathrm{P_{\alpha\beta}({\mathrm{\Delta_{31}>0}},{\mathrm{\delta_{CP}}}, \theta_{13})}}
={\mathrm{P_{\alpha\beta}({\mathrm{\Delta_{31}<0}},{\mathrm{\delta_{CP}'}}, \theta_{13}')}} \nonumber \\
{\mathrm{{\bar P}_{\alpha\beta}({\mathrm{\Delta_{31}>0}},{\mathrm{\delta_{CP}}},\theta_{13})}}
={\mathrm{{\bar P}_{\alpha\beta}({\mathrm{\Delta_{31}<0}},{\mathrm{\delta_{CP}'}}, \theta_{13}')}}
\label{eq:sgndelmsq}
\eea
Note that a combined effect of the $({\mathrm{\delta_{CP}}}, \theta_{13})$ ambiguity
and the sign($\Delta_{31}$) degeneracy gives rise to a four-fold degeneracy.
From the expression
for ${\mathrm{ P^{m}_{\mu e} }}$, it has been determined
\cite{barg} that the condition for this ambiguity to be resolved is
\bea                                                                           \sin 2\theta_{13}
>
\frac
{{\mathrm{2\alpha \tan \theta_{23} \sin 2\theta_{12} \sin {\hat A} \Delta}}}
{{\mathrm{{\hat A}[\frac{\sin(1-{\hat A})\Delta}{1-{\hat A}}
- \frac{\sin(1+{\hat A})\Delta}{1+{\hat A}}]}}}
\eea

\noindent
This places constraints on $\theta_{13}$ as well as the
baseline, which must be large enough for the denominator to be
large (thus weakening the
constraint on $\theta_{13}$). Figure {\ref{fig8}}(b) gives the
orbit diagram for the energy and
baseline earlier discussed.
It is seen that the orbit ellipses for positive and negative
$\Delta_{31}$ are well-separated for $\sin^2 2\theta_{13}=0.001$.
Thus the degeneracy is lifted even for very small values of
$\theta_{13}$. From Figure {\ref{fig10}}(b),
it is clear that ${\mathrm{P^{m}_{\mu \tau}}}$ is also
free from this ambiguity for large baselines.
However,
${\mathrm{P^{m}_{\mu \mu}}}$ shows a mass hierarchy
degeneracy for $\sin^2 2\theta_{13}=0.01$, as seen in
Figure {\ref{fig9}}(b).

\item
Currently $\theta_{23}$ is determined from
${\mathrm{P^{v}_{\mu \mu}}}$ in atmospheric
(SK) and accelerator (K2K) experiments. Since this
is a function of $\sin^2 2\theta_{23}$, these measurements cannot
differentiate $\theta_{23}$ from $\pi/2-\theta_{23}$.
If the probability is a function
of $\sin^2 \theta_{23}$ or $\cos^2 \theta_{23}$
(e.g. the leading contributions in \pmue and ${\mathrm P}_{e \tau}$ respectively),
then for a fixed value of
$\theta_{13}$,
${\mathrm{P_{\alpha\beta}({\mathrm{\theta_{23}}},\theta_{13})}}
\neq {\mathrm{P_{\alpha\beta}({\mathrm{\pi/2-\theta_{23}}},\theta_{13})}}$. However, different values
of $\theta_{13}$ and the CP phase ${\mathrm{\delta_{CP}}}$
can make them equal{\footnote{This degeneracy may be present even if ${\mathrm{\delta_{CP}}}=0$.}},
leading to an ambiguity in the determination
of the latter two parameters known as the {\bf{$(\theta_{23},
\pi/2-\theta_{23})$ degeneracy}}:
\bea
{\mathrm{P_{\alpha\beta}({\mathrm{\theta_{23}}},{\mathrm{\delta_{CP}}}, \theta_{13})}}
={\mathrm{P_{\alpha\beta}({\mathrm{\pi/2-\theta_{23}}},{\mathrm{\delta_{CP}'}}, \theta_{13}')}} \nonumber \\
{\mathrm{{\bar P}_{\alpha\beta}({\mathrm{\theta{23}}},{\mathrm{\delta_{CP}}},\theta_{13})}}
={\mathrm{{\bar P}_{\alpha\beta}({\mathrm{\pi/2-\theta_{23}}},{\mathrm{\delta_{CP}'}}, \theta_{13}')}}
\label{eq:theta23}
\eea
The presence of this ambiguity along with the two previously discussed degeneracies
will give rise to an eight-fold degeneracy \cite{barg} in such probabilities.
However, if a probability is a function of $\sin^2 2\theta_{23}$ (e.g. ${\mathrm {P_{\mu \tau}}}$),
then for a fixed $\theta_{13}$, ${\mathrm{P_{\alpha\beta}({\mathrm{\theta_{23}}},\theta_{13})}}
={\mathrm{P_{\alpha\beta}({\mathrm{\pi/2-\theta_{23}}},\theta_{13})}}$ to start with.
Therefore, different values of $\theta_{13}$ and ${\mathrm{\delta_{CP}}}$
give rise to different values of the probability, and the $(\theta_{23},
\pi/2-\theta_{23})$ degeneracy will not lead to any further ambiguity in the determination
of $\theta_{13}$. So the total degeneracy in this case is
four-fold{\footnote{Note that probabilities that are
functions of $\sin^2 2\theta_{23}$, though
only 4-fold degenerate with
respect to $\theta_{13}$ and ${\mathrm{\delta_{CP}}}$,
cannot ascertain whether $\theta_{23}<\pi/4$ or
$\theta_{23}>\pi/4$. This information can
be obtained from a comparison of probabilities which are functions of $\sin^2 \theta_{23}$ and $\cos^2 \theta_{23}$
(for {\it e.g.}, the golden and silver channels discussed in \cite{cerv,doni}).}}.

The CP trajectories for ${\mathrm{P^{m}_{\mu e}}}$,
${\mathrm{P^{m}_{\mu \mu}}}$ and ${\mathrm{P^{m}_{\mu \tau}}}$
with complementary values of $\theta_{23}$ and
our sample L and E are given in Figure {\ref{fig11}}(a) and (b)
and Figure {\ref{fig12}}.
From Figure {\ref{fig11}}(a) and (b),
we see that the different $\theta_{23}$ orbits do not intersect,
i.e. this degeneracy is resolved in
${\mathrm{P^{m}_{\mu e}}}$ and ${\mathrm{P^{m}_{\mu \mu}}}$
for a long baseline like the one discussed, even though in principle
it could exist for these probabilities as they are functions of
$\sin^2 \theta_{23}$.
However, in the case of ${\mathrm{P^{m}_{\mu \tau}}}$,
as discussed above, this degeneracy is absent
since its approximate matter expression contains only
$\sin^2 2\theta_{23}$ in its dominant terms.
This feature is evident in Figure {\ref{fig12}}.
We see that all points on the orbits with complementary
$\theta_{23}$ and equal $\theta_{13}$ coincide,
so no two independent values of ${\mathrm{\delta_{CP}}}$
can exist which give the same probability.

\item
It can also be observed from the analytic expressions for ${\mathrm X_{\alpha \beta}}$
and ${\mathrm Y_{\alpha \beta}}$ (Eq.~\ref{eq:deltacoeff}) that
the magic baseline scenario \cite{barg,magic} discussed for
${\mathrm{P^{m}_{\mu e}}}$
is equally valid for ${\mathrm{P^{m}_{\mu \mu}}}$ and ${\mathrm{P^{m}_{\mu \tau}}}$ when $\theta_{23}=\pi/4$.
This method of degeneracy resolution requires the baseline to satisfy ${\mathrm{{\hat A}}} \Delta=n\pi$,
in which case all coefficients of ${\mathrm{\delta_{CP}}}$ vanish. The energy and
parameter-independent condition on L
becomes $\rho {\mathrm L} = 32532$ Km gm/cc.
From the ${\mathrm{\rho L}}$ vs L curve (Figure {\ref{fig1}}),
it is seen that this corresponds to L $\simeq$ 7600 Km.
An identical analysis applies to ${\mathrm{P^{m}_{\mu \mu}}}$
and ${\mathrm{P^{m}_{\mu \tau}}}$ with maximal $\theta_{23}$,
with only a small correction if $\theta_{23}$ is non-maximal.

\end{itemize}

In conclusion, we find that a study of probabilities at a baseline and energy
like the one discussed demonstrates, in general, a breaking of most of the parameter degeneracies
which can confuse the determination of $\theta_{13}$ from oscillation measurements.
For ${\mathrm {P_{\mu e}}}$, this is true for all three classes of degeneracies mentioned above.
In the case of the muon survival probability, Figure {\ref{fig9}}(b) shows
that sign$(\Delta_{31})$ may still be undetermined for specific values of
$\theta_{13}$ and the CP phase
${\mathrm{\delta_{CP}}}$, for which the same value of (${\mathrm {P_{\mu \mu}}}$,${\mathrm {\bar P}_{\mu \mu}}$)
 are obtained with opposite signs of $\Delta_{31}$. For example, for
$\sin^2 2\theta_{13}=0.01$, we have checked that ${\mathrm{\delta_{CP}}}=0.305\pi$ and ${\mathrm{\delta_{CP}}}=0.337\pi$
give the same point in (${\mathrm {P_{\mu \mu}}}$,${\mathrm {\bar P}_{\mu \mu}}$) space
for positive and negative $\Delta_{31}$ respectively for  L $=$ 9700 Km and E $=$ 5.23 GeV.
In ${\mathrm {P_{\mu \tau}}}$, all degeneracies involving a measurement of $\theta_{13}$
and ${\mathrm{\delta_{CP}}}$ are lifted for the energy and baseline discussed.
Also, we note that the maximum total degeneracy of \pmutau
is four-fold even in principle, since
its analytic expression in matter (to second order in the expansion parameters)
is a function of $\sin 2\theta_{23}$.
\\\\

\section{Determining the Mass Hierarchy via Atmospheric
           $\nu_{\mu}$ in a charge discriminating Detector}

In the discussion in Sections 2.1-2.3, we have
shown that large matter effects in neutrino
oscillations are not necessarily confined to
\numutonue  or \nuetonutau conversions, but can be searched for in
\numutonutau oscillation and
\numutonumu survival probabilities.
We have examined their origin by studying the
inter-relations of all the
three matter
probabilities,
${\mathrm{P^{m}_{\mu e}}}$,
${\mathrm{P^{m}_{\mu \tau}}}$ and
${\mathrm{P^{m}_{\mu \mu}}}$, and identified baseline
and energy ranges where they act coherently to give
observationally large effects.
The effects discussed are strongly sensitive to the sign of
${\mathrm{\Delta_{31}}}$, as is apparent from the plots shown in
Figure~\ref{fig4}.
It is useful to recall Eq.~\ref{eq:deltapmumueqn} and emphasize that
contrary to what one generally assumes,
 ${\mathrm {\Delta P_{\mu \mu}}}$ is not necessarily
negative, i.e. \pmumum is not always less than or equal to
${\mathrm P}^{\mathrm{v}}_{\mu \mu}$.
This would be true if large matter effects
resided only in ${\mathrm {P^{m}_{\mu e}}}$, since
${\mathrm {\Delta P_{\mu e}}}$
is positive over the relevant range of energies and baselines.
However,
as shown earlier, matter effects can not only lead to a
substantial enhancement of ${\mathrm {P^{m}_{\mu e}}}$, but
can also cause an appreciable drop or rise in ${\mathrm {P^{m}_{\mu \tau}}}$,
i.e. ${\mathrm {\Delta P_{\mu \tau}}}$ can be
significant.
Depending on the extent of this, {\it
\pmumum may be larger or smaller than ${\mathrm P}^{\mathrm{v}}_{\mu \mu}$}.
In this section, we focus on the detection of matter effects
in muon and anti-muon survival rates in atmospheric neutrinos.


Atmospheric neutrinos, while not a controlled source in the sense of the
 beam experiments described above, offer a compensating advantage: a very broad
 L/E band extending about  5 orders of magnitude
(1 to $10^5$ Km/GeV, with neutrino energies,
${\mathrm E}_{\nu} \sim 1$ GeV and above,
L from a few Km to about 12500 Km).
The longer baseline lengths allow matter effects to develop,
and
offer possibilities for determination of the mass hierarchy using a
detector capable of identifying muon charge.
Other physics possibilities with a charge discriminating detector using either
atmospheric neutrinos or a neutrino beam as source have been discussed in
\cite{nufac,pet,cpt,debchou,probir}.

\begin{table}[htb]
\begin{center}
\begin{tabular}
{||c|c|c|c|c|c||} \hline \hline &&&&&
\\
\hspace{0.5cm}
{\sf {E (GeV)}} $\Rightarrow$ \hspace{0.5cm} & \hspace{0.5cm} 2 - 3 \hspace{0.5cm} & \hspace{0.5cm} 3 - 5 \hspace{0.5cm} & \hspace{0.5cm} 5 - 7 \hspace{0.5cm} &  \hspace{0.5cm}
7 - 10 \hspace{0.5cm} & \hspace{0.5cm} {{2 - 10}} \hspace{0.5cm}
\\
\hspace{0.5cm} {\sf {L (Km)}} $\Downarrow$ \hspace{0.5cm}
&{\sf{${\mathrm{N_{mat}, N_{vac}}}$}} & {\sf{${\mathrm{N_{mat},
N_{vac}}}$}}  & {\sf{${\mathrm{N_{mat}, N_{vac}}}$}}
&{\sf{${\mathrm{N_{mat}, N_{vac}}}$}} &{\sf{${\mathrm{N_{mat},
N_{vac}}}$}}
 \\ &&&&& \\
\hline
2000 - 4000 & 99, 102 & 37,  41 & 11,  12 & 23,  23 & 170, 178
\\\hline
4000 - 6000 & 55,  59 & 92,  93 & 17,  19 & 3,  3 & 167, 174
\\\hline
6000 - 8000 & 71, 71 & 47,  48 & 45,  45  & 10,  12 & 173, 176
\\\hline
8000 - 9700 & 49, 49  & 47,  48 & 27,  25 & 22,  23  & 145, 145
\\\hline
9700 - 10500 &
25,  26 & 20,  21 & 6,  5  & 12,  13   & 63,  65
\\\hline
10500 - 12500 & 60,  58  & 54,  55   & 10,  11 & 27,  26 & 151,  150
\\
\hline
2000 - 12500 &
359, 365 & 297, 306 & 116, 117
& 97, 100 & 869,  888
\\
\hline\hline
\end{tabular}
\end{center}
\caption[]{\footnotesize{Number of $\mu^+$ events in matter and in
vacuum in restricted bins of E and L for $\Delta_{31}=0.002$ eV$^2$
and $\sin^2 2\theta_{13}=0.1$, $\sin^2 2\theta_{23}=1.0$.} }
\label{table1}
\end{table}
\begin{table}[htb]
\begin{center}
\begin{tabular}
{||c|c|c|c|c|c||} \hline \hline &&&&&
\\
\hspace{0.5cm}
{\sf {E (GeV)}} $\Rightarrow$ \hspace{0.5cm} & \hspace{0.5cm} 2 - 3 \hspace{0.5cm} & \hspace{0.5cm} 3 - 5 \hspace{0.5cm} & \hspace{0.5cm} 5 - 7 \hspace{0.5cm} &  \hspace{0.5cm}
7 - 10 \hspace{0.5cm} & \hspace{0.5cm} {{2 - 10}} \hspace{0.5cm}
\\
\hspace{0.5cm} {\sf {L (Km)}} $\Downarrow$ \hspace{0.5cm} &
{\sf{${\mathrm{N_{mat}, N_{vac}}}$}} & {\sf{${\mathrm{N_{mat},
N_{vac}}}$}}  & {\sf{${\mathrm{N_{mat}, N_{vac}}}$}}  &
{\sf{${\mathrm{N_{mat}, N_{vac}}}$}} &
{\sf{${\mathrm{N_{mat}, N_{vac}}}$}} \\ &&&&& \\
\hline
2000 - 4000 & 229,  233 & 99,  93 & 31,  28 & 55,  53 & 414, 407
\\
\hline
4000 - 6000 & 134,  139 & 191,  215 & 42,  45 & 10,  8 & 377, 407
\\\hline
6000 - 8000 & 168,  169 & 115,  113 & {\bf 81,  109}  & {\bf
21,  30} & 385, 421
\\\hline
8000 - 9700 & 120,  118  & 129,  115 & {\bf 59,  63} & {\bf
43,  59}  & 351, 356
\\\hline
9700 - 10500 &
63,  62 & 39,  51 & 24,  13  & 29,  32   & 155,  158
\\\hline
10500 - 12500 & 136,  138  & 118,  138   & 29,  28  & 70,  69 & 353,
373 \\
\hline
2000 - 12500 & 850, 859 & 691, 725 & 266, 287
& 228, 251 & 2035, 2122
\\
\hline\hline
\end{tabular}
\end{center}
\caption[]{\footnotesize{Number of $\mu^-$ events in matter and in
vacuum in restricted bins of E and L for $\Delta_{31}=0.002$ eV$^2$
and $\sin^2 2\theta_{13}=0.1$, $\sin^2 2\theta_{23}=1.0$ (see text
for details). }} \label{table2}
\end{table}
Prior to giving the results of our calculations, we briefly describe some
of the inputs we have used.
The prototype for our calculations is a 100 kT Iron
detector, with detection and charge discrimination capability
for muons provided by a magnetic field of about 1.2 Tesla.
Sensitive elements are assumed to be Glass Spark Resistive Plate Chambers
(RPC). We have assumed a (modest) 50\% efficiency of the detector for muon
detection. This incorporates the kinematic cuts
as well as the detection efficiency.
The L/E resolution will depend on event kinematics and
on muon track detection capabilities{\footnote{For our study, we
do not use any resolution function. For the most part in what follows,
we draw our conclusions based on event rates summed over
fairly large ranges in L and/or E.}}.
In the calculations presented here, we have used the Bartol\cite{bartol}
atmospheric flux and set a muon detection threshold of $2$ GeV.
These specifications have been culled from the MONOLITH \cite{monolith}
and INO \cite{ino} proposals and hence are realistic to the best of
our knowledge.
The main systematic errors in this experiment are those that arise
in the determination of the energy and the direction (and hence
pathlength) of the neutrino in the initial state.
As is true in all neutrino experiments using extra terrestrial sources,
the statistical errors are expected to dominate the systematic errors.
So, in this discussion, we have not taken systematic errors into account.

In order to demonstrate the important qualitative features
of our results, we have selectively provided both tables
as well as plots of event rates
versus L and L/E for positive $\Delta_{31}$.
Atmospheric neutrinos offer the advantage of
being able to appropriately select ranges in
L and E from a large spectrum of L and E values.

\begin{figure}[!h]
\vskip 1cm
{\centerline{
\hspace*{2em}
\epsfxsize=11cm\epsfysize=7.5cm
                     \epsfbox{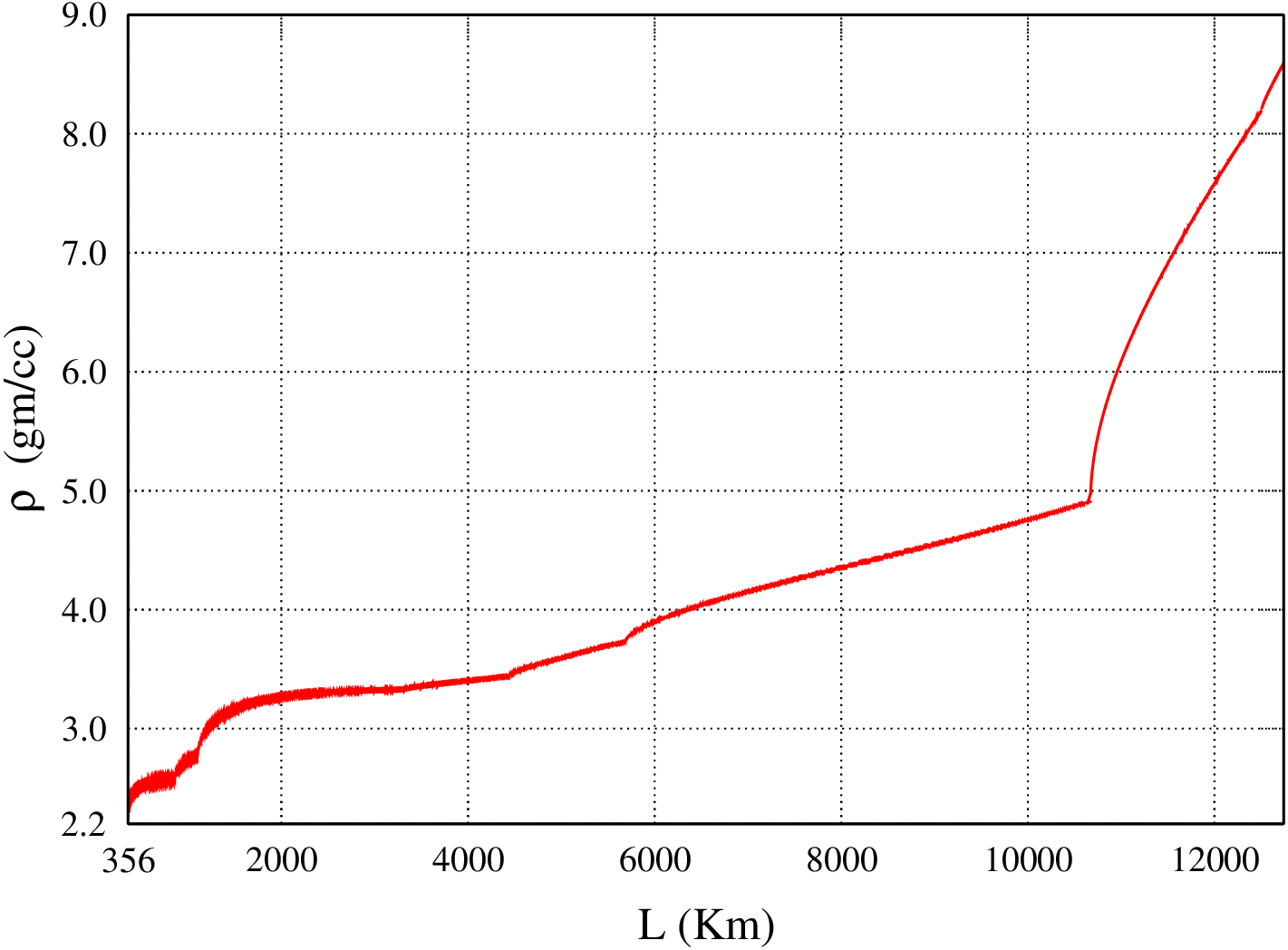}
}
\caption{\footnotesize The average density ${\mathrm{\rho}}$ for
each baseline L, obtained from the PREM density profile, plotted vs L.}
\label{fig13}
}
 \end{figure}

An approximate feel for where significant matter effects
are likely to show up in muon survival rates can be obtained
using our earlier discussion in Sections 2.1-2.3, especially
that which leads to
Eq.~\ref{eq:eres}, Eq.~\ref{eq:muecondtn},
Eq.~\ref{eq:mutaucondtn} and Eq.~\ref{eq:mumucondtn}.
In particular, the resonance condition (Eq.~\ref{eq:eres})
gives the following constraint on the product of
density and energy
\be
{\mathrm{\rho E_{res}}} = 1.315 \times 10^{4}~\Delta_{31}
\cos 2 \theta_{13}~{\mathrm{GeV~gm/cc}}
\label{eq:rhoe1}
\ee
From Eq.~\ref{eq:rhoe1}, using the value
$\Delta_{31}=0.002$ eV$^2$ and approximating to the case
of a constant average density ${\mathrm{\rho}}$ for any given baseline,
one obtains the expression{\footnote{Note that the resonance condition is not very sensitive to the
value of $\theta_{13}$, since the dependence is through
a cosine.
However, it does depend on what value of $\Delta_{31}$ we use. For example,
for $\Delta_{31} = 0.002$ eV$^2$, ${\mathrm{\rho E_{res}}} = 26.3 \times \cos 2 \theta_{13}$,
which gives ${\mathrm{\rho E_{res}}}=24.96$ for $\sin^2 2\theta_{13}=0.1$
and ${\mathrm{\rho E_{res}}}=26.18$ for $\sin^2 2\theta_{13}=0.01$.
However, for $\Delta_{31} = 0.0015$ eV$^2$,
${\mathrm{\rho E_{res}}} = 19.7 \times \cos 2 \theta_{13}$,
giving ${\mathrm{\rho E_{res}}}=18.7$ for $\sin^2 2\theta_{13}=0.1$
and ${\mathrm{\rho E_{res}}}=19.6$ for $\sin^2 2\theta_{13}=0.01$.}}
\be
{\mathrm{\rho E_{res}}} \simeq {\mathrm{25~GeV~gm/cc}}.
\label{eq:rhoe}
\ee
In Figure {\ref{fig13}}, we plot the average density,
${\mathrm{\rho}}$ as a function of the pathlength, L for the earth.
A rough measure of ${\mathrm{E_{res}}}$ for any given baseline
may thus be obtained using this plot and Eq.~\ref{eq:rhoe}.
This is also useful in understanding the broad features of
the tables representing our actual calculations, and in selecting baseline
and energy ranges for closer scrutiny, even though the constant density approximation
fails once baselines are very long.

We now proceed with the discussion of our event rate calculations
which, as stated earlier, have been performed
by solving the full three flavour neutrino
propagation equation using PREM \cite{prem} density profile of the earth.
We have scanned the energy range
2-10 GeV{\footnote{Higher values of energies are also possible, but the
falling flux factor considerably diminishes the event rates. Therefore we take
E upto 10 GeV only.}} and the L range of 2000-12500 Km for evidence of matter effects.
We have assumed $\Delta_{21} = 8.3 \times 10^{-5}$ eV$^2$,
$\sin^2\theta_{12} = 0.27$\cite{keg}, ${\mathrm{\delta_{CP}}}=0$ and an
exposure of 1000 kT-yr for the tables and all plots.
Note that for baselines $>$ 10500 Km, passage is through the core of the earth
\cite{akhm}.


In Table {\ref{table1}}, we show the
$\mu^{+}$ event rates in matter and in
vacuum for
various energy and baseline bins.
As expected, matter effects are {\it negligible} for anti-muons if
$\Delta_{31}$ is positive.

 \begin{figure}[!h]
{\centerline{\hspace*{2em}
\epsfxsize=10cm\epsfysize=8.0cm
                     \epsfbox{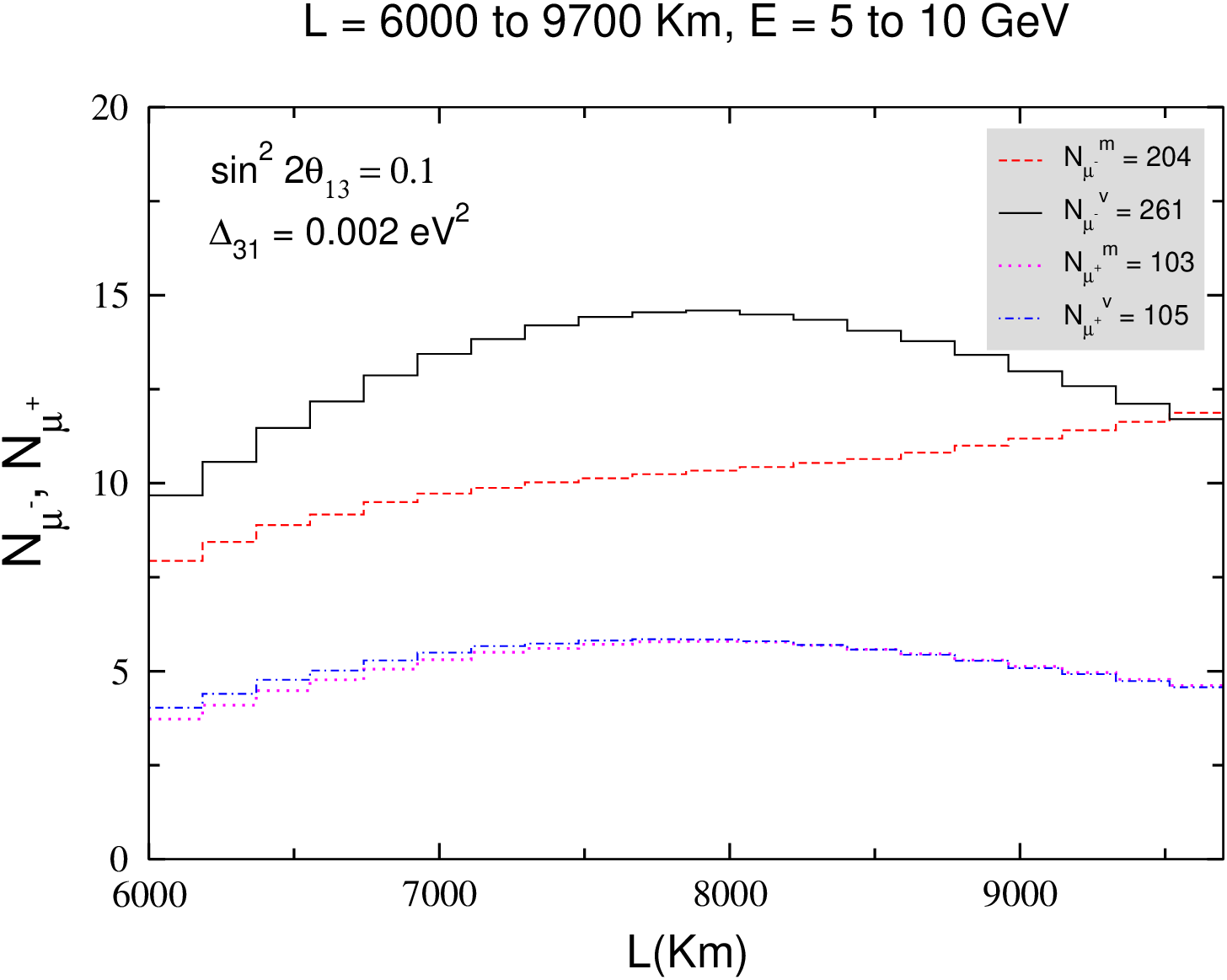}
}
\caption[]{\footnotesize
{Atmospheric
muon and anti-muon events plotted against L
for energy range of 5 GeV to 10 GeV and baselines
between 6000 Km to 9700 Km.
The numbers in the legend correspond to the integrated
number of muon and anti-muon
events for the restricted range of L and E
in matter and in vacuum for a given value of
$\sin^2 2\theta_{13}$.
}
}
\label{fig14}
}
 \end{figure}

Table {\ref{table2}} shows the atmospheric event rates for the same
bin choices, but for muons.
We note that while there are visible differences
between matter and vacuum rates, they are not
always significantly large. A careful
{\it selection of energy ranges and
baselines} is necessary to
extract a statistically significant signal.
Our plots below reflect these selections.
In particular, in Table {\ref{table2}},
a $\sim 4\sigma$ signal for matter effects
can be extracted from the energy bins 5-10 GeV
and the L bins 6000-9700 Km (shown in bold text in the table), as discussed below.
Consolidating the event rates over large bins
in energy and baseline will lead to a dilution,
or possibly a wash-out of the signal, as is evident
from this table.
For example, the right-most vertical column shows the energy
integrated events for the range 2-10 GeV and the last row
lists the events integrated over the 2000-12500 Km
baseline range. In both cases, one notes a {\it dilution} of
the signal for matter effects. We also note that for neutrinos passing through the core \cite{akhm},
the difference between the
matter and vacuum rates is not appreciable except
in the energy range 3-5 GeV for these values
of parameters\footnote{
Varying $\sin^2 \theta_{23}$ and keeping $\sin^2 2\theta_{13}$
fixed does not change the muon event rates significantly.
For instance, for $\sin^2 \theta_{23}=0.6$ and
$\sin^2 2\theta_{13}=0.1$,
we get 116 events in matter as compared to 139 events in vacuum
in the energy bin E = 3 - 5 GeV and baseline range L = 10500 - 12500 Km.
However, as expected, decreased values of
$\theta_{13}$ do diminish the statistical significance of the signal.
For $\sin^2 \theta_{23}=0.4$ and $\sin^2 2\theta_{13}=0.05$,
we get 132 events in matter while in
vacuum we get 140 events in the energy bin
E = 3 - 5 GeV and baseline range L = 10500 - 12500 Km.}
(Table {\ref{table2}}, sixth row).
These effects may be
looked for in
neutrino factory experiments.

 \begin{figure}[!h]
{\centerline{\hspace*{2em}
\epsfxsize=10cm\epsfysize=8.0cm
                     \epsfbox{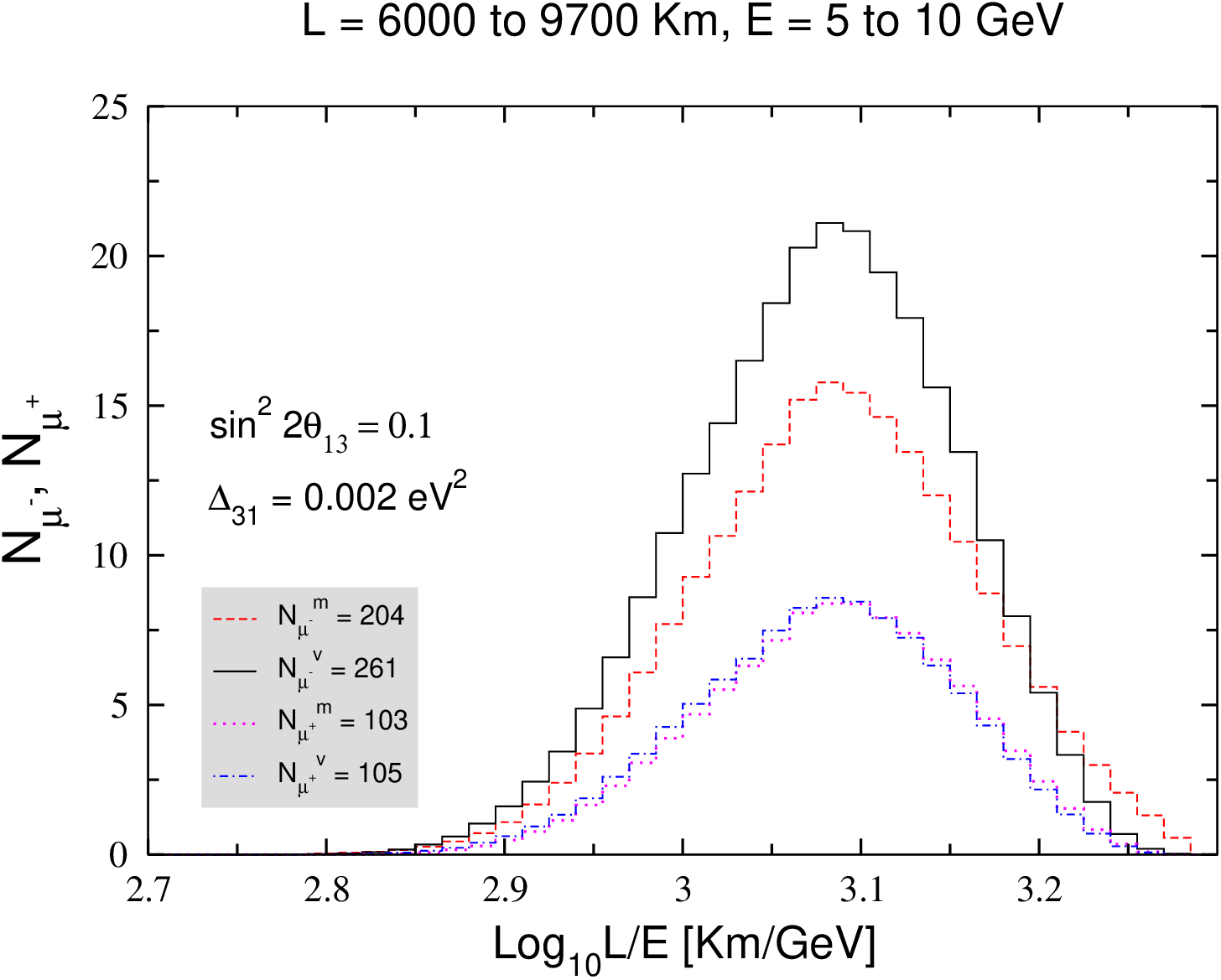}
}
\caption[]{\footnotesize
{
The total event rate for muons and anti-muons in matter and
in vacuum plotted against ${\mathrm{{Log_{10}(L/E)}}}$ for the
restricted choice of L and E range (see Figure \ref{fig14}).
The numbers in the legend correspond to the integrated
number of muon and anti-muon
events for the restricted range of L and E
in matter and in vacuum for a given value of
$\sin^2 2\theta_{13}$.
}
}
\label{fig15}
}
 \end{figure}

In Figure {\ref{fig14}} we show event rates for muons and anti-muons versus L
for the baseline and energy bins mentioned above (L $=$ 6000-9700 Km, E $=$ 5-10 GeV).
We have assumed $\Delta_{31} = 0.002$ \evsq and  $\sin^2 2 \theta_{13} = 0.1$
for this plot.
The effect is large for this choice of energies and baselines
because it is a combination of the
two effects visible in the bottom panels of
Figure \ref{fig4}.
The fall in \pmumum at 9700 Km between 6-15 GeV (which
 persists over a range of baselines) obtains a large
contribution from \pmutaum while the effect shown at 7000
Km arises primarily due to ${\mathrm {P^{m}_{\mu e}}}$.
Both work to lower the muon survival rate below its vacuum value.
In contrast to an expected vacuum oscillation rate of 261 events,
one expects 204 events in matter if the sign of $\Delta_{31}$ is positive.

That we have indeed chosen the bins for which matter effects and sensitivity to the mass hierarchy are maximum,
can be demonstrated by a computation of the sensitivity with which
the expectation for matter effects with Normal Hierarchy (NH) differs from that for Inverted Hierarchy (IH) or for vacuum oscillations.
To find the sensitivity, we define
\begin{equation}
\rm{\sigma_{NH-other}} = \mathrm{\frac{\vert N_{NH} - N_{other} \vert}{\sqrt{N_{NH}}}},
\label{eq:sigma}
\end{equation}
where $\rm{N_{other}}=\rm{N_{IH}}$ when computing the sensitivity to mass hierarchy $\rm{\sigma_{NH-IH}}$
and $\rm{N_{other}}=\rm{N_{vac}}$ for the sensitivity to matter effects $\rm{\sigma_{NH-vac}}$.
We split the 24 bins in Table {\ref{table2}} into two sets, with
fixed values of parameters $\Delta_{31}=0.002$ eV$^2$, $\sin^2
2\theta_{23}=1.0$, $\sin^2 2\theta_{23}=0.1$. The 4 highlighted bins
form one set and the other 20 bins form the second. The
sensitivities are calculated using Eq. \ref{eq:sigma}, where the
number of events is taken to be the sum of events over each of the
sets, thus assuming each set to form a single bin. Now since the
number of $\mu^-$ events in vacuum will be nearly equal to the
number in matter with inverted hierarchy, a signal for matter effect
is equivalent to a signal for the mass hierarchy when using only
$\mu^-$ events. This is brought out in Table {\ref{table3}}, where
$\rm{\sigma_{NH-IH}}$ is seen to be close to $\rm{\sigma_{NH-vac}}$.
Note that this property is specific to a charge-discriminating
detector. Hence in the subsequent discussion, we describe the values
of $\rm{\sigma_{NH-vac}}$ only as a measure of sensitivity to matter
effects as well as to the mass hierarchy, when working with only
$\mu^-$ events.

For the sum of $\mu^-$ events in the 4 highlighted bins, a large
value of $\rm{\sigma_{NH-vac}}=4.0 \sigma$ is obtained, as seen in
the second row of Table {\ref{table3}} for the standard value
$\sin^2 2\theta_{13}=0.1$. For the 20 non-highlighted bins, the
corresponding values of event numbers are $\rm{N_{vac}}=1860$ and
$\rm{N_{NH}}=1831$, giving a small $\rm{\sigma_{NH-vac}}=0.7
\sigma$. If the sum of events of all 24 bins is taken,
$\rm{N_{vac}}=2121$ and $\rm{N_{NH}}=2035$, corresponding to a
signal of $\rm{\sigma_{NH-vac}}=1.9 \sigma$. Thus the 4 highlighted
bins give a 4$\sigma$ signal for matter effect, which also indicates
their sensitivity to the mass hierarchy. The above values are for
$\sin^2 2\theta_{13}=0.1$. Table {\ref{table3}} also gives the
values of NH-vacuum and NH-IH sensitivity for the 4 highlighted bins
for two other values of $\theta_{13}$.

In calculating the sensitivities shown in Tables {\ref{table3}} and
{\ref{table3a}}, we have assumed that the neutrino mass-squared
differences and mixing angles will be measured with good precision.
Hence no marginalization over these parameters was done in
calculating the sensitivities. The ability of a detector to rule out
the wrong hierarchy hypothesis, however, does depend on the
precision with which the neutrino parameters were measured at that
point. Lower precision leads to reduced sensitivity.

Summing over the four highlighted bins in Table {\ref{table2}},
leads to integrating over large ranges in energy and in pathlength.
Hence the detector resolution is not crucial in determining the
number of events and the sensitivity. However, the sensitivity is
likely to be better if the detector has good energy and pathlength
resolution. The dependence of the sensitivity on detector resolution
is currently under study \cite{us3}.

Figure {\ref{fig15}}, for the same energy/baseline ranges and
parameter values shows the event distributions for muons and
antimuons versus L/E.

For a charge-discriminating detector, the data for the number of $\mu^+$ events would also be available separately.
If the number of $\mu^+$ events is taken into account and the quantity
($\mathrm{N_{\mu^{-}}}-\mathrm{N_{\mu^{+}}}$) is considered as an observable, then the
sigma sensitivity is defined as follows:
\begin{equation}
\rm{\sigma_{NH-IH}} = \mathrm{\frac{(N^{NH}_{\mu^{-}}-N^{NH}_{\mu^{+}}) - (N^{IH}_{\mu^{-}}-N^{IH}_{\mu^{+}})}
{\sqrt{(N^{NH}_{\mu^{-}}+N^{NH}_{\mu^{+}})}}}
\label{eq:sigma2}
\end{equation}
Using this, the NH-IH sensitivity is seen to improve in comparison to the sensitivity obtained using only $\mu^-$ events,
as seen by comparing Table {\ref{table3a}} with Table {\ref{table3}}.
However, the NH-vacuum sensitivity suffers a decrease by this method.


\begin{table}
\begin{center}
\begin{tabular}{||c | c | c | c | c | c ||}
\hline
\hline
 &&&&&\\
\hspace{0.5cm} $\sin^2 2\theta_{13}$ \hspace{0.5cm}&\hspace{0.5cm}
$\mathrm{N_{vac}}$ \hspace{0.5cm}&
\hspace{0.5cm}$\mathrm{N^{NH}_{mat}}$\hspace{0.5cm} &
$\mathrm{N^{IH}_{mat}}$ \hspace{0.5cm}&\hspace{0.5cm}
$\rm{\sigma_{NH-vac}}$ \hspace{0.5cm}& \hspace{0.5cm}$\rm{\sigma_{NH-IH}}$\hspace{0.5cm}\\
&&&&&\\
\hline
        $0.05$ & 260 & 227  & 264  & 2.2$\sigma$ & 2.5$\sigma$  \\ \hline
        $0.1$  & 261 & 204 & 262 & 4.0$\sigma$ & 4.1$\sigma$  \\ \hline
        $0.2$  & 263 & 163 & 261 & 7.8$\sigma$ & 7.7$\sigma$
          \\
\hline
\hline
\end{tabular}
\caption{\footnotesize{Number of $\mu^-$ events in vacuum and in
matter (Normal Hierarchy and Inverted Hierarchy) and corresponding
values of $\sigma$ sensitivity computed by comparing NH-vacuum and
NH-IH for 3 different values of $\sin^2 2\theta_{13}$ for the 4
highlighted bins in the E and L range E $=$ 5 - 10 GeV, L $=$ 6000 -
9700 Km in Table {\ref{table2}}.
$\Delta_{31}=0.002$ eV$^2$ and $\sin^2 2\theta_{23}=1.0$.
}} \label{table3}
\end{center}
\end{table}


\begin{table}
\begin{center}
\begin{tabular}{||c | c | c | c | c | c | c | c | c ||}
\hline \hline &&&&&&&&\\
 \hspace{0.1cm}$\sin^2
2\theta_{13}$\hspace{0.1cm} &
\hspace{0.1cm}$\mathrm{N^{vac}_{\mu^{-}}}$  \hspace{0.1cm}&
\hspace{0.1cm} $\mathrm{N^{NH}_{\mu^{-}}}$ \hspace{0.1cm} &
\hspace{0.1cm} $\mathrm{N^{IH}_{\mu^{-}}}$ \hspace{0.1cm} &
 \hspace{0.1cm}$\mathrm{N^{vac}_{\mu^{+}}}$  \hspace{0.1cm}&  \hspace{0.1cm}$\mathrm{N^{NH}_{\mu^{+}}}$  \hspace{0.1cm}&
 \hspace{0.1cm}$\mathrm{N^{IH}_{\mu^{+}}}$  \hspace{0.1cm}&
\hspace{0.1cm}$\rm{\sigma_{NH-vac}}$
 \hspace{0.1cm}& \hspace{0.1cm}
$\rm{\sigma_{NH-IH}}$ \hspace{0.1cm}\\
&&&&&&&&\\
        \hline
        $0.05$ & 260 & 227 & 264 & 106 & 104 & 94 & 1.7$\sigma$ & 2.6$\sigma$  \\ \hline
        $0.1$  & 261 & 204 & 262 & 105 & 104 & 85 & 3.2$\sigma$ & 4.4$\sigma$  \\ \hline
                $0.2$  & 263 & 163 & 261 & 105 & 102 & 70 & 6.0$\sigma$ & 8.0$\sigma$
          \\
\hline
\hline
\end{tabular}
\caption{\footnotesize{Number of $\mu^-$ and $\mu^+$ events in
vacuum and in matter (Normal Hierarchy and Inverted Hierarchy) and
corresponding values of NH-vacuum and NH-IH $\sigma$ sensitivity
computed by considering the quantity
($\mathrm{N_{\mu^{-}}}-\mathrm{N_{\mu^{+}}}$) as an observable, for
the same values of $\theta_{13}$ and E and L ranges as in Table
{\ref{table3}}. $\Delta_{31}=0.002$ eV$^2$ and $\sin^2
2\theta_{23}=1.0$. This depicts the gain in NH-IH sensitivity if the
$\mu^-$ as well as $\mu^+$ events are taken into account and the
difference is taken to be the observable.} }
\label{table3a}
\end{center}
\end{table}


\begin{table}
\begin{center}
\begin{tabular}{||c |c | c | c | c ||}
\hline \hline
&&&&\\
 \hspace{0.5cm}$\sin^2 2\theta_{13}$  \hspace{0.5cm}&\hspace{0.5cm}
$\mathrm{N_{vac}}$ \hspace{0.5cm}&
\hspace{0.5cm}$\mathrm{N^{NH}_{mat}}$\hspace{0.5cm} &\hspace{0.5cm}
\hspace{0.5cm}$\mathrm{N_{Down}}$\hspace{0.5cm} &
\hspace{0.5cm}$\rm{\sigma_{NH-vac}}$\hspace{0.5cm}
\\&&&&
        \\
        \hline
        $0.05$ & 260 & 227  & 410  & 1.8$\sigma$   \\ \hline
        $0.1$  & 261 & 204 & 410 & 3.3$\sigma$   \\ \hline
        $0.2$  & 263 & 163 & 410 & 6.6$\sigma$
          \\
\hline
\hline
\end{tabular}
\caption[]{\footnotesize{
Number of $\mu^-$ events in vacuum and in
matter (Normal Hierarchy), number of Down (unoscillated) $\mu^-$
events
 and values of NH-vacuum $\sigma$ sensitivity computed using ratio of Up/Down events
for the same values of $\theta_{13}$ and E and L ranges as
in Table {\ref{table3}}.
$\Delta_{31}=0.002$ eV$^2$ and $\sin^2 2\theta_{23}=1.0$.
}} \label{table3b}
\end{center}
\end{table}

\begin{figure}[!h]
{\centerline{\hspace*{2em}
\epsfxsize=10cm\epsfysize=8.0cm
                     \epsfbox{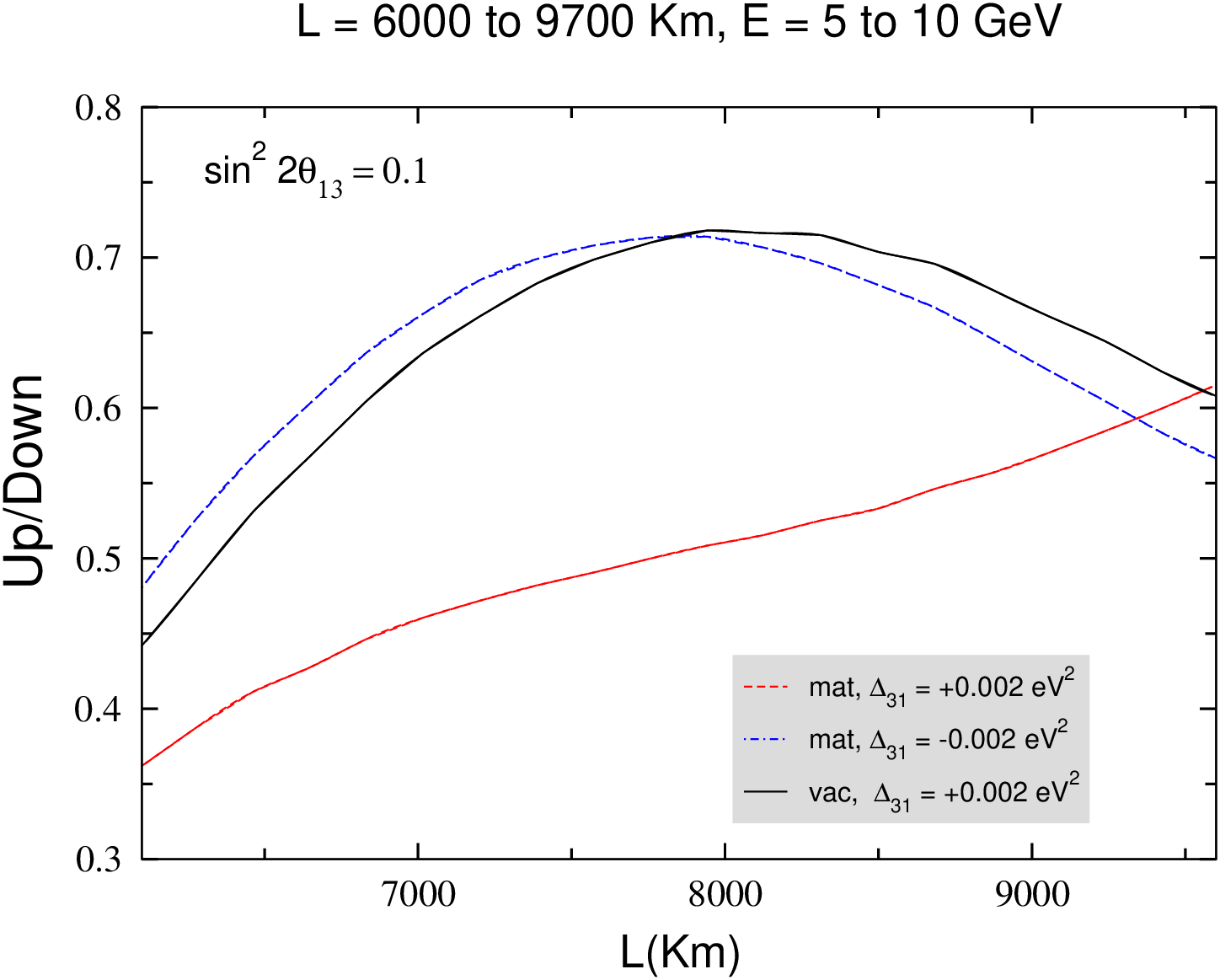}
}
\caption[]{\footnotesize
{Ratio of atmospheric upward and downward
going muon ($\mu^{-}$) events
plotted against L
for energy range of 5 GeV to 10 GeV and baselines
between 6000 Km to 9700 Km.
}
}
\label{fig16}
}
\end{figure}

\begin{figure}[!h]
{\centerline{\hspace*{2em}
\epsfxsize=10cm\epsfysize=8.0cm
                     \epsfbox{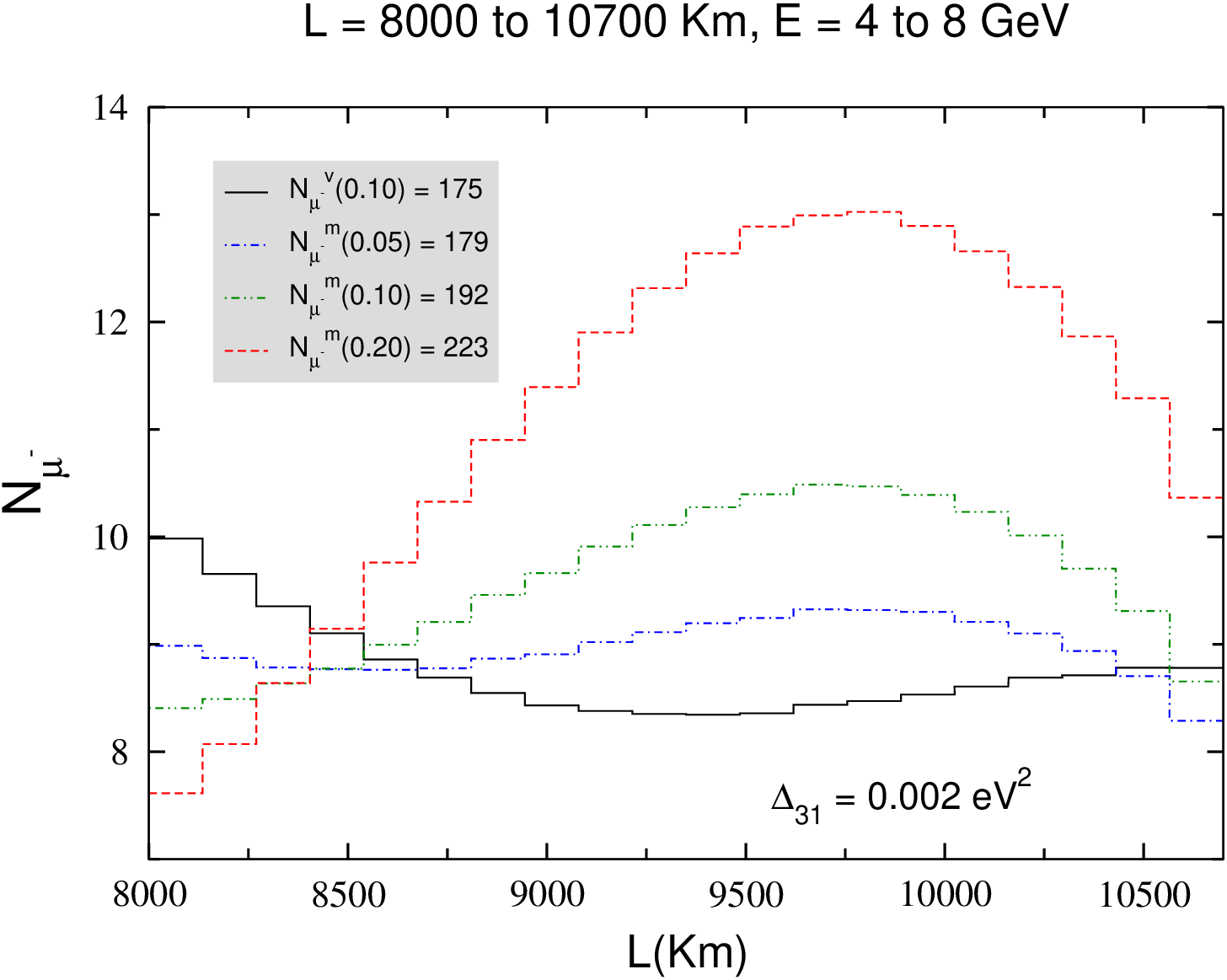}
}
\caption[]{\footnotesize
{Atmospheric muon events plotted against L
for energy range of 4 GeV to 8 GeV and baselines
between 8000 Km to 10700 Km.
The numbers in the legend correspond to the integrated
number of muon events for the restricted range of L and E
in matter and in vacuum for various values of
$\sin^2 2\theta_{13}$ (in parenthesis).
}
}
\label{fig17}
}
\end{figure}
 \begin{figure}[!h]
{\centerline{\hspace*{2em}
\epsfxsize=10cm\epsfysize=8.0cm
                     \epsfbox{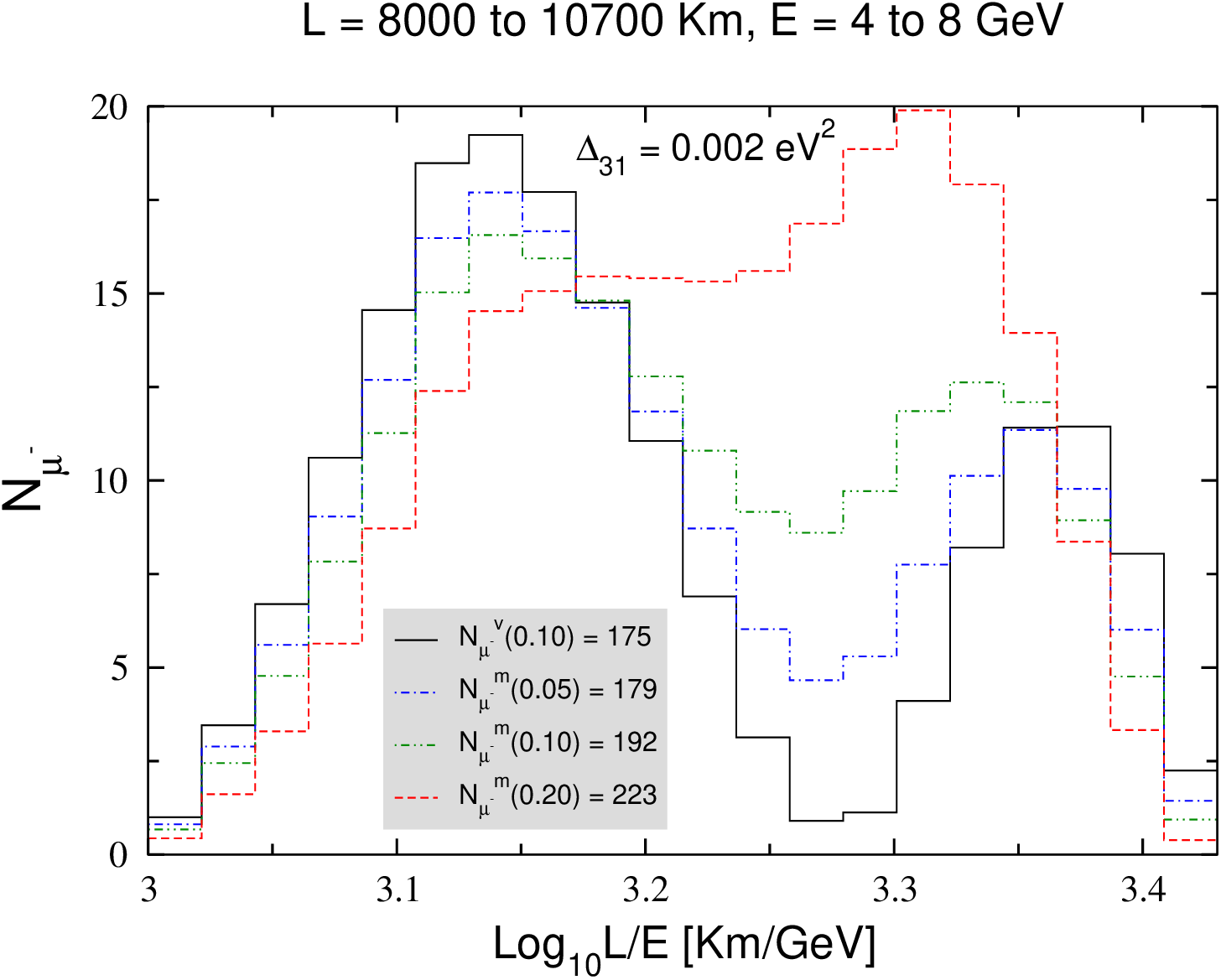}
}
\caption[]{\footnotesize
{
The total event rate for muons in matter and
in vacuum plotted against ${\mathrm{{Log_{10}(L/E)}}}$ for the
restricted choice of L and E range (see Figure \ref{fig17}).
The numbers in the legend correspond to the integrated
number of muon events for the restricted range of L and E
in matter and in vacuum for various values of
$\sin^2 2\theta_{13}$ (in parenthesis).
}
}
\label{fig18}
}
\end{figure}
\begin{figure}[ht]
\centerline{
\epsfxsize=7.5cm\epsfysize=6.5cm\epsfbox{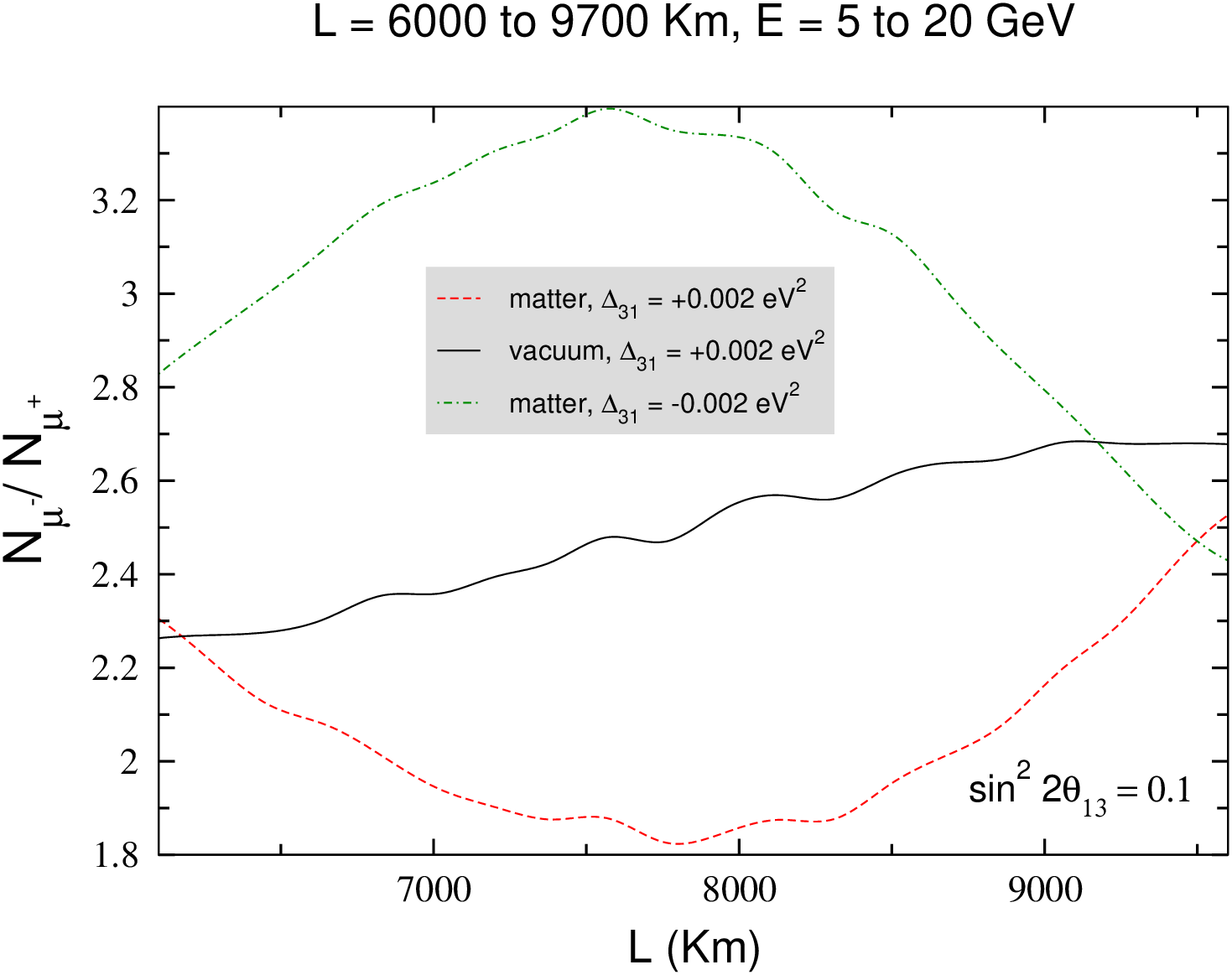}
        \hspace*{1.5ex}
\epsfxsize=7.5cm\epsfysize=6.5cm
                     \epsfbox{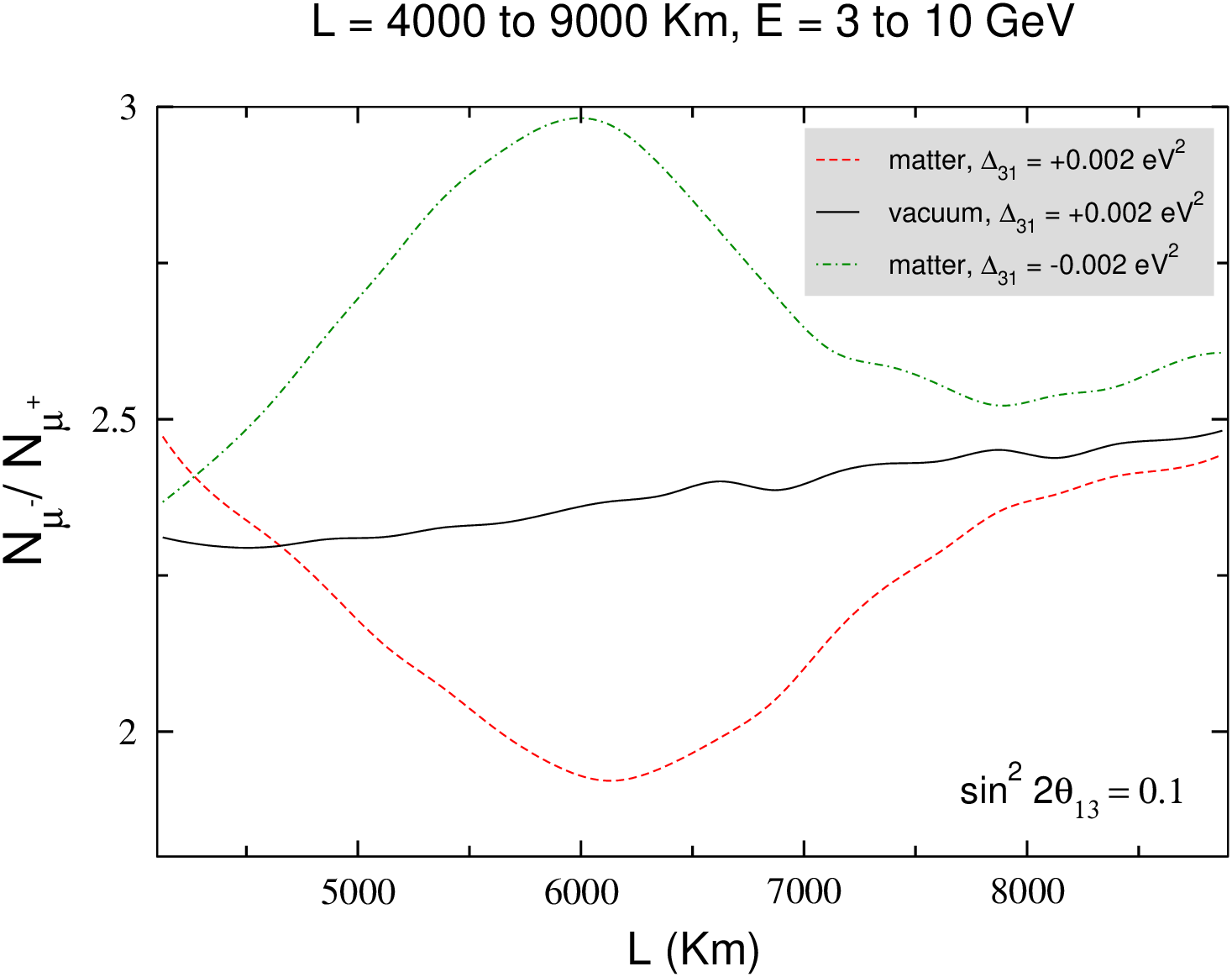}
}
\hskip 5cm
{\bf (a)}
\hskip 7cm
{\bf (b)}
\caption
{\footnotesize
Figure {\bf (a)} and {\bf (b)} depict the ratio of atmospheric muon and anti-muon events
plotted against L
for two different wide energy ranges (5 - 20 GeV and 3 - 10 GeV) and two baseline ranges
(6000 - 9700 Km and 4000 - 9000 km respectively).
}
\label{fig19}
\end{figure}

\begin{figure}[!h]
{\centerline{\hspace*{2em}
\epsfxsize=8cm\epsfysize=7.0cm
                     \epsfbox{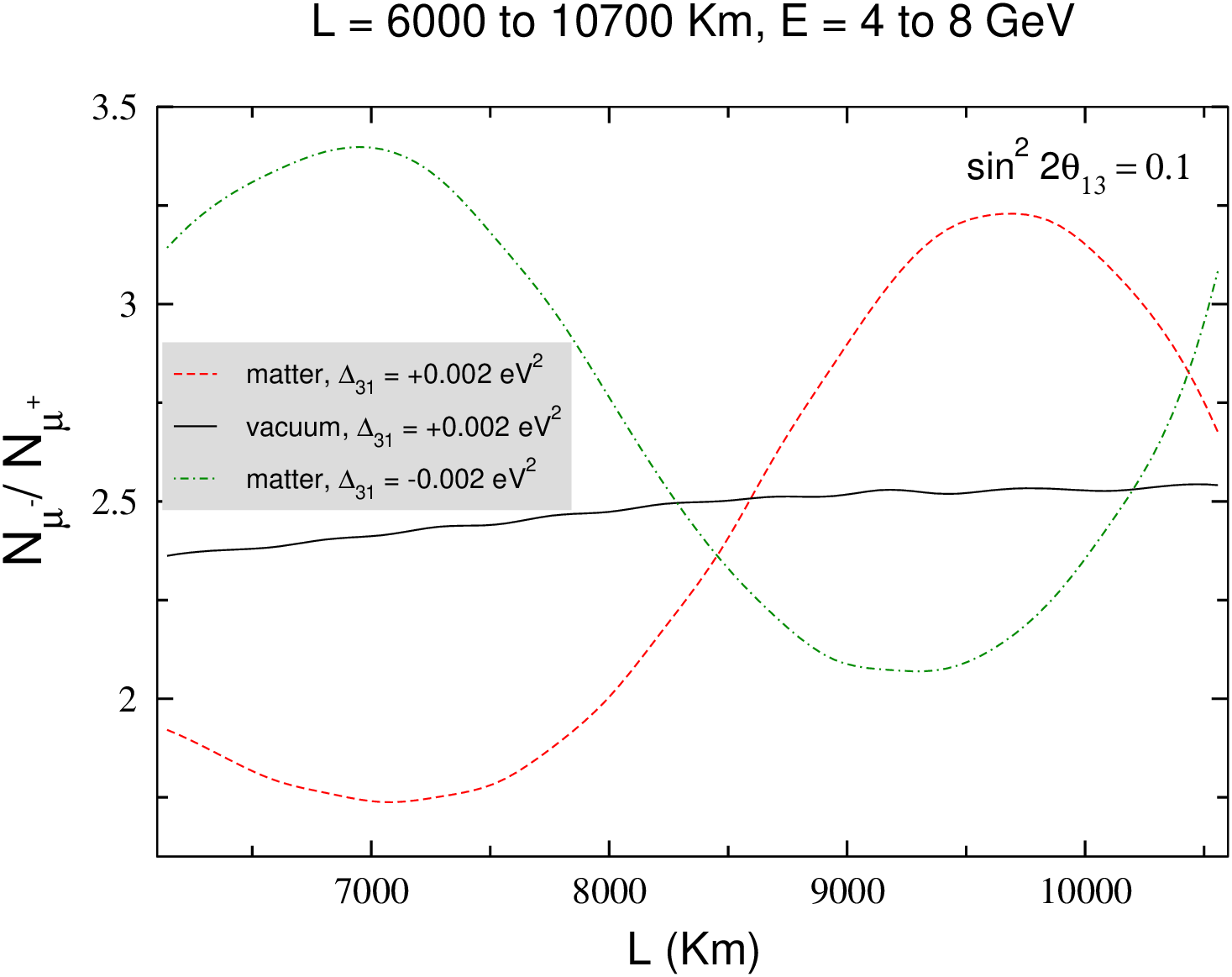}
}
\caption[]{\footnotesize{
Ratio of atmospheric muon and anti-muon
events plotted against L for energy range of
4 GeV to 8 GeV and baselines
between 6000 Km to 10700 Km.
For the restricted energy range considered here, the ratio for
the case of matter oscillations with NH is greater than that of
the vacuum ratio, when L is greater than 9000 Km. This is because
of the increase of $\mathrm{P_{\mu \mu}}$ caused by the sharp fall
in $\mathrm{P_{\mu \tau}}$.
}}
\label{fig20}
}
\end{figure}

The above assumes that the cosmic ray fluxes will be well measured
in ten years' time and the atmospheric neutrino fluxes can be predicted
with much smaller errors than currently available. If the uncertainty
in atmospheric neutrino flux prediction remains high, then one can
use Up/Down event ratios to cancel the normalization uncertainty of the
flux predictions.
Since the atmospheric neutrino fluxes depend only on the modulus of
the cosine of the zenith angle, the muon event rates expected in
case of no oscillations can be directly obtained from the experiment,
by measuring the rates of downward going muons, binned according to
the same energy and the same value of $|\cos \theta|$.
Thus the downward going neutrinos
provide the necessary information on unoscillated fluxes.
Figure {\ref{fig16}} shows the
Up/Down muon event ratio
versus L for L $=$ 6000-9700 Km, E $=$ 5-10 GeV in vacuum
and in matter for both signs of $\Delta_{31}$.
For this range of energies and baselines,
the downward event rates are calculated to be 410 for muons and 164 for anti-muons
(same in matter and in vacuum).
As mentioned earlier, the upward event rate for muons is 204 in matter and 261 in vacuum.
For matter oscillations, this gives ${\mathrm{Up/Down}} = 0.50 \pm 0.04$,
which differs from the corresponding ratio of $0.64$ for vacuum
oscillations by $\sim 3.5 \sigma$.
Table {\ref{table3b}} lists the values of NH-vacuum sensitivity using the ratio of Up/Down events for the E and L ranges
discussed earlier, for three different values of $\theta_{13}$.

As an example of how the muon survival rate can {\it rise}
above the expected vacuum value, we show in
Figure {\ref{fig17}}
the muon event rate for the baseline range of
8000-10700 Km
and the energy range of 4-8 GeV.
The solid curve is the vacuum event rate for $\sin^2 2\theta_{13} = 0.1$,
while the three others are the muon survival rates incorporating matter
effects for values of $\sin^2 2\theta_{13} = 0.05$,
0.1 and 0.2 respectively{\footnote{Note that the vacuum curve will
not change appreciably with a change in $\theta_{13}$,
so it serves as the reference plot for vacuum.}}.
The curves exhibit the rise above the vacuum
survival rate visible in the energy range 4-7 GeV in
the bottom panel of Figure \ref{fig4}(a).
In addition, the curves demonstrate the sensitivity to $\theta_{13}$
which is exhibited in the probability plot given in
Figure {\ref{fig6}}(b) in the energy range 4-7 GeV.
For values of $\sin^2 2 \theta_{13}$
close to the present CHOOZ bound,
a $>3 \sigma$ signal for matter effect is visible in Figure {\ref{fig17}}.
A similar $\theta_{13}$
sensitivity cum event rate plot is also given in
Figure {\ref{fig18}} with the muon rate now plotted versus L/E.
Here it is seen that ${\mathrm{N_{\mu^-}}}$ is particularly
sensitive to $\theta_{13}$ in the range $3.21 - 3.34$ of
${\mathrm{{Log_{10}(L/E)}}}$.
The value for vacuum oscillations is $23$ and it rises to $43$ when
matter effects are included, even for as small a value as $\sin^2 2\theta_{13}
= 0.05$.
Treating this ${\mathrm{{Log_{10}(L/E)}}}$ range as a single bin,
$\rm{\sigma_{NH-vac}}=3.0$ for $\sin^2 2\theta_{13} = 0.05$,
indicating a large signal for matter effect in this bin. Table
{\ref{table4}} gives the values of $\rm{\sigma_{NH-vac}}$ for this
${\mathrm{{Log_{10}(L/E)}}}$ bin for three different values of
$\theta_{13}$, and Table {\ref{table4a}} gives the corresponding
values of the sensitivity using the Up/Down event ratios. This
effect may be looked for provided the L/E resolution is sufficient
to observe it. {\it{We note that a significant rise of the muon
survival rate in matter over its vacuum values is a signal of the
size of the matter effect in \pmutau overcoming that due to
${\mathrm {P_{\mu e}}}$, in spite of the fact that the latter is
close to resonance values}} (Figure \ref{fig4}(a), in the energy
range 4-7 GeV).


\begin{table}
\begin{center}
\begin{tabular}{||c | c | c | c ||}
\hline \hline &&&\\
 \hspace{0.5cm}$\sin^2 2\theta_{13}$ \hspace{0.5cm}& \hspace{0.5cm}$\mathrm{N_{vac}}$ &\hspace{0.5cm}
$\mathrm{N^{NH}_{mat}}$\hspace{0.5cm} &
\hspace{0.5cm}$\rm{\sigma_{NH-vac}}$\hspace{0.5cm}
\\&&&
        \\
        \hline\hline
        $0.05$ & 23 & 43  & 3.0$\sigma$  \\ \hline
        $0.1$ & 23 &  63 &  5.0$\sigma$ \\ \hline
        $0.2$ & 24 & 104 &  7.8$\sigma$
          \\ \hline \hline
\end{tabular}
\caption[]{\footnotesize{Number of $\mu^-$ events in vacuum and in
matter (Normal Hierarchy)
 and corresponding values of NH-vacuum $\sigma$ sensitivity for 3 different values of $\sin^2 2\theta_{13}$
for the $\mathrm{Log_{10} (L/E)}$ range 3.21 - 3.34 shown in Figure {\ref{fig18}}
(E and L restricted between E $=$ 4 - 8 GeV, L $=$ 8000 - 10700 Km).
%
$\Delta_{31}=0.002$ eV$^2$ and $\sin^2 2\theta_{23}=1.0$.
}} \label{table4}
\end{center}
\end{table}


\begin{table}
\begin{center}
\begin{tabular}{||c | c | c | c | c ||}
\hline
\hline
&&&&\\
\hspace{0.5cm}$\sin^2 2\theta_{13}$ \hspace{0.5cm}&\hspace{0.5cm}
$\mathrm{N_{vac}}$\hspace{0.5cm} &\hspace{0.5cm}
$\mathrm{N^{NH}_{mat}}$ \hspace{0.5cm}&\hspace{0.5cm}
$\mathrm{N_{Down}}$ \hspace{0.5cm}&
\hspace{0.5cm}$\rm{\sigma_{NH-vac}}$\hspace{0.5cm}
\\&&&&
        \\
        \hline\hline
        $0.05$ & 23 & 43  & 156 & 2.7$\sigma$   \\ \hline
        $0.1$  & 23 & 63  & 156 & 4.3$\sigma$   \\ \hline
        $0.2$  & 24 & 104 & 156 & 6.1$\sigma$
          \\
\hline
\hline
\end{tabular}
\caption[]{\footnotesize{
Number of $\mu^-$ events in vacuum and in
matter (Normal Hierarchy), number of Down (unoscillated) $\mu^-$
events
 and values of NH-vacuum $\sigma$ sensitivity computed using ratio of Up/Down events
for the same values of $\theta_{13}$ and $\mathrm{Log_{10} (L/E)}$ range as
in Table {\ref{table4}}.
$\Delta_{31}=0.002$ eV$^2$ and $\sin^2 2\theta_{23}=1.0$.
}} \label{table4a}
\end{center}
\end{table}

Further demonstration of the sensitivity to the sign of $\Delta_{31}$
and of the effects discussed in this paper can be gleaned from plots
of the ratio of muon to anti-muon  event rates
[${\mathrm{N_{\mu^-}/N_{\mu^+}}}$] \cite{pet}.
In Figure {\ref{fig19}}(a) we show this ratio for the baseline
range 6000-9700 Km and the energy range 5-20 GeV.
The dominant effect here is due to the reduction in \pmumum
stemming from  a large enhancement of ${\mathrm {P^{m}_{\mu e}}}$,
(as evident in the probability plots for 7000 Km in Figure \ref{fig4}(b) above)
and this is clearly visible in the event rates.
Comparing the curves for positive and negative $\Delta_{31}$,
we see that the peak to peak difference between them is
about 90\%.

Selecting  different (lower) ranges in both baselines and energy,
we show the rate ratio for
4000 - 9000 Km and 3 - 10 GeV in Figure {\ref{fig19}}(b).
As evident from the discussion of the previous section,
the \pmutaum effects develop after 8000 Km for the most part,
hence this choice is primarily suited to demonstrate the effect
of \pmuem on ${\mathrm {P^{m}_{\mu \mu}}}$. The influence of \pmutaum
is, however, visible at the higher baselines,
as it draws the $\Delta_{31}>0$ curve close to the vacuum curve
because, as discussed above, the \pmuem and \pmutaum effects work in the
opposite direction in this region.
The dip-peak separation for the curves for $\Delta_{31}>0$ and
$\Delta_{31}<0$ is about 60\% in this case.

Finally, in an attempt to show an example of a baseline/energy range
where both effects, i.e.the effect where \pmutaum plays a major role
by raising the rate ratio above its vacuum value, and the effect where
the resonant rise in \pmuem dominates over \pmutaum to bring about a
decrease in ${\mathrm{N_{\mu^-}/N_{\mu^+}}}$ compared
to its value in vacuum,
we choose the baseline range 6000-10700 Km and the energy range
4-8 GeV (Figure {\ref{fig20}}).
For $\Delta_{31}>0$, the dip in the ratio is maximum around 7000 Km
and originates from the drop in the probability visible in
the bottom panel of Figure \ref{fig4}(b),
while the rise above the vacuum curve, which is maximum
around 9700 Km, is a manifestation of
{\bf the increase in ${\rm P_{\mu \mu}}$
    consequent to the large decrease in ${\rm P_{\mu \tau}}$ occurring
    in the range 4-8 GeV},
shown in
Figure \ref{fig4}(a) and \ref{fig6}(b).




\section{Conclusions}

To conclude, we summarize the salient points of our study:

\begin{itemize}

\item
In addition to ${\mathrm {P_{\mu e}}}$, interesting and appreciable matter effects
can arise in \pmutau at baselines $>$ 7000 Km.
These can synergistically combine
to give correspondingly large effects in ${\mathrm {P^{m}_{\mu \mu}}}$.
As a result,
observably large signals may appear in a charge discriminating detector capable
of measuring muon survival rates, providing a useful handle
on the sign of $\Delta_{31}$.

\item
The resonance amplified matter effects in \pmutaum may be observable in
future experimental facilities with intense $\nu_{\mu}$ beams
and detectors  capable of $\tau$ identification.

\item
We find that both \pmutaum and \pmumum are sensitive to $\theta_{13}$.
We have identified baselines and energies where
the matter effects are amplified by the resonance
and the sensitivity to $\theta_{13}$ is high.
We have also studied the $\theta_{23}$ dependence of \pmutaum and ${\mathrm {P^{m}_{\mu \mu}}}$.

\item
We have performed a detailed study of parameter degeneracies
at very long baselines, using numerical results with a realistic
earth density profile.
We find that for these baselines and at energies for which the
criterion ${\mathrm {1.27 \Delta_{31} L/E}} = n\pi/2$ is satisfied,
\pmuem is largely free of the degeneracies
which could obscure measurements of $\theta_{13}$ and sign$(\Delta_{31})$.
The $({\mathrm{\delta_{CP}}}, \theta_{13})$ degeneracy
is lifted for \pmutaum and ${\mathrm {P^{m}_{\mu \mu}}}$ also.
The sign$(\Delta_{31})$ degeneracy vanishes for \pmutaum at such
baselines and energies,
but can be present
in ${\mathrm {P^{m}_{\mu \mu}}}$ for
some specific values of $\theta_{13}$ and ${\mathrm{\delta_{CP}}}$.
We also note that the determination of
$\theta_{13}$ from ${\mathrm {P^{m}_{\mu \tau}}}$
does not suffer from the ($\theta_{23}$,$\pi/2 - \theta_{23}$) degeneracy,
${\mathrm {P^{m}_{\mu \tau}}}$  being a function of $\sin2\theta_{23}$.
Further, for the baselines and energies relevant for us this degeneracy
 disappears from \pmumum and ${\mathrm {P^{m}_{\mu e}}}$ as well.

\item
In the second part of the paper we perform detailed calculations
of muon survival rates for atmospheric neutrinos, using a 100 kT charge-discriminating
iron calorimeter as prototype. Our choice here is guided by our motivation
to explore whether it is possible to realistically determine the sign
of $\Delta_{31}$ in an atmospheric neutrino experiment, over a timescale
shorter than that anticipated for neutrino factories.

\item
While atmospheric neutrinos make available a wide range of baselines
and energies, we find that in order to detect matter effects using
total event rates a careful screening and selection of L and E is
necessary to overcome the lack of a high intensity beam. We use our
earlier discussion of matter effects in ${\mathrm {P^{m}_{\mu e}}}$,
\pmutaum and \pmumum as a guide to identify fairly broad L and E
ranges where the sensitivity to the sign of $\Delta_{31}$ is
significantly large. As an example, \pmutaum and \pmuem combine
constructively in the range 5-10 GeV and 6000-9700 Km to yield a
potentially large signal $(\sim 4\sigma)$ for matter effect for
$\Delta_{31}>0$ ($\sim 3.5\sigma$ for the Up/Down ratio). We also
identify regions where an appreciable sensitivity to $\theta_{13}$
exists. For a large mass iron calorimeter type detector, a $3\sigma$
discrimination is seen in the $3.2 - 3.35$ range of
${\mathrm{{Log_{10}(L/E)}}}$ for $\sin^2 2\theta_{13}$ as low as
0.05.

\item
We have discussed the possibility of detection of matter effects and
the sign of $\Delta_{31}$ using a charge discriminating iron calorimeter.
Although these effects appear only for $\nu_{\mu}$ for
$\Delta_{31}$ positive (and only for $\bar{\nu}_{\mu}$
for $\Delta_{31}$ negative), it may also be possible to look for them
using high statistics mega-ton water Cerenkov
detectors like Hyper-Kamiokande in Japan and UNO
in the US \cite{yit,uno}.
These effects may also be searched for
in the accumulated SK data.

\end{itemize}

\vskip 1cm \noindent{\bf Acknowledgments :} We would like to thank
S.~Choubey, M.~V.~N.~Murthy and P.~Roy for suggestions and
discussions. We also acknowledge useful communications from
M.~Lindner, E.~Lisi, S.~Palomares-Ruiz, S.~Petcov, A.~Takamura and
W.~Winter. S.G. acknowledges support from the Alexander von Humboldt
Foundation.




\begin{thebibliography}{99}

\bibitem{k}
Y.~Fukuda {\it et al.} (Kamiokande Collaboration), Phys.\ Lett.\ B
{\bf 335}, 237 (1994).

\bibitem{sk1}
Y.~Fukuda {\it et al.} (Super-Kamiokande Collaboration), Phys.\
Rev.\ Lett.\  {\bf 81}, 1562 (1998) [hep-ex/9807003].

\bibitem{sk2}
Y.~Fukuda {\it et al.} (Super-Kamiokande Collaboration), Phys.\
Rev.\ Lett.\  {\bf 82}, 2644 (1999) [hep-ex/9812014].

\bibitem{nu04a}
E.~Kearns, {\it Proceedings of the XXIst International Conference on
Neutrino Physics and Astrophysics (Neutrino-2004), Paris}, see
http://neutrino2004.in2p3.fr/.

\bibitem{mac}
M.~Ambrosio {\it et al.} (MACRO Collaboration), Phys.\ Lett.\ B {\bf
566}, 35 (2003)
[hep-ex/0304037];\\
M.~Ambrosio {\it et al.} (MACRO Collaboration), Eur.\ Phys.\ J.\ C
{\bf 36}, 323 (2004).

\bibitem{soudan}
W.~W.~M.~Allison {\it et al.} (SOUDAN-2 Collaboration), Phys.\
Lett.\ B {\bf 449}, 137 (1999) [hep-ex/9901024].

\bibitem{imb}
R.~Becker-Szendy {\it et al.} (IMB Collaboration), {\it Proceedings
of the 16th International Conference On Neutrino Physics And
Astrophysics (Neutrino-94), Eilat} [Nucl.\ Phys.\ Proc.\ Suppl.\
{\bf 38}, 331 (1995)].

\bibitem{k2k}
M.~H.~Ahn {\it et al.} (K2K Collaboration), Phys.\ Rev.\
Lett.\  {\bf 90}, 041801 (2003) [hep-ex/0212007];\\
T.~Nakaya, {\it Proceedings of the XXIst International Conference on
Neutrino Physics and Astrophysics (Neutrino-2004), Paris} [Nucl.\
Phys.\ Proc.\ Suppl.\ {\bf 143}, 96 (2005)].

\bibitem{nu04b}
C.~Cattadori, {\it Proceedings of the XXIst International Conference
on Neutrino Physics and Astrophysics (Neutrino-2004), Paris} [Nucl.\
Phys.\ Proc.\ Suppl.\ {\bf 143}, 3 (2005)].

\bibitem{btc}
B.~T.~Cleveland {\it et al.}, Astrophys.\ J.\  {\bf 496}, 505
(1998).

\bibitem{jna}
J.~N.~Abdurashitov {\it et al.} (SAGE Collaboration), J.\ Exp.\
Theor.\ Phys.\ {\bf 95}, 181 (2002) [Zh.\ Eksp.\ Teor.\ Fiz.\  {\bf
122}, 211 (2002)]
[astro-ph/0204245]; \\
V.~Gavrin, {\it Proceedings of the VIIth International conference on
Topics in Astroparticle and Underground Physics (TAUP03), Seattle},
see http://www.int.washington.edu/talks/WorkShops/TAUP03/Parallel/.

\bibitem{wha}
W.~Hampel {\it et al.} (GALLEX Collaboration), Phys.\ Lett.\ B {\bf
447}, 127 (1999).

\bibitem{mal}
M.~Altmann {\it et al.} (GNO Collaboration), Phys.\ Lett.\ B {\bf
490}, 16 (2000) [hep-ex/0006034]; \\
E.~Bellotti, {\it Proceedings of the VIIth International conference
on Topics in Astroparticle and Underground Physics (TAUP03),
Seattle}, see
http://www.int.washington.edu/talks/WorkShops/TAUP03/Parallel/.

\bibitem{sfu}
S.~Fukuda {\it et al.} (Super-Kamiokande Collaboration), Phys.\
Lett.\ B {\bf 539}, 179 (2002) [hep-ex/0205075].

\bibitem{qra}
Q.~R.~Ahmad {\it et al.} (SNO Collaboration), Phys.\ Rev.\ Lett.\
{\bf 87}, 071301 (2001)
[nucl-ex/0106015]; \\
{\bf 89}, 011301 (2002)
[nucl-ex/0204008]; \\
{\bf 89}, 011302 (2002) [nucl-ex/0204009].

\bibitem{sna}
S.~N.~Ahmed {\it et al.} (SNO Collaboration), Phys.\ Rev.\ Lett.\
{\bf 92}, 181301 (2004) [nucl-ex/0309004].

\bibitem{keg}
K.~Eguchi {\it et al.} (KAMLAND Collaboration), Phys.\ Rev.\ Lett.\
{\bf 90}, 021802 (2003)
[hep-ex/0212021]; \\
T.~Araki {\it et al.} (KAMLAND Collaboration),
  Phys.\ Rev.\ Lett.\  {\bf 94}, 081801 (2005)
  [hep-ex/0406035].

\bibitem{msw}
L.~Wolfenstein,
Phys.\ Rev.\ D {\bf 17}, 2369 (1978);\\
S.~P.~Mikheev and A.~Y.~Smirnov,
Sov.\ J.\ Nucl.\ Phys.\  {\bf 42}, 913 (1985)
[Yad.\ Fiz.\  {\bf 42}, 1441 (1985)].

\bibitem{sruba}
S.~Goswami, {\it Proceedings of the XXIst International Conference
on Neutrino Physics and Astrophysics (Neutrino-2004), Paris} [Nucl.\
Phys.\ Proc.\ Suppl.\ {\bf 143}, 121 (2005)], hep-ph/0409224.


\bibitem{bahcall}
J.~N.~Bahcall, M.~C.~Gonzalez-Garcia and C.~Pena-Garay,
JHEP {\bf 0408}, 016 (2004)
[hep-ph/0406294].

\bibitem{maltoni}
  M.~Maltoni, T.~Schwetz, M.~A.~Tortola and J.~W.~F.~Valle,
  New J.\ Phys.\  {\bf 6}, 122 (2004)
  [hep-ph/0405172].

\bibitem{concha}
M.~C.~Gonzalez-Garcia, {\it Proceedings of the Nobel Symposium 2004
on Neutrino Physics, Sweden}, see
http://www.physics.kth.se/nobel2004/program.html, hep-ph/0410030.

\bibitem{lin}
  P.~Huber, M.~Lindner, M.~Rolinec, T.~Schwetz and W.~Winter,
  Phys.\ Rev.\ D {\bf 70}, 073014 (2004)
  [hep-ph/0403068]; \\
  Nucl.\ Phys.\ Proc.\ Suppl.\  {\bf 145}, 190 (2005)
  [hep-ph/0412133].

\bibitem{mil} E. Ables {\it et al.} (MINOS Collaboration), Fermilab Report No. P875,
1995;\\
P. Adamson {\it et al.} (MINOS Collaboration), The MINOS Technical
Design Report No. Nu-MI-L-337, 1998;
see http://www-numi.fnal.gov/ ;\\
 A.~Marchionni (MINOS Collaboration), {\it Proceedings of the
7th International Workshop on Neutrino Factories and Superbeams
(NuFact05), Frascati}, Report no. FERMILAB-CONF-05-429-AD-E;\\
M.~Thomson, {\it Proceedings of the XXIst International Conference
on Neutrino Physics and Astrophysics (Neutrino-2004), Paris} [Nucl.\
Phys.\ Proc.\ Suppl.\ {\bf 143}, 249 (2005)] ;\\
 P.~Shanahan, Eur.\
Phys.\ J.\ C {\bf 33}, s834 (2004).
%
\bibitem{pap}
P.~Aprili {\it et al.} (ICARUS Collaboration), Report No. CERN-SPSC-2002-027; \\
A.~Bueno, {\it Proceedings of the XXIst International Conference on
Neutrino Physics and Astrophysics (Neutrino-2004), Paris} [Nucl.\
Phys.\ Proc.\ Suppl.\ {\bf 143}, 262 (2005)]
; \\
J.~Lagoda, {\it Proceedings of the 4th International Conference on
non-accelerator new physics (NANP2003), Dubna}, see
http://nanp.ru/2003/program.html.
%

\bibitem{ddu}
D.~Duchesneau (OPERA Collaboration), eConf {\bf C0209101}, TH09
(2002) [Nucl.\ Phys.\ Proc.\ Suppl.\  {\bf 123}, 279 (2003)]
[hep-ex/0209082]; \\
D.~Autiero, {\it Proceedings of the XXIst International Conference
on Neutrino Physics and Astrophysics (Neutrino-2004), Paris}, see
http://neutrino2004.in2p3.fr/; \\
M.~Dracos, {\it Proceedings of the 4th International Conference on
non-accelerator new physics (NANP2003), Dubna}, see
http://nanp.ru/2003/program.html.
%

\bibitem{yit}
Y.~Hayato, {\it Proceedings of the XXIst International Conference on
Neutrino Physics and Astrophysics (Neutrino-2004), Paris}
[Nucl.\ Phys.\ Proc.\ Suppl.\ {\bf 143}, 269 (2005)];\\
Y.~Itow {\it et al.},
hep-ex/0106019; \\
see http://neutrino.kek.jp/jhfnu/.


\bibitem{day}
D.~S.~Ayres {\it et al.} (NOVA Collaboration), Report No.
FERMILAB-PROPOSAL-0929, hep-ex/0503053;
Also see http://www-nova.fnal.gov/ ;\\
A.~Weber, Eur.\ Phys.\ J\ C\ {\bf 33}, s843-s845 (2004);
\\
M.~Messier, {\it Proceedings of the XXIst International Conference
on Neutrino Physics and Astrophysics (Neutrino-2004), Paris}, see
http://neutrino2004.in2p3.fr/.
%


\bibitem{reactors}
E.~A.~K.~Abouzaid {\it et al.}, see
http://www-library.lbl.gov/docs/LBNL/565/99/PDF/LBNL-56599.pdf
;\\
%
K.~Anderson {\it et al.},
[hep-ex/0402041].
\bibitem{kaska}
F.~Suekane (KASKA Collaboration), {\it Proceedings of the 5th
Workshop on Neutrino Oscillations and their Origin (NOON2004),
Tokyo},
see http://www-sk.icrr.u-tokyo.ac.jp/noon2004/, hep-ex/0407016;\\
F.~Suekane, K.~Inoue, T.~Araki and K.~Jongok, {\it Proceedings of
the 4th Workshop on Neutrino Oscillations and their Origin
(NOON2003), Kanazawa}, see
http://www-sk.icrr.u-tokyo.ac.jp/noon2003/, hep-ex/0306029.


%
\bibitem{mhs}
M.~H.~Shaevitz and J.~M.~Link,
hep-ex/0306031;\\
Also see http://theta13.lbl.gov/ and
http://braidwood.uchicago.edu/;\\
L.~Oberauer, {\it Proceedings of the XXIst International Conference
on Neutrino Physics and Astrophysics (Neutrino-2004), Paris} [Nucl.\
Phys.\ Proc.\ Suppl.\ {\bf 143}, 277 (2005)].

%
\bibitem{map}
M.~Apollonio {\it et al.} (CHOOZ Collaboration),
 Phys.\ Lett.\ B {\bf 466}, 415 (1999)[hep-ex/9907037];\\
%
M.~Apollonio {\it et al.} (CHOOZ Collaboration), Eur.\ Phys.\ J.\ C
{\bf 27}, 331 (2003) [hep-ex/0301017].
%
\bibitem{dcho}
T.~Lasserre, {\it Proceedings of the 5th workshop on Neutrino
Oscillations and their Origin (NOON2004), Tokyo}, see http://www-sk.icrr.u-tokyo.ac.jp/noon2004/~;\\
See also http://doublechooz.in2p3.fr/~.
%
\bibitem{uno}
C.~K.~Jung, {\it Proceedings of the International Workshop on Next
Generation Nucleon Decay and Neutrino Detector (NNN99), Stony
Brook}, edited by M.~V.~Diwan and C.~K.~Jung [AIP Conf.\ Proc.\ {\bf 533}, 29 (2000)], hep-ex/0005046;\\
Also see http://ale.physics.sunysb.edu/uno/publications.shtml/
; \\
K.~Nakamura, {\it Proceedings of the Conference on Neutrinos and
Implications for Physics Beyond the Standard Model, C.~N.~Yang
Institute for theoretical physics, SUNY, Stony Brook (2002)}, see
http://insti.physics.sunysb.edu/itp/conf/neutrino/talks/nakamura.pdf.

%
\bibitem{bern}
J.~Bernabeu, S.~Palomares-Ruiz, S.~T.~Petcov,
Nucl.\ Phys.\ B {\bf 669}, 255 (2003)
[hep-ph/0305152].
%
\bibitem{kaj}
T.~Kajita, {\it Proceedings of the 5th workshop on Neutrino
Oscillations and their Origin (NOON2004), Tokyo},
see http://www-sk.icrr.u-tokyo.ac.jp/noon2004/;\\
 H.~Gallagher,
{\it Proceedings of the XXIst International Conference on Neutrino
Physics and Astrophysics (Neutrino-2004), Paris} [Nucl.\ Phys.\
Proc.\ Suppl.\ {\bf 143}, 79 (2005)].
%
\bibitem{smi}
A. Yu. Smirnov, {\it Proceedings of the 5th workshop on Neutrino
Oscillations and their Origin (NOON2004), Tokyo}, see
http://www-sk.icrr.u-tokyo.ac.jp/noon2004/.
%
\bibitem{nufac}
C.~Albright {\it et al.},
hep-ex/0008064; \\
M.~Apollonio {\it et al.},
hep-ph/0210192 and references therein; \\
C.~Albright {\it et al.} (Neutrino Factory/Muon Collider
Collaboration), physics/0411123.
%
\bibitem{us1}
R.~Gandhi, P.~Ghoshal, S.~Goswami, P.~Mehta and S.~Uma~Sankar,
Phys.\ Rev.\ Lett.\ {\bf 94}, 051801 (2005)
[hep-ph/0408361].
%
\bibitem{ban}
M.~C.~Banuls, G.~Barenboim and J.~Bernabeu,
Phys.\ Lett.\ B {\bf 513}, 391 (2001)
[hep-ph/0102184].
%
\bibitem{berna}
J.~Bernabeu and S.~Palomares-Ruiz, {\it Proceedings of the
International Europhysics Conference on High-Energy Physics
(HEP2001), Budapest (2001)}, hep-ph/0112002.
%
\bibitem{tab}
T.~Tabarelli de Fatis,
Eur.\ Phys.\ J.\ C {\bf 24}, 43 (2002)
[hep-ph/0202232].
%
\bibitem{pet}
  S.~Palomares-Ruiz and S.~T.~Petcov,
  Nucl.\ Phys.\ B {\bf 712}, 392 (2005)
  [hep-ph/0406096].

\bibitem{indu}
D.~Indumathi and M.~V.~N.~Murthy,
Phys.\ Rev.\ D {\bf 71}, 013001 (2005)
[hep-ph/0407336].

%
\bibitem{sawg}
R.~Gandhi, P.~Mehta and S. Uma Sankar, A note prepared for solar and
atmospheric working group of American Physical Society, Report No.
HRI-P-04-10-001.

%
\bibitem{bnl-hs}
M.~Diwan {\it et al.}, hep-ex/0211001; \\
%
M.~V.~Diwan {\it et al.},
Phys.\ Rev.\ D {\bf 68}, 012002 (2003)
[hep-ph/0303081].
%

\bibitem{rubbia}
  A.~Rubbia,
  Nucl.\ Phys.\ Proc.\ Suppl.\  {\bf 147}, 103 (2005)
  [hep-ph/0412230].

\bibitem{bar}
V.~D.~Barger, K.~Whisnant, S.~Pakvasa and R.~J.~N.~Phillips,
Phys.\ Rev.\ D {\bf 22}, 2718 (1980).


\bibitem{ki2}
C.~W.~Kim, W.~K.~Sze and S.~Nussinov,
Phys.\ Rev.\ D {\bf 35}, 4014 (1987).


\bibitem{bil}
S.~M.~Bilenky,
Fiz.\ Elem.\ Chast.\ Atom.\ Yadra {\bf 18}, 449 (1987);\\
S.~M.~Bilenky and S.~T.~Petcov,
Rev.\ Mod.\ Phys.\  {\bf 59}, 671 (1987)
[Erratum-ibid.\  {\bf 61}, 169 (1989)].
%
\bibitem{zag}
H.~W.~Zaglauer and K.~H.~Schwarzer,
Z.\ Phys.\ C {\bf 40}, 273 (1988).
%
\bibitem{pant}
J.~Pantaleone,
Phys.\ Lett.\ B {\bf 292}, 201 (1992).
%
\bibitem{fogli}
G.~L.~Fogli, E.~Lisi, D.~Montanino, G.~Scioscia,
Phys.\ Rev.\ D {\bf 55}, 4385 (1997)
[hep-ph/9607251].
%
\bibitem{oh2}
T.~Ohlsson and H.~Snellman,
J.\ Math.\ Phys.\  {\bf 41}, 2768 (2000)
[Erratum-ibid.\  {\bf 42}, 2345 (2001)]
[hep-ph/9910546]; \\
ibid.,
Phys.\ Lett.\ B {\bf 474}, 153 (2000)
[Erratum-ibid.\  {\bf 480}, 419 (2000)]
[hep-ph/9912295].
%
\bibitem{xin}
Z.~z.~Xing,
Phys.\ Lett.\ B {\bf 487}, 327 (2000)
[hep-ph/0002246].
%
\bibitem{fre1}
M.~Freund, M.~Lindner, S.~T.~Petcov, A.~Romanino,
Nucl.\ Phys.\ B {\bf 578}, 27 (2000)
[hep-ph/9912457]; \\
%
M.~Freund, P.~Huber, M.~Lindner,
Nucl.\ Phys.\ B {\bf 585}, 105 (2000)
[hep-ph/0004085];\\
%
M.~Freund, M.~Lindner, S.~T.~Petcov, A.~Romanino,
Nucl.\ Instrum.\ Meth {\bf A451}, 18 (2000);\\
M.~Freund,
Phys.\ Rev.\ D {\bf 64}, 053003 (2001)
[hep-ph/0103300].
%
\bibitem{oh1}
T.~Ohlsson,
Phys.\ Scripta {\bf T93}, 18 (2001).
%
\bibitem{otasato}
T.~Ota and J.~Sato,
Phys.\ Rev.\ D {\bf 63}, 093004 (2001)
[hep-ph/0011234].
%
\bibitem{shrock}
See {\it for e.g.}
I.~Mocioiu and R.~Shrock,
Phys.\ Rev.\ D {\bf 62}, 053017 (2000)
[hep-ph/0002149]; \\
I.~Mocioiu and R.~Shrock,
JHEP {\bf 0111}, 050 (2001)
[hep-ph/0106139].
%
\bibitem{barger}
V.~D.~Barger, S.~Geer, R.~Raja and K.~Whisnant,
Phys.\ Rev.\ D {\bf 62}, 013004 (2000)
[hep-ph/9911524]; \\
ibid.,
Phys.\ Lett.\ B
{\bf 485}, 379 (2000)
[hep-ph/0004208].
%
\bibitem{kim}
K.~Kimura, A.~Takamura and H.~Yokomakura,
Phys.\ Lett.\ B {\bf 537}, 86 (2002)
[hep-ph/0203099]; \\
ibid.,
Phys.\ Rev.\ D {\bf 66}, 073005 (2002)
[hep-ph/0205295]; \\
Phys.\ Lett.\ B {\b 544}, 286 (2002)
[hep-ph/0207174].
%
\bibitem{bcr}
B.~Brahmachari, S.~Choubey and P.~Roy,
Nucl.\ Phys.\ B {\bf 671}, 483 (2003)
[hep-ph/0303078].
%
\bibitem{har}
P.~F.~Harrison, W.~G.~Scott and T.~J.~Weiler,
Phys.\ Lett.\ B {\bf 565}, 159 (2003)
[hep-ph/0305175].
%
\bibitem{akh}
E.~K.~Akhmedov, R.~Johansson, M.~Lindner, T.~Ohlsson and T.~Schwetz,
JHEP {\bf 0404}, 078 (2004)
[hep-ph/0402175].
%
\bibitem{prem}
A.~M.~Dziewonski and D.~L.~Anderson,
Phys. Earth Planet Int. 25 297(1981);
see http://solid\_earth.ou.edu/prem.html;\\
We use the parametrization given in
R.~Gandhi, C.~Quigg, M.~H.~Reno and I.~Sarcevic,
Astropart.\ Phys.\  {\bf 5}, 81 (1996)
[hep-ph/9512364].
%
\bibitem{fre3}
See the first reference in \cite{fre1}.
%



\bibitem{lind}
M.~Lindner,
Nucl.\ Phys.\ Proc.\ Suppl.\  {\bf 118}, 199 (2003)
[hep-ph/0210377];\\
M.~Lindner,
Springer Tracts Mod.\ Phys.\  {\bf 190}, 209 (2003)
[hep-ph/0209083].
%
\bibitem{barg}
V.~Barger, D.~Marfatia and K.~Whisnant,
Phys.\ Rev.\ D {\bf 65}, 073023 (2002)
[hep-ph/0112119]; \\
ibid.,
Phys.\ Rev.\ D {\bf 66}, 053007 (2002)
[hep-ph/0206038]; \\
ibid.,
Phys.\ Lett.\ B {\bf 560}, 75 (2003)
[hep-ph/0210428].
%
\bibitem{mina}
H.~Minakata and H.~Nunokawa,
JHEP {\bf 0110}, 001 (2001)
[hep-ph/0108085]; \\
T.~Kajita, H.~Minakata, H.~Nunokawa,
Phys.\ Lett.\ B {\bf 528}, 245 (2002)
[hep-ph/0112345]; \\
H.~Minakata, H.~Nunokawa and S.~J.~Parke,
Phys.\ Rev.\ D {\bf 66}, 093012 (2002)
[hep-ph/0208163].
%
\bibitem{yasu}
J.~Burguet-Castell, M.~B.~Gavela,
J.~J.~Gomez Cadenas, P.~Hernandez, O.~Mena,
Nucl.\ Phys.\ B {\bf 608}, 301 (2001)
[hep-ph/0103258]; \\
  M.~Aoki, K.~Hagiwara and N.~Okamura,
  Phys.\ Lett.\ B {\bf 606}, 371 (2005)
  [hep-ph/0311324];\\
O.~Yasuda,
  New J.\ Phys.\  {\bf 6}, 83 (2004)
  [hep-ph/0405005].
%
\bibitem{cerv}
A.~Cervera, A.~Donini, M.~B.~Gavela, J.~J.~Gomez Cadenas, P.~Hernandez, O.~Mena and S.~Rigolin,
Nucl.\ Phys.\ B {\bf 579}, 17 (2000)
[Erratum-ibid.\ B {\bf 593}, 731 (2001)]
[hep-ph/0002108].
%
\bibitem{doni}
A.~Donini, D.~Meloni and P.~Migliozzi,
Nucl.\ Phys.\ B {\bf 646}, 321 (2002)
[hep-ph/0206034].
%
\bibitem{magic}
P.~Huber and W.~Winter,
Phys.\ Rev.\ D {\bf 68}, 037301 (2003)
[hep-ph/0301257].
%
\bibitem{us2}
R.~Gandhi, P.~Ghoshal, S.~Goswami, P.~Mehta and S.~Uma Sankar,
work in progress.
%
\bibitem{cpt}
{\sl
Tests of CPT using atmospheric neutrinos in such a detector have been discussed
in}
A.~Datta, R.~Gandhi, P.~Mehta and S.~Uma Sankar,
Phys.\ Lett.\ B
{\bf 597}, 356-361 (2004)
[hep-ph/0312027].
%
\bibitem{debchou}
  D.~Choudhury and A.~Datta,
  JHEP {\bf 0507}, 058 (2005)
  [hep-ph/0410266].
%
\bibitem{probir}
S.~Choubey and P.~Roy,
Phys.\ Rev.\ Lett.\  {\bf 93}, 021803 (2004)
[hep-ph/0310316].
%
\bibitem{bartol}
V.~Agrawal, T.~K.~Gaisser, P.~Lipari and T.~Stanev,
Phys.\ Rev.\ D {\bf 53}, 1314 (1996)
[hep-ph/9509423].
%
\bibitem{monolith}
N.~Y.~Agafonova {\it et al.}  (MONOLITH Collaboration), Report No.
LNGS-P26-2000; http://castore.mi.infn.it/$\sim$monolith/.
%
\bibitem{ino}
See http://www.imsc.res.in/$\sim$ino and working group reports and talks therein.
%
\bibitem{akhm}
{\sl Core effects in atmospheric neutrinos have been discussed in}
S.~T.~Petcov, Phys.\ Lett.\ B {\bf 434}, 321 (1998)
[hep-ph/9805262];\\
E.~K.~Akhmedov, A.~Dighe, P.~Lipari, A.~Y.~Smirnov,
Nucl.\ Phys.\ B {\bf 542}, 3 (1999)
[hep-ph/9808270]; \\
M.~Chizhov, M.~Maris, S.~T.~Petcov,
hep-ph/9810501; \\
M.~V.~Chizhov and S.~T.~Petcov,
Phys.\ Rev.\ D {\bf 63}, 073003 (2001)
[hep-ph/9903424]; \\
J.~Bernabeu, S.~Palomares-Ruiz, A.~Perez, S.~T.~Petcov,
Phys.\ Lett.\ B {\bf 531}, 90 (2002)
[hep-ph/0110071].
%
\bibitem{us3}
R.~Gandhi, P.~Ghoshal, S.~Goswami, P.~Mehta and S.~Uma Sankar,
work in progress.


\end{thebibliography}
\end{document}